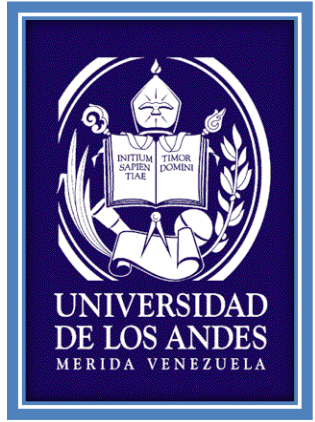

# Modelado Multi-agentes de Sistemas Dinámicos
## *-Un enfoque Auto-organizado, Emergente, Homeostático y Autopoiético-*

**Nelson Josué Fernández Parada**

# Modelado Multi-agentes de Sistemas Dinámicos
## *-Un enfoque Auto-organizado, Emergente, Homeostático y Autopoiético-*

**Nelson Josué Fernández Parada**

Tesis presentada como requisito parcial para optar al título de:
**Ph.D. en Ciencias Aplicadas**

Director:
Ph.D. José Lizandro Aguilar
Co-director:
Ph.D. Oswaldo Terán

Centro de Microelectrónica y Sistemas Distribuidos

Universidad de los Andes
Facultad de Ingeniería, Doctorado en Ciencias Aplicadas
Mérida, Venezuela
2015

# Dedicatoria

*A quienes conforman mi bello hogar: Yois, Tere y Samito.*

*A quienes conforman mi más amorosa familia: Benji, Chechi, Alix, Dora, Oswaldo, Sonia, Javier, Yamile; Jairo, Andrés, Diego, Santi; Alejandro, Jessica y Camilo Iván.*

*A mis amigos y colegas más cercanos: Carlos Gershenson, Alberto Ramírez, Diego Lizcano, Silvia Álvarez, Carlos Ernesto Maldonado, Adriana Maldonado, Andrés Gonzáles, Alí Coronel, Christopher Stephens, Tom Froese y Mario Cosenza.*

*Y finalmente, a la computadora y a la computación universal...*

# Agradecimientos

# RESUMEN

*Esta tesis presenta una propuesta de modelado de sistemas dinámicos (SD) usando múltiples agentes. El modelado se desarrolla desde una perspectiva que permite evaluar las propiedades que confiere a los SDC orden, cambio, adaptabilidad, equilibrio y autonomía. La base teórica la constituyen la teoría de sistemas multi-agentes, la teoría de grafos, y en especial, la teoría de la información. Además de la especificación de un SD como una red computacional de agentes, se logró el establecimiento de métricas que permiten caracterizar y evaluar su complejidad inherente. El resultado fue la definición, formalización y medición de las propiedades de emergencia, auto-organización, complejidad, homeostasis y autopoiesis. La valoración del modelo de SD propuesto se hizo en sistemas discretos (Redes Booleanas Aleatorias y Autómatas Celulares) y sistemas ecológicos. En particular, el aporte de esta tesis fue el desarrollo de un enfoque metodológico de modelado de SD con un conjunto mayor de propiedades. Estas propiedades fueron la emergencia, auto-organización, complejidad, homeostasis y autopoiesis, a partir de las cuales se ahondó en los aspectos de cambio, regularidad, adaptabilidad, equilibrio dinámico y autonomía en SD. Lo anterior se logró desde una base sencilla de cálculo e interpretación basada en la teoría de la información, que no requiere de conocimiento matemático avanzado, lo cual hizo más práctica su aplicación en diversos campos como las ciencias computacionales, el urbanismo, y en especial, la ecología.*

***Palabras clave:*** *Ciencias de la complejidad, teoría de la información, caos, criticalidad, autonomía, sistemas ecológicos.*

# ABSTRACT

*This thesis presents the theoretical, conceptual and methodological aspects that support the modeling of dynamical systems (DS) by using several agents. The modeling approach permits the assessment of properties representing order, change, equilibrium, adaptability, and autonomy, in DS. The modeling processes were supported by a conceptual corpus regarding systems dynamics, multi-agent systems, graph theory, and, particularly, the information theory. Besides to the specification of the dynamical systems as a computational network of agents, metrics that allow characterizing and assessing the inherent complexity of such systems were defined. As a result, properties associated with emergence, self-organization, complexity, homeostasis and autopoiesis were defined, formalized and measured. The validation of the underlying DS model was carried out on discrete systems (boolean networks and cellular automata) and ecological systems. The central contribution of this thesis was the development of a methodological approach for DS modeling. This approach includes a larger set of properties than in traditional studies, what allows us to go deepen in questioning essential issues associated with the DS field. All this was achieved from a simple base of calculation and interpretation, which does not require advanced mathematical knowledge, and facilitates their application in different fields of science.*

**Keywords**: *Complex sciences, information theory, chaos, criticality, autonomy, ecological systems.*

# Contenido

# Lista de figuras

# Lista de tablas

# Introducción

Desde una perspectiva amplia, se puede estimar que todo lo que nos rodea presenta algún grado de dinamismo, es decir de cambio. Aunque puede resultar ser una cuestión de escala (nivel de abstracción), el cambio se da incluso en entidades cómo las rocas, que lo hacen a escalas de tiempo geológico.

El cambio en los diversos sistemas naturales y artificiales, ha llamado la atención de la comunidad científica durante mucho tiempo. A pesar que cada día la renovada y aumentada capacidad computacional nos ha permitido acercarnos a su entendimiento, los aspectos relativos a su complejidad ha limitado la posibilidad de su predicción. Mucho de ello se debe a que en los sistemas cambiantes, los comportamientos globales vienen de la diversidad de interacciones surgidas cómo resultado de la naturaleza de sus elementos. Por ello, la ciencia tradicional y su enfoque reduccionista necesita hoy en día complementarse, ahondar en las interacciones, para acercarse más al entendimiento de un sistema dinámico con rasgos de complejidad.

Si bien conocemos en diferentes áreas y disciplinas mucho acerca de las partes, hoy tenemos la posibilidad de conocer y ahondar más en las interacciones. El panorama que surge se hace muy interesante. Uno donde podemos armonizar diversos aspectos del universo, de la vida y la computación, tales como la materia, la energía, la cultura, la tecnología, entre muchos otros. Sobre esta base se puede llegar a descifrar los aspectos ocultos del comportamiento y dinamismo en los problemas actuales en el campo de la biología, ecología, sociología y economía.

Cómo fundamento para el entendimiento de la complejidad, tenemos que el dinamismo surge desde las interacciones auto-organizantes de las partes. Las interacciones permiten mantener la integridad del sistema, gracias a la autonomía y la autorregulación de las partes. Estos rasgos, renombrados cómo: emergencia, auto-organización autopoiesis y homeostasis, son los que le proporcionan a los sistemas dinámicos la suficiente robustez y posibilidad de cambio ante las variaciones internas y externas. Es decir, le dan la complejidad necesaria para mantener su adaptación y evolución en un contexto de variedad requerida (Ross Ashby, 1956). En ese contexto, se hace posible una refinación y redefinición de las propiedades y conceptos necesarios para la descripción apropiada de un sistema dinámico compuesto de múltiples entidades y en diferentes niveles de abstracción. Este enfoque resulta promisorio, por la posibilidad de contribuir al entendimiento de algunos de los retos actuales, conocidos cómo las 3 principales emergencias en la historia de la humanidad, planteados en (Gershenson, 2013a): (i) la transición entre lo no vivo y lo vivo, (ii) el surgimiento de la conciencia y (iii) la naturaleza del espíritu humano (cómo el máximo nivel de conciencia).

# Motivación

En el estudio de los sistemas dinámicos con múltiple número de agentes, han sido apreciables los esfuerzos por generar nuevas nociones, formalismos y metodologías que nos permitan describir los conceptos de emergencia, auto-organización y complejidad (Aguilar, 2014; Bar-Yam, 2004a; Gershenson and Heylighen, 2003; Heylighen, 2011; Koza, 1995; Shalizi et al., 2004). Sin embargo, ha surgido la necesidad de valorar otros aspectos complementarios, cómo la homeostasis y la autopoiesis. Es decir, existe la posibilidad de valorar un número mayor de propiedades en los *SD*, con el fin de lograr un mejor entendimiento del resultado global de las interacciones que dan origen a su estructura y dinámica.

Sobre la anterior base, la investigación doctoral aquí desarrollada buscó lograr una mejor interpretación de las dinámicas colectivas en los *SD*, y ahondar en el entendimiento de los aspectos relativos a su complejidad. Para ello, se enfocó en generar métricas precisas y formales que representaran adecuadamente los conceptos de emergencia, auto-organización, complejidad, homeostasis y autopoiesis ($EACHA$), para fines de modelado de *SD*. En este sentido, se buscó generar una contribución al conocimiento de los *SD* que permitiera clarificar el significado de los conceptos que $EACHA$ describiesen. Para tal fin, se diseñaron aplicaciones a fenómenos computacionales y naturales que nos posibilitaran estudiar y analizar el sentido práctico de las formalizaciones a desarrollar. El alcance de los resultados obtenidos en esta tesis, se orientó a brindar una visión de los *SD* que fuese más allá de los límites del reduccionismo. Nuestro esfuerzo estuvo dirigido a ayudar a responder a las preguntas esenciales en cuanto a la complejidad en *SD*, que la comunidad científica se viene planteando. Todo ello, a través de un enfoque interdisciplinar, soportado por una red científica colaborativa internacional.

# Objetivos

*General*

Desarrollar un enfoque metodológico para sistemas dinámicos complejos, que considere la exploración, generación y modelado de los aspectos de auto-organización, emergencia, complejidad, homeostasis y autopoiesis.

*Específicos*

- Definir formalmente un sistema multi-agente como un sistema dinámico

- Proponer nuevas nociones, con sus formalismos, para la caracterización de los sistemas multi-agente como sistemas dinámicos.

- Desarrollar casos de estudio inspirados, entre otros, en un sistema socio-ecológico de referencia.

# Antecedentes

En décadas recientes, el estudio de los sistemas complejos ha incrementado el entendimiento de un amplio número y rango de fenómenos. Sus fundamentos matemáticos se hallan en las teorías del caos, de la información y de la computación. Se han tratado de explicar desde el estudio de los fractales, los mapas iterativos y la termodinámica. Igualmente, se han desarrollado modelos y técnicas de modelado de gran utilidad, que se han basado en técnicas analíticas y simulación computacional. Se incluyen allí, la mecánica estadística, la dinámicas estocástica, los autómatas celulares y el modelado multi-agentes, por mencionar algunos. (Bar-Yam, 2003; Mitchell, 2009).

Los conceptos usados para clarificar el estudio de los *SD*, tales como emergencia, adaptabilidad, auto-organización y complejidad, han sido usados en diferentes contextos con diferentes significados. Los tópicos abordan casos como:

- El análisis de la medidas de complejidad (Crutchfield and Young, 1989; Edmonds, 1999; Lloyd, 2001), en donde se inspeccionan desde la filosofía del modelado, hasta las formas de medir la complejidad, su perspectiva estadística, y la clasificación temática de las diferentes medidas.
- El análisis de sistemas naturales (incluidos ecosistemas) desde la teoría de la computación y la información, para cuantificar la complejidad estructural. En específico, el estudio del fenómeno de adaptación desde simulaciones computacionales de sistemas naturales, da el entendimiento para el diseño y construcción de sistemas complejos (Crutchfield, 2011; Koza, 1995).
- El estudio de la irreductibilidad de la complejidad, explicada desde el contenido de información algorítmica, dada cuando en una cadena de bits la información contenida es igual a su tamaño. En estos casos, la cadena de bits es incompresible y no tiene redundancia (Chaitin, 2007).
- El análisis de la estructura y función de redes complejas de ámbitos biológicos o sociales, partiendo de variadas técnicas de modelado y sobre la base de indicadores estadísticos para su caracterización (Newman, 2003).

- El estudio de la auto-organización como la generación de la propia estructura en organismos biológicos (Camazine et al., 2001), considerando que la auto-organización y la emergencia pueden coexistir como fenómenos separados dado que cada una enfatiza en diferentes aspectos del sistema (De Wolf and Holvoet, 2005), la auto-organización de sistemas fuera del equilibrio (Nicolis and Prigogine, 1977), y la apreciación de la complejidad como una condición intermedia entre el caos y el orden, relacionándose así con la auto-organización (Heylighen, 2011).
- La caracterización de la dinámica auto-organizante de estructuras complejas en la sociedad y la naturaleza, en aspectos relacionados como la evolución de la complejidad, la evolución de la optimización y el estudio de casos como los medios de transporte (Schweitzer, 1997a).
- La evaluación de las complejidades urbanas y ecológicas que surgen de dinámicas locales a globales (Bottom-Up), desde estas perspectivas, las ciudades son observadas como organismos vivos (Batty, 2012; Batty, 1971). Al mismo tiempo, en la ecología se valora la contribución del enfoque de los sistemas complejos a las dinámicas de acoplamiento del habitat, a las fluctuaciones poblacionales, y a las interacciones en redes tróficas. Desde este punto, se ve como el estudio de la complejidad en ecología es un campo naciente (Parrott, 2010).
- El modelado individual basado en dinámicas colectivas de sistemas físicos de partículas que siguen el movimiento Browiniano (Schweitzer, 1997b).
- Los intentos de unificación desde la teoría de la información del concepto de auto-organización, y no sólo eso, sino la relación de ésta última con la teoría del caos (Haken, 1989). Desde la perspectiva de Haken (1989) y Prokopenko (2009a), se aborda la existencia de 3 aspectos de la auto-organización para la vida artificial: (a) la evolución del sistema desde la interacción de sus componentes a estados de mayor organización, (b) la manifestación de esta organización en una coordinación y manifestación global, (c) la expresión de patrones no impuestos por influencias externas.
- Las necesidades de un nuevo paradigma para la complejidad desde varias teorías, que satisfagan diversas aplicaciones en distintos sistemas (Morin, 1992).
- El análisis de la universalidad de la complejidad y su irreductibilidad computacional, que limita la predictibilidad del sistema, por cuanto existe gran diversidad de interacciones locales que brindan apreciable riqueza en el comportamiento del mismo (Wolfram and Gad-el-Hak, 2003; Wolfram, 1984).

Como punto de coincidencia, muchos de los anteriores autores tocan una diversidad de nociones, en especial de complejidad, emergencia y auto-organización. Se refieren de manera apreciable a la ambigüedad, confusión, e incluso abuso, de la terminología, en discursos no científicos (Edmonds, 1999).

En el caso de los sistemas multi-agentes, han existido trabajos seminales que han permitido consolidar su teoría. Entre ellos se destacan los aportes teóricos y prácticos de Wooldridge and Jennings (1995), para el diseño y construcción de agentes inteligentes. En el marco de la teoría de agentes se define que es uno y el uso de formalismos matemáticos para su representación y la de su razonamiento. Se abordan también las arquitecturas y modelado desde la ingeniería del software, y el caso no menos importantes de los lenguajes y sistemas de programación y experimentación de agentes. Por su parte, Ferber and Gutknecht (1998), proponen un meta-modelo de organización artificial entre agentes llamado AALAADIN. Basados en AALAADIN, proponen la plataforma MADKIT para apoyar el modelado con agentes heterogéneos. Ante el auge tomado por los sistemas multi-agentes, se ha analizado su vinculación con terminos como situabilidad, flexibilidad y autonomía, para clarificar sus nociones y sus aspectos metodológicos de representación ( Jennings et al. 1998; Teran 2001). Ferber et al. (2004) han avanzado en el modelado con la propuesta de principios generales para sistemas multi-agentes de organización centralizada (OCMAS en Inglés).

Trabajos más actuales se han esforzado en involucrar las propiedades de complejidad, auto-organización y emergencia en los sistemas multi-agentes. Muchos de ellos han buscado el desarrollo de arquitecturas, plataformas de modelado y modelos específicos, para que sea viable representar, reproducir y medir la auto-organización y la emergencia en sistemas multi-agentes. (Aguilar et al., 2007; Aguilar, José, Besembel, I., Cerrada, M., Hidrobo, F., Narciso, 2008; Perozo, Niriaska, Aguilar, José, Terán, O, Molina, 2012a, 2012b).

# Organización de la tesis

La tesis tiene una organización preferencialmente secuencial, de manera que los elementos de un capitulo precedente brinda elementos para el entendimiento de los demás.

En la siguiente sección se dan los elementos principales de la teoría de los *SD*, y las nociones básicas de auto-organización, emergencia, complejidad, homeostasis y autopoiesis (*AECHA*). Los elementos de los *SD* incluyen a la complejidad como una mirada complementaria para su estudio, que va más allá del reduccionismo tradicional; por su parte, las nociones *AECHA* sientan la base conceptual de la formalización.

En el capítulo 2 se discute como puede llegarse a la especificación de un sistema dinámico como una red computacional de agentes. Para ello se propone integrar los elementos de *AECHA*, y se finaliza con varios formalismos que también integran la teoría de *SD*, de multi-agentes y de grafos.

En el capítulo 3 se presentan los formalismos desarrollados para cada noción de *AECHA*. Allí se muestra la riqueza de los formalismos, que dio origen a diversos indicadores. En ese capítulo las nociones derivadas de la teoría de la información para *AECHA*, se constituyen como el eje central de la tesis.

El capítulo 4 presente aplicaciones en sistemas discretos, como en las redes booleanas aleatorias, y en autómatas celulares elementales; además, contiene una aplicación urbana en un modelo de tráfico que se basa en tres reglas de autómata celular elementales, de gran utilidad para la planeación en ciudades vivas y auto-organizantes.

El capítulo 5 discute y analiza el sentido ecológico de la complejidad, a partir de tres aplicaciones a sistemas ecológicos, ecosistemas acuáticos, y el análisis de comunidades de mamíferos. Se destaca que este capítulo presenta importantes consideraciones para incluir a la complejidad como otro indicador ambiental, útil para el de los ecosistemas.

El capítulo 6 finaliza la tesis con las conclusiones y trabajos futuros.

# CAPITULO 1: SISTEMAS DINÁMICOS; AUTO-ORGANIZACIÓN, EMERGENCIA, COMPLEJIDAD, HOMESTASIS Y AUTOPOIESIS.

## Resumen

*Este capítulo presenta los aspectos teóricos y conceptuales básicos acerca de los sistemas dinámicos complejos (SD). Desde una perspectiva histórica, se discuten el elemento de su incertidumbre; a partir de allí, se ahonda en la inclusión de la dimensión de la complejidad. Con lo anterior como base, se muestran nuestras consideraciones acerca de las características de complejidad, emergencia, auto-organización, homeostasis y autopoiesis en SD. Este último apartado tiene como fin brindar una visión lo más clara posible de cada una de ellas, con el objeto de contribuir a su desambiguación y sentar las bases para su posterior modelado matemático.*

## 1.1 Teoría de los Sistemas Dinámicos (*TDS*)

En esta tesis observamos la *TDS* como el área de las matemáticas usadas para describir el comportamiento de sistemas dinámicos complejos (Mitchell, 2009). En consecuencia las generalizaciones y nociones que se presentan en apartados subsiguientes se enfocan en este tipo de descripción.

### 1.1.1 Visión General

La teoría de sistemas dinámicos (*TSD*), concierne con la descripción y predicción de sistemas que exhiben un comportamiento complejo en un nivel macroscópico, que emerge desde las acciones colectivas de interacciones de sus componentes. La palabra *dinámica* es referida al cambio, y los sistemas dinámicos (*SD*), efectivamente, cambian en el tiempo de alguna forma. Existen variados ejemplos de *SD,* como lo son: el sistema solar, el corazón y el cerebro de las criaturas vivas, la población mundial, o el cambio climático. No obstante, ejemplos como las rocas también son un tipo de *SD,* nada más que sus cambios se dan a escalas de tiempo geológico (Mitchell, 2009).

La *TSD* describe en términos generales las formas en las que el sistema puede cambiar, qué tipo de comportamientos macroscópicos son posibles, y qué tipos de predicciones se pueden hacer acerca de estos comportamientos. Actualmente esta teoría está en boga debido a los fascinantes resultados obtenidos de uno de sus productos intelectuales, el estudio del caos (Crutchfield, 2011).

### 1.1.2  Aspectos Históricos

El origen de la *TSD* se remonta a tiempos aristotélicos, dado que este filósofo fue autor de varias teorías de movimiento, una de ellas aceptada ampliamente por más de 1.500 años. Esta teoría básicamente definió dos principios (Gyekye, 1974), (i) el movimiento de la tierra difería del movimiento de los cielos y (ii) los objetos terrestres se movían de diferente forma en dependencia de lo que estuvieran hechos. Posteriormente, los aportes de Galileo sobre el movimiento de los planetas desde bases experimentales, contribuyeron a sentar muchos principios que hemos conocido en diversos cursos de física. Después de la muerte de Galileo, en ese mismo año, nació Newton, la persona más importante en la historia de la dinámica, dado que inventó una rama de las matemáticas que describe el movimiento y el cambio (Madhusudan, 2010). Los físicos de la época llamaron al estudio del movimiento *Mecánica* (Verlinde, 2011), la cual podía explicarse en términos de acciones combinadas de *máquinas* simples como palancas, poleas, ruedas y ejes. En particular, un trabajo clave en esta área es el de Newton, conocido como Mecánica Clásica. La mecánica clásica se divide en dos ramas (Christianson, 1998): (i) la cinemática o cómo las cosas se mueven, y (ii) la dinámica que se ocupa por explicar el movimiento de las cosas como producto de las leyes de la cinemática. Newton pudo explicar el *cambio* en el movimiento elíptico de orbitas planetarias en términos de la fuerza, llamada gravedad. Desde la mecánica newtoniana, que produjo una figuración del "mecanismo del universo", Pierre Sión Laplace vio la importancia que tenía esta figuración para la predicción; por lo que mencionó que desde las leyes de Newton, y al tener la velocidad y posición de una partícula, era posible, en principio, predecir todo (Rivas Lado, 2001).

### 1.1.3  La Incertidumbre de la Previsibilidad

Dos descubrimientos en el *siglo XX* mostraron que esa capacidad predictiva definida por Laplace no era posible. El primero fue el descubrimiento del *principio de incertidumbre*, en mecánica cuántica, de Werner Heisenber en 1927, que estableció que no se puede hacer medición de la posición y el *momentum* (masa y velocidad) de una partícula al mismo tiempo. El segundo fue el descubrimiento del Caos; los Sistemas Caóticos son aquellos en que minúsculas incertidumbres en la medición de la posición inicial, resultan en grandes errores en términos de predicción a largo plazo. Esto es conocido como la sensibilidad a las condiciones iniciales. Comportamientos caóticos han sido observados en desordenes cardiacos, turbulencias de fluidos, circuitos electrónicos, grifos que gotean, entre otros. El matemático Francés Henri Poincaré, fundador de la topología algebraica, demostró la sensibilidad de las condiciones iniciales en el problema de los tres cuerpos, que trataba de predecir las posiciones futuras de varias masas que se atraían arbitrariamente una a la otra bajo las leyes de Newton. Este hecho marcó uno de los momentos importantes en la historia de la *TSD*, pues puso en duda la presuposición científica de previsibilidad, y demostró los límites de la previsibilidad heredada hasta ese momento y evidenciada en los trabajos de Galileo, Newton y Laplace (Gershenson, 2011).

En nuestros días, la conciencia de falta de previsibilidad de muchos de los *SD* ha llevado a considerar no solo las partes, sujetos de especial interés desde el método científico tradicional, sino también las múltiples interacciones que existen entre ellas. Esta gran cantidad de interacciones son las que crean sistemas entretejidos cuya dinámica es difícil de separar. Son las interacciones la característica más importante de muchos *SD*, dado que el comportamiento de un elemento depende de otro, y así el de muchos otros, de forma tal que el rasgo que sobresale es la interdependencia entre elementos del sistema. Esto es lo que hace que los *SD* se tornen difíciles de predecir, que se tornen complejos, y por ende, que se clasifiquen como tal (Bar-Yam, 2003; Gershenson, 2011).

### 1.1.4 Implicaciones Matemáticas

Típicamente, se considera que un sistema está "resuelto" cuando se llega a un conjunto finito de expresiones finitas, que pueden ser usadas para predecir su estado en un tiempo $t$, dado el estado del sistema en algún tiempo inicial $t_0$, o anterior $t_1$. Las soluciones halladas de esta forma, no son generalmente apropiadas o posibles para muchos *SD*. En este sentido, la contribución central de la *TDS* a la ciencia moderna es que las soluciones exactas no son necesarias para entender y analizar los procesos no lineales. Así, en lugar de hallar soluciones exactas o descripciones estadísticas ajustadas, el énfasis de la *TDS* está en describir la geometría y la estructura topológica del conjunto de soluciones. Es decir, la *TDS* da una visión geométrica de un proceso estructural de elementos, como atractores, cuencas y separatrices; hecho que la distingue del enfoque puramente probabilístico de la mecánica estadística, en la que la estructura geométrica no es considerada. En adición, la TDS también direcciona las preguntas acerca de qué estructuras genéricas se hallan, esto es, qué tipos de comportamientos son típicos a través del espectro de los *SD* (Mitchell, 2009; Mitchell et al., 1994).

## 1.2 Noción General de *SD*

A continuación se hace referencia a las descripciones matemáticas que permiten de manera idealizada representar muchos de los fenómenos reales que acontecen a nuestro alrededor y que tienen como característica principal su cambio y evolución en el tiempo. Desde esta consideración básica, se han definido dos componentes para un *SD* (Scheinerman, 1996): (i) un vector de estado que describe su estado actual, y (ii) una función-regla por medio de la cual, dado el estado actual, se puede obtener el estado del sistema en el siguiente instante de tiempo. Desde esta perspectiva, los *SD* se han definido como: sistemas que cambian o evolucionan su estado ($q$) en el tiempo ($t$), en razón a una regla de evolución ($R$) que lo demarca. Esta regla puede tener la estructura $R: Q \times T \rightarrow Q$, donde $Q$ es el conjunto de estados posibles y $T$ el tiempo. Así, $R$ conduce a un estado $q \in Q$. De acuerdo a como se especifique la regla se pueden dar diferentes tipos de *SDs*.

Desde nuestra perspectiva, un tanto más amplia e inscrita en el marco de la *TDS* expuesta en el ítem 1.1., un *SD* puede ser observado como un sistema dependiente del tiempo, tal que las interacciones de sus elementos según sus estados locales, determinan los estados globales del sistema.

La tabla 1-1 lista los más conocidos tipos de *SD* reportados en la literatura, acorde con la variación del tiempo, su linealidad o no, y su naturaleza determinista.

**Tabla 1-1 Tipos de Sistemas Dinámicos (*SD*)**

| Tipo de *SD* | Característica | Formalismo/Modelo de Ejemplo |
|---|---|---|
| Discreto | El tiempo transcurre en lapsos pequeños, por lo que el cambio se da en instantes separados de tiempo. $t \in \mathbb{R}$ | Ecuación logística * $q_{t+1} = rq_t(1 - q_t)$ Ej. Crecimiento poblacional ($r$), limitado por la capacidad de carga ($K$) del sistema, que corresponde con la cantidad de elementos que puede soportar el mismo. |
| Continuo | El tiempo transcurre de manera continua, y así el cambio. $t \in \mathbb{Z}$ | Ecuación diferencial* $\frac{dx}{dt} = rq(1 - x)$ Ej. Modelo de predador-presa de Lokta-Volterra |
| Lineales | Efecto proporcional a la causa. Opera el principio de superposición (Si se conocen dos soluciones para un sistema lineal, la suma de ellas es también una solución). $f(x + y) = f(x) + f(y)$ | Ecuación lineal* $q_{t+1} = rq_t$ Ej. Sistema Masa-Resorte (Oscilador Armónico Simple) |
| No lineales | Efecto **no** proporcional a la causa. Opera el principio de **no** superposición, es decir: $f(x + y) \neq f(x) + f(y)$. | Péndulo doble, movimiento descrito por ecuaciones de Lagrange. |
| Deterministas | Regla $R$ puede ser funciones iterativas. El determinismo en *SD* permite generar cierta predictibilidad sobre el sistema basado en las relaciones causa-efecto presentes en él, no obstante, dicha predictibilidad tiene sus límites. | $R$ está dada por $R(q(t), r) \rightarrow q(t + \tau)$, donde $q$ es un estado del conjunto de estados posibles $Q$, $r$ un parámetro de la regla y $\tau$ es un lapso de tiempo. De esta manera, dado un estado $q(0)$, la evolución estará determinada $\forall t$. |

| | | Ej. La planificación de una línea de producción. |
|---|---|---|
| Estocásticos | Abarcan eventos aleatorios; su evolución puede ser estimada de manera probabilística. | La variable de estado $q(t)$ es aleatoria, cuyo dominio son los reales, de acuerdo con una función de probabilidad. Ej. el tiempo de funcionamiento de una máquina entre avería y avería, su tiempo de reparación, el tiempo que necesita un operador humano para realizar una determinada operación |

*Donde $t$ es tiempo, $q$ es la variable de estado y $r$ parámetro que afecta la variable $q$.

## 1.3 La Complejidad en Sistemas Dinámicos: Propiedades de Auto-organización, Emergencia, Complejidad, Homeostasis y Autopoiesis.

El punto central del estudio de la complejidad en los Sistemas Dinámicos (*SD*), está en entender las leyes y mecanismos por los cuales un comportamiento coherente global y difícil de entender, emerge desde las actividades colectivas de componentes locales, relativamente simples. Dada la diversidad de sistemas que se incluyen en la categoría de complejos, el descubrimiento de cualquier rasgo común o ley universal se hace valiosa para conformar marcos teóricos generales (Mitchell, 2009).

Actualmente, para observar a un *SD* bajo la óptica de su complejidad, se requiere de mayores conceptos y profundizar en sus aspectos formales. Se requiere, también, de sofisticadas herramientas, tanto analíticas como computacionales. En este sentido, la representación de un *SD* con un número mayor de propiedades que faciliten su análisis se hace apreciable (Bar-Yam, 2003; Bersini et al., 2010; Luisi, 2007). Propiedades como el caos (sensibilidad a las condiciones iniciales), la no linealidad (no proporcionalidad entre causa y efecto), la criticalidad auto-organizada (orden desde condiciones críticas), la auto-organización en sistemas fuera del equilibrio, han estado entre las más consideradas en este campo (Bak et al., 1988; Langton, 1990; Nicolis and Prigogine, 1989).

Ahora bien, tradicionalmente la emergencia y la auto-organización han constituido los fenómenos más utilizados para explicación de la complejidad en *SD*. Sin embargo, tales propiedades podrían no ser suficientes para entender, entre otros, la adaptabilidad y la evolución espacio-temporal de los *SD*. Así, se ha observado como pertinente ahondar en otras propiedades referidas al equilibrio y a la autonomía, como lo son las propiedades de homeostasis y de autopoiesis (Fernández et al., 2010a). La *homeostasis* (equilibrio dinámico, en general) y la *autopoiesis* (auto-producción, auto-mantenimiento) constituyen propiedades emergentes que dan soporte al proceso de auto-organización. El sustento de ello está en que: (i) la capacidad auto-organizante del *SD* de adaptarse a ambientes cambiantes, acorde con sus necesidades de sobrevivencia y/o mejoramiento de su funcionamiento colectivo al interior del *SD*, se logra sobre la base del establecimiento de un equilibrio dinámico u homeostasis. Es decir, la auto-organización se respalda en mecanismos de autorregulación (Martius et al., 2007). (ii) Adicionalmente, la capacidad auto-organizante de los *SD* se apoya en procesos autopoiéticos (Maturana and Varela, 1980) que le confieren la propiedad de autonomía. La autonomía, es inherente a sistemas cerrados que se auto-referencian, se auto-constituyen en una dinámica de continua producción de sí mismos. A través de la homeostasis y la autopoiesis, se puede desarrollar, conservar, producir y sustentar la propia organización en el tiempo (Garciandía, 2005), desde un enfoque de auto-mantenimiento del *SD*.

Desde la anterior perspectiva, es posible definir un *SD* como un sistema con rasgos de emergencia, auto-organización, complejidad, homeostasis y autopoiesis (*EACHA*) (Fernández et al., 2010a). A continuación presentamos una síntesis de las nociones básicas de *EACHA* que, desde nuestra visión, contienen los elementos para su posterior formalización.

### 1.3.1   Emergencia

La emergencia es uno de los conceptos que más mal entendido y mal usado ha sido en décadas recientes. Las razones para ello son variadas e incluyen: polisemia (múltiples significados), moda o deseo de impresionar (buzzwording), confusión, Platonismo, y aún misticismo. Podemos decir que la controversia de su concepto ha durado décadas. Aun así, el concepto de emergencia puede ser definido desde ciertas miradas (Anderson, 1972) (Aguilar, 2014). Ahora bien, en general las propiedades de un sistema son emergentes, si ellas no están presentes en sus componentes. En otras palabras, propiedades globales que son producidas por las interacciones locales son emergentes. La emergencia se da a una escala dada, y no puede ser descrita sobre la base de las propiedades de una escala inferior (Fernández et al., 2011).

Por ejemplo, la temperatura de un gas se puede decir que es emergente (Shalizi, 2001), pues sus moléculas no poseen esta propiedad: la temperatura es una propiedad colectiva. Igualmente, en una pieza de oro emergen propiedades tales como la conductividad, maleabilidad, color, brillo; las cuales no pueden ser descritas a partir de las propiedades de los átomos de oro (Anderson, 1972). De una manera amplia e informal, la emergencia puede ser vista como las diferencias que son observados, en los sistemas, a diferentes escalas (Prokopenko et al., 2009).

Se estima que algunas personas podrían percibir dificultades en describir fenómenos a diferentes escalas (Gershenson, 2013b), pero ello puede ser la consecuencia de tratar de encontrar una única "verdadera" descripción de un fenómeno. Dado que los fenómenos no dependen de las descripciones que hacemos de ellos, podemos tener múltiples y diferentes descripciones del mismo fenómeno. Como alternativa, es más informativo manejar diferentes descripciones a la vez, lo que actualmente es más apropiado y necesario cuando se estudian sistemas complejos.

### 1.3.2 Auto-organización

Desde la mitad del siglo pasado, la auto-organización ha sido vista como un evento de cambio de la propia organización, sin que una entidad externa influya en el proceso (Ross Ashby, 1947a). Su base está en las reglas locales que conducen al sistema a producir una estructura o comportamientos específicos (Georgé et al., 2009).

Este proceso viene desde la dinámica interna del sistema, definida por las interacciones entre los elementos. Su resultado es un grado determinado de regularidad, que puede derivar en la expresión de un conjunto de patrones (Ebeling, 1993; Eriksson and Wulf, 1999). La auto-organización es un proceso por el cual se alcanza dinámicamente la función y el comportamiento global del sistema, a partir de las interacciones de sus elementos. Tal función o comportamiento no es impuesta por uno o pocos elementos, ni determinada jerárquicamente. Así, la auto-organización se adquiere autónomamente por la interacción de los elementos que producen, a su vez, retro-alimentaciones que *regulan* el sistema (Gershenson, 2007). Desde esta perspectiva, se puede ver la auto-organización como un proceso dinámico, adaptativo, que favorece el mantenimiento de la estructura del sistema, en el que se pueden expresar patrones de comportamiento para hacer frente a los cambios ambientales (Perozo, 2011; Perozo, Niriaska, Aguilar, José, Terán, O, Molina, 2012b).

En la naturaleza, la auto-organización ha sido usada para describir los procesos básicos por los cuales se forman enjambres, cardúmenes de peces, parvadas de patos o el tráfico vehicular. En general ha sido descrita en muchos sistemas, donde las interacciones locales además de generar un orden, derivan en la manifestación de un comportamiento global (Camazine et al., 2001; Gershenson, 2007).

Cabe destacar que la generación de patrones como un fenómeno emergente vía auto-organización, ha constituido un punto de discusión importante. Esto ha llevado a que la auto-organización y la emergencia hayan sido utilizadas de manera intercambiable. Sin embargo, autores como De Wolf and Holvoet (2005) han expresado que se puede dar auto-organización con y sin emergencia; así como la producción de auto-organización y emergencia al mismo tiempo; o de la emergencia sin la auto-organización.

En general, casi cualquier sistema dinámico puede ser visto como auto-organizante: sí su dinámica conduce a un atractor, que puede ser visto como u estado más probable. Podemos decidir llamar ese atractor cómo "organizado". Así, los sistemas dinámicos con un atractor (o más) tenderán a él, de tal manera que el sistema incrementará, por sí mismo, su propia organización (Ross Ashby, 1947b). Sobre esta base, y de manera práctica, podríamos describir casi que cualquier sistema como auto-organizante, más que cómo auto-organizado o no (Gershenson and Heylighen, 2003). Lo auto-organizante muestra más el rasgo dinámico, activo de la auto-organización.

La conveniencia de tener una medida de auto-organización que capture la naturaleza de la dinámica local en una escala determinada, es especialmente relevante para nuestro tiempo (Ay et al., 2012; Prokopenko, 2009b). Esto es especialmente relevante en el naciente campo de la auto-organización guiada u orientada (AOG) (Ay et al., 2012; Prokopenko, 2009b). La AOG puede describir cómo direccionar la dinámica auto-organizante de un sistema hacia una configuración deseada (Gershenson, 2012a). Esta configuración deseada puede no siempre ser el atractor natural de un sistema controlado. El mecanismo para guiar la dinámica, y el diseño de tal mecanismo, se beneficiará desde la medición y caracterización de la dinámica de un sistema en una forma precisa y concisa.

### 1.3.3 Complejidad

Hay docenas de nociones y medidas de complejidad, propuestas en diferentes áreas con diferentes propósitos (Edmonds, 1999; Lloyd, 2001). Etimológicamente, complejidad viene del latín *plexus*, lo cual significa *entrelazado*. Algo complejo es algo difícil de separar. Esto significa que los componentes del sistema son interdependientes, de manera que su futuro está parcialmente determinado por sus interacciones (Gershenson, 2013b). Por lo tanto, estudiar los componentes por separado – como se acostumbra desde el enfoque reduccionista- no es suficiente para describir la dinámica de los sistemas dinámicos complejos. Esto, además, dificulta la predictibilidad de los *SD*.

Un ejemplo de complejidad clásico se halla en el conocido "Juego de la Vida" de Jhon Conway (Berlekamp et al., 1982), en el cual, 4 reglas simples generan dinámicas ricas como estructuras estables, móviles y/u oscilatorias, que son difíciles de predecir.

Tradicionalmente se ha pensado que la complejidad se refiere a aquella condición que limita al modelista para formular el comportamiento completo del *SD*, en un lenguaje dado. En realidad, lo complejo se debe a las interacciones en el *SD* que generan información nueva y relevante—no presente en las condiciones iníciales ni de frontera— y que influyen en el desarrollo del *SD*. La complejidad, por tanto, puede estar asociada a los aspectos emergentes y auto-organizantes del *SD*.

Dada la dificultad existente en la adecuada interpretación de la noción de complejidad, y más aún en su medición, tener medidas globales de complejidad es una tarea actual, de gran importancia y utilidad.

Una medida útil de complejidad debe estar disponible para permitirnos responder preguntas como: ¿Es un desierto más o menos complejos que una paramo? ¿Cuál es la complejidad de diferentes brotes de influenza? ¿Cuáles organismos son más complejos: predadores o presas; parásitos u hospederos; individuos o sociedades? ¿Cuál es la complejidad de situaciones tan disimiles cómo diferentes como los géneros musicales y los diferentes regímenes de tráfico? ¿Cuál es la complejidad requerida de una compañía para enfrentar la complejidad de un mercado? ¿Cuándo un sistema es más o menos complejo?

### 1.3.4 Homeostasis

Originalmente, el concepto de homeostasis fue desarrollado para describir funciones internas de regulación fisiológica, tales como la temperatura o los niveles de glucosa. Probablemente, la primera persona que reconoció el mantenimiento de un ambiente interno casi constante para la vida fue Bernard (Bernard, 1859). Subsecuentemente, Cannon (Cannon, 1932) acuño el término *homeostasis* ( del griego *hómoios* (similar) y *estasis* (estado, estabilidad). Canon, definió la homeostasis cómo la habilidad de un organismo para mantener estados estacionarios de operación durante cambios internos y externos. Sin embargo, esto no implicaba un estado inmóvil o inactivo. Aunque algunas condiciones pueden variar, las principales propiedades de un organismo pueden ser mantenidas.

Más tarde, el cibernético británico *William Ross Ashby* propuso, en forma alternativa, que la homeostasis implicaba una reacción adaptativa, para mantener "variables esenciales" dentro de un rango (Ross Ashby, 1960a, 1947b). Estas variables son esenciales, porque en ellas se basa gran porcentaje del funcionamiento del sistema, de forma que si fallan, el sistema lo hará igualmente.

Dos conceptos relacionados con la homeostasis son los conceptos de ultra-estabilidad y adaptación homeostática (Di Paolo, 2000). En ellos se basó Ashby para explicar la generación de comportamiento y aprendizaje en máquinas y sistemas vivos. La ultra-estabilidad se refiere a la operación normal de un sistema dentro de una "zona de viabilidad" para hacer frente a los cambios ambientales. Esta zona de viabilidad está definida por los rangos entre los valores mínimos y máximos de las variables esenciales del sistema. Sí al operar, un valor de estas variables sobrepasan los límites de esta zona, el sistema cambia, se regula para encontrar nuevos parámetros que hacen estas variables retornen a su zona de viabilidad. Es allí donde nace la expresión equilibrio dinámico, pues el sistema se ajusta constantemente para estar en una zona de funcionamiento óptimo.

Un *SD* tiene una alta capacidad homeostática para mantener su dinámica cerca a cierto estado o estados (atractores). Cuando las perturbaciones ambientales ocurren, el sistema se adapta para enfrentar los cambios dentro de su zona de viabilidad sin que el sistema pierda su integridad (Ross Ashby, 1947a).

La homeostasis puede ser vista como un proceso dinámico de auto-regulación y adaptación, por el cual el sistema adapta su comportamiento en el tiempo (Williams, 2006). En este sentido, los procesos de homeostasis, emergencia y auto-organización guardan relación, porque corresponden con una respuesta específica a las presiones ambientales, en cuyo caso se puede decir que el sistema muestra adaptación (Prokopenko et al., 2009).

Es de resaltar que para a pesar de haberse descrito gran cantidad de procesos donde opera la homeostasis, la profundización en sus aspectos formales para su medición y valoración es un gran y atractivo campo de estudio (Fernández et al., 2012b, 2010a).

### 1.3.5 Autopoiesis

La Autopoiesis (del griego *auto-propio-* y *poiesis-creación, producción-*), fue propuesto como un concepto para definir lo vivo y distinguirlo de lo no vivo. Recientemente Maturana (Maturana, 2011), en respuesta a Froese (Froese and Stewart, 2010), aclara que la noción de autopoiesis fue creada para connotar y describir los procesos moleculares que toman lugar en la realización de los seres vivos cómo entidades *autónomas*. El significado de la palabra fue usada, desde sus inicios, para describir redes cerradas de producción molecular (Maturana and Varela, 1980). La noción de autopoiesis surgió de una serie de preguntas relacionadas con la dinámica interna de los *seres vivos*, las cuales Maturana comenzó a considerar en los 60s. Entre ellas estaban: ¿Cuál debería ser la constitución de un sistema que veo cómo vivo, cómo resultado de su operación? ¿Qué pasó hace 3.800 billones de años para que podamos decir que la vida comenzó allí?

En la autopoiesis, los organismos vivos ocurren cómo entidades dinámicas, moleculares, autónomas y discretas. Estas entidades están en continua realización de su auto-producción. Así, la autopoiesis describe la dinámica interna de un ser vivo en el dominio molecular. Maturana resalta que los seres vivos son sistemas dinámicos en continuo cambio. Las interacciones entre los elementos de un sistema autopoiético, regulan la producción y la regeneración de los componentes del sistema, lo que le da el potencial de desarrollarse, preservarse y producir *su propia organización* (Varela et al., 1974).

Desde la autonomía expresada en la autopoiesis se puede tener que una bacteria puede producir otra bacteria por división celular, mientras que un virus requiere de un hospedero para producir otro virus. La producción de una nueva bacteria, se da por las interacciones entre los elementos de ella. La producción de un nuevo virus, depende de las interacciones de los elementos con un sistema externo. Así, la bacteria es más autopoiético que un virus. En este sentido, la autopoiesis está relacionada estrechamente con la autonomía (Moreno et al., 2008; Ruiz-Mirazo and Moreno, 2004). Por ejemplo, una bacteria puede producir otra bacteria por división celular, mientras que un virus requiere de un hospedero para producir otro virus. La producción de una nueva bacteria, se da por las interacciones entre los elementos de ella. La producción de un nuevo virus, depende de las interacciones de los elementos con un sistema externo. Así, la bacteria es más autopoiética por su mayor autonomía que un virus. En esta comparación es apropiado resaltar que el virus despliega las propiedades de autoreferencia y heteroreferencia al utilizar la maquinaria de las células eucarióticas para sus fines reproductivos. Esto visiblemente desencadena una red de procesos para la producción de su material constitutivo.

Desde la anterior perspectiva, la autopoiesis está más relacionada con la autonomía (Moreno et al., 2008; Ruiz-Mirazo and Moreno, 2004). Consideramos que la autonomía es una condición emergente que surge del proceso autopoiético y lo representa en gran medida.

La autonomía, además, está siempre limitada en sistemas abiertos, dado que su estado dependerá de las interacciones con el ambiente. Desde esta perspectiva, se puede observar a la autopoiesis como un proceso con rasgos homeostáticos que contribuye a los procesos auto-organizantes de la estructura del sistema, y que es expresión de la autonomía del sistema. Normalmente, las diferencias en autonomía entre sistemas, cómo en el ejemplo de bacterias y virus, pueden ser claramente identificadas, lo que nos promueve a generar escalas de autonomía al evaluar la respuesta de los organismos frente a los cambios de su medio.

El concepto de autopoiesis se ha extendido a otras áreas más allá de la biología (Froese and Stewart, 2010; Froese et al., 2007; Luisi, 2003; Seidl, 2004). Algunos aspectos formales de la autopoiesis fueron abordados por Varela et al. (1974), quienes proponen pruebas operacionales para determinar sistemas autopoiéticos. A pesar de la oposición de Maturana a la formalización de la autopoiesis en términos matemáticos, (Varela, 1979) y (Urrestarazu, 2011) desarrollaron algunos aspectos formales. Recientemente, Gershenson y Fernández (2011) propusieron una medida de autopoiesis sobre la base de la complejidad comparativa del sistema y su ambiente.

## 1.4 Síntesis

Desde una perspectiva histórica, en este capítulo hemos dado una mirada a la teoría de los *SD* sobre una base más amplia; una que nos permite entender el porqué de su dificultad de predicción. Dificultad que subyace en las interacciones relevantes del sistema, que lo auto-organiza y que nos produce nueva información. Nueva información que emerge también, cuando el sistema afronta los cambios ambientales. Cambios que muestran la respuesta autónoma del sistema y que definen su grado de adaptabilidad. Todo ello ahora queda inmerso en una visión de *SD* con rasgos de complejidad. La mayoría son sistemas abiertos, donde el control completo es difícilmente alcanzable, debido a la impredictibilidad que generan las interacciones. Ante lo anterior, surge la posibilidad de adaptarnos como solución.

Adicionalmente en el capítulo se presentan, las nociones, y sus características, de auto-organización (incremento del orden-regularidad), emergencia (nuevas propiedades a escala global), complejidad (interconexión), homeostasis (auto-regulación, ultra-estabilidad) y autopoiesis (auto-producción, auto-determinación), lo que sientan las bases y la motivación para su posterior formalización.

# CAPÍTULO 2: ESPECIFICACIÓN DE SISTEMAS DINÁMICOS COMO REDES COMPLEJAS DE AGENTES

## Resumen

*El capítulo 1 consideró la dimensión de la complejidad en SD y la consideración de un número mayor de propiedades para su análisis. Desde esta base, el capítulo 2 parte de la concepción tradicional de SD, para proponer una noción con propiedades que van más allá de la emergencia y la auto-organización. Se incluyen, entonces, la homeostasis y la autopoiesis. La asociación de tales propiedades genera una interdependencia entre los aspectos funcionales y estructurales de los SD que también son abordados. Posteriormente, al considerar las interacciones entre elementos, que pueden ser descritos como agentes, se muestran los formalismos para la representación de los SD como redes computacionales de agentes. Estos formalismos son la base para el estudio de la complejidad en SD, lo que posibilita la fusión de las teorías multi-agentes y de redes. Desde este enfoque hibrido, un sistema multi-agentes es visto como una red computacional con rasgos a determinar, tales como su emergencia y su auto-organización, entre otros. Con este aporte se espera dar un mejor entendimiento de lo que sucede a partir de las interacciones entre los componentes de los SD.*

## 2.1 Nociones de *SD*: de lo Tradicional a la Inclusión de lo Complejo.

A continuación se presentan dos perspectivas para la definición de *SD*, una tradicional y otra que incluye la complejidad. Esta última está inscrita en el marco de la *TDS* expuesta en el ítem 1.1., y se proviene de lo expresado en Fernández, Aguilar, et al. (2010).

### 2.1.1 Noción Tradicional de *SD*

Como base general se tiene que, los *SD* cuentan con características básicas como lo son, el cambio o evolución de su estado (q) en el tiempo, en razón a una regla de evolución ($R$) que lo demarca. Desde una base determinista, se podría estimar que $R$ está dada por $R(q(t), r) \rightarrow q(t + \tau)$, donde q es un estado del conjunto de estados posibles $Q$, y $r$ un parámetro de la regla. Cabe destacar que r tiene uno o más puntos críticos ($r_c$), el cual o los cuales, son entendidos como el valor de r en el que se genera el cambio o transición de estado de $q_i$ a $q_j$. Por su parte, $R$ puede estar descrita por funciones iterativas, ecuaciones diferenciales o instrucciones iterativas, según sea su naturaleza discreta, continua o algorítmica, respectivamente.

Sobre la anterior base, ha sido posible estudiar gran cantidad de *SDs* existentes en nuestro medio. No obstante, en ciertos *SD*s con gran cantidad de componentes que presentan alta conectividad, plasticidad en sus interacciones y roles muy fluidos, las reglas de su cambio y evolución pueden ser no bien entendidas por sus estudiosos. Más aún, cuando a partir de estas reglas que evolucionan de simples a complejas y con la auto-organización como método, muchos *SD* pueden llevar a cabo tareas que tiene un grado de dificultad considerable. Es allí donde el observador puede apreciar la dimensión emergente de su orden y comportamiento global (Aguilar et al., 2007).

En las anteriores condiciones, se hace pertinente observar un *SD* bajo la óptica de su complejidad. Un evento inherente al funcionamiento, que requiere de mayores conceptualizaciones y profundizaciones en cuanto a sus aspectos formales, como a continuación se propone.

### 2.1.2 Una Definición de *SD* que Considera la Complejidad o Sistemas Dinámicos Complejos (*SDC*).

La noción base de *SD* en la que se fundamentará este manuscrito parte de la presentada en (Fernández et al., 2010a). En ella se incluye los rasgos de complejidad, por lo que podemos definir un Sistema Dinámico Complejo (*SDC*) y expresa lo siguiente:

…"*Dado un contexto particular de observación[1], un SDC corresponde con un fenómeno unitario que presenta un conjunto de componentes homogéneos o heterogéneos*

---

[1] *Corresponde a los aspectos propios del observador que, influyen en el proceso de modelación y el consecuente modelo. Estos aspectos son: (i) su cultura, entendida como el conjunto de significados a partir de los cuales el sujeto comprende el mundo y (ii) el lingüístico, como aquel lenguaje especializado que se construye sobre la base de la cultura del sujeto. A partir de ellos el observador determina el sistema unitario a partir de las relaciones de retro-alimentación y de recursividad circular entre la dupla sistema-entorno. Es de anotar que el entorno se refiere, por ejemplo, a otros miembros de la comunidad de conocimientos, modelos de simulación, teorías, o al mundo físico y social. Mayores ampliaciones de estos elementos pueden hallarse en Fernández et al (2010).*

*interconectados. Estos componentes, pueden observarse a diversas escalas espacio-temporales y generan, a partir de sus interacciones, diferentes tipos de propiedades, organizaciones, patrones y/o comportamientos propios. Todo ello, con el fin de alcanzar el propósito global que el sistema tiene en sí mismo y sobre la base de su complejidad"...*

Un análisis de mayor profundidad de la anterior noción, lleva a considerar la existencia de características y elementos importantes como:

- **La interconexión de los elementos** define la topología de sus *interacciones*, que de acuerdo con el caso y eventos de reforzamiento o debilitamiento, conducirán o no al establecimiento de posteriores *interrelaciones*. Las interacciones, en este sentido, las entendemos como contactos de corta duración y conexión; mientras que las interrelaciones las entendemos como fenómenos de asociación o combinación de elementos de naturaleza más perdurable. La dinámica de las interconexiones es uno de los elementos básicos de la complejidad del *SDC*, pues es su dinámica la que conduce a cambios en la organización propia no *impuesta*. Al mismo tiempo, las interconexiones entre las partes están acordes con los cambios ambientales y las propias posibilidades de los elementos para *adaptarse* a ellas.

- **Las adaptaciones en el** *SDC* se dan de manera individual y colectiva; en el tiempo y en el espacio, y sobre la base de los rangos de adaptabilidad de las partes. Desde esta visión, el *SDC* se auto-organiza a partir de la topología de las interacciones e interrelaciones entre los elementos que constituyen una estructura determinada. Los elementos tienen como rasgo, una potencialidad desconocida de establecer relaciones entre sí. Ellas se establecen sobre la base de su adaptabilidad; otra razón más de la complejidad del *SDC*.

- **Las escalas espacio-temporales** en las que se ocurren los elementos, corresponden a la ventana de observación de espacio y/o tiempo, donde se definen las características de los elementos y los parámetros del entorno que condicionan sus interacciones. En estas escalas es que operan las variables y procesos que explican los patrones observados en determinado nivel. Las escalas espacio-temporales son un rasgo básico de dinamismo y complejidad sistémica, pues requieren del análisis de su cambio, variabilidad y heterogeneidad, para la explicación y posible predicción, o no, de los patrones observados. Ejemplo de este tipo de escala en *SDCs* son el nivel de observación local de las interacciones, y el nivel superior o global en el que son observables las propiedades emergentes. No obstante, pueden existir otros niveles intermedios en el *SDC*. Por otra parte, las relaciones entre niveles se da también por mecanismos de retro-alimentación, en el sentido local-global y global-local. De esta última forma, las propiedades emergentes tienen también influencias en las interacciones del nivel local.

- **La generación de diferentes tipos de organizaciones**, patrones y comportamientos **propios** a partir de las interacciones locales, en los diferentes *niveles* del *SDC* a escalas espacio-temporales definidas, determina claramente el carácter emergente del *SDC*. Desde esta perspectiva, se visualiza que determinado fenómeno considerado como emergente puede ser, además, producto de la auto-organización. Igualmente, los patrones (*Pt*) definen la unidad global del *SDC*, pues son producto de cualidades sintetizadoras que representan, en parte, la emergencia del *SD*.

- **La propia organización (auto-organización)** incluye, a su vez, la capacidad del *SDC* para regular su comportamiento en términos adaptativos, en busca de un equilibrio dinámico u **homeostasis**. Un hecho importante, debido a que el flujo de intercambios entre los elementos y el entorno puede llevar a la falta de regularidad en el *SDC*. La homeostasis abarca, articula e integra estados del *SDC* que presentan diferentes condiciones complementarias como las dinámicas-estacionarias, las de estabilidad-inestabilidad, las de síntesis-degradación, entre otras. La funcionalidad de la homeostasis es mantener la auto-organización, ante el constante flujo de materia, energía y/o información en el *SDC*, en límites que posibiliten el mantenimiento de su autonomía, integridad, compatibilidad y demás atributos esenciales (Zona de Viabilidad del *SDC*). Todo este fenómeno se da en el marco de la adaptabilidad del *SDC*.

- Consecuente con lo anterior, se estima que un mecanismo homeostático de importancia en un *SDC* complejo es la **autopoiesis**; un proceso en el que a partir de las interacciones entre los componentes del *SDC* logra el auto-mantenimiento, en el contexto de la relación síntesis-degradación del *SDC*. En esta propiedad se refleja parte de la *autonomía* del *SDC*. La autopoiesis favorece la auto-organización del *SDC*, por medio de mecanismos de auto-regulación que mantienen la propia estructura del *SDC*, en consonancia con un entorno variable.

Los elementos develados de emergencia, auto-organización, homeostasis y autopoiesis en la noción propuesta, son un medio por el cual el *SDC logra* persistir espacial y/o temporalmente situación que no es otra cosa que su propósito esencial. Al mismo tiempo le confieren una complejidad determinada, de acuerdo con el tipo de *SDC*.

Se puede estimar que la persistencia de un *SDC* en un medio cambiante se da en términos de la conservación, adaptabilidad y evolución de sus elementos, y del *SDC* como un todo. Estas últimas propiedades vienen de las interacciones relevantes[2] que generan la auto-organización de la estructura, la emergencia de comportamientos, y en definitiva, la complejidad del *SDC*. Complejidad que le confiere al *SDC* su robustez (Conservación) ante la variación de su ambiente, pero también la posibilidad de cambiar de forma adaptativa y evolutiva ante condiciones más drásticas.

### 2.1.3 Interdependencia Funcional de la Auto-organización, Homeostasis y Autopoiesis, y la Generación del Espacio Estructural del *SDC*.

Normalmente, se ha estudiado cómo la estructura de un sistema determina su función. Sin embargo, en muchos sistemas complejos hay una retroalimentación donde la funcionalidad genera ciertas propiedades en el espacio estructural de un *SDC*. La interdependencia funcional de auto-organización, homeostasis y autopoiesis, se puede observar en la Figura 2-1. En ella se observa como la auto-organización y la autopoiesis, sobre la base de mecanismos homeostáticos de auto-regulación, son las que desde el espacio funcional generan el *SDC* en el espacio estructural. Desde este punto de partida se puede estimar que la relación entre el espacio funcional y estructural se hace cíclica, por cuanto la estructuras también determinan la función.

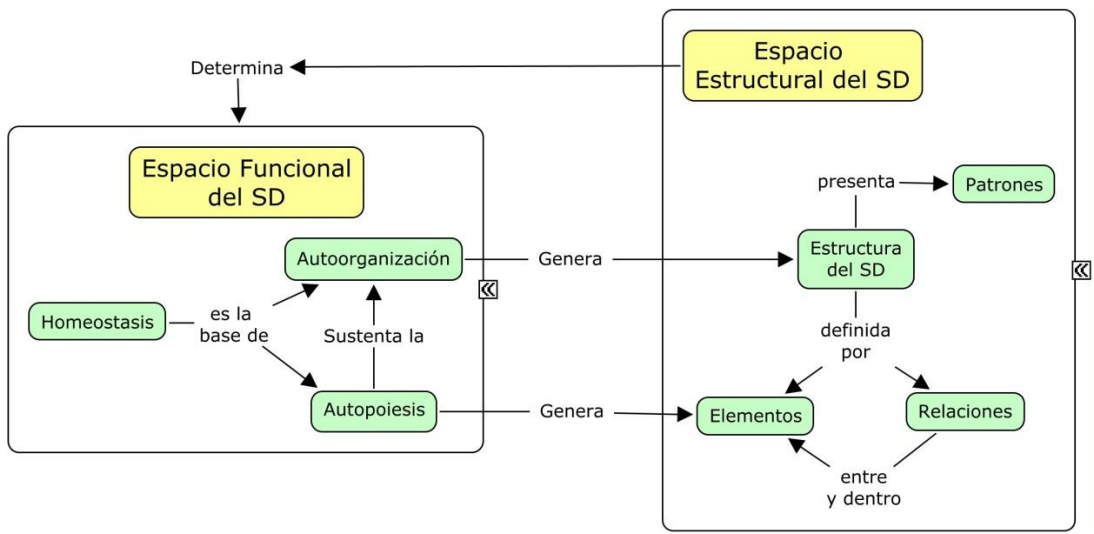

**Figura 2-1 Relación entre los Espacios Funcionales y Estructurales en *SDC*** (Fernández et al., 2012a)

---

[2] Son relevantes porque producen información adicional que no estaba en las condiciones iniciales o de frontera.

Entre los aspectos de mayor importancia en el espacio estructural, se hallan la estructura y los patrones. La estructura estará referida a los elementos y el tipo de relación entre ellos ($Estructura = \{elementos, relación\}$). La estructura puede emerger como resultado de: (i) una acción auto-organizante que define el tipo de relación entre los elementos y (ii) por procesos autopoiéticos, de manera que se procura así misma, en términos de autonomía. Al mismo tiempo, la estructura expresa en patrones, la mayoría de los casos definibles u observables, de manera directa o indirecta (Fernández et al., 2011).

Históricamente, la estructura ha tenido un papel preponderante en el estudio y explicación de los *SDC*, dada su condicional aparentemente "palpable". No obstante, vale la pena realizar algunas precisiones al respecto. La primera es que la estructura, en muchas ocasiones, no aparece de manera explícita, sino que por su carácter *emergente*, resulta ser una condición a determinar. En este aspecto, suele suceder que lo que el observador percibe del *SDC* son algunos patrones de orden global. Los patrones se generan a partir de procesos de retroalimentación, entre otros, y sobre la base de la adaptabilidad del *SDC*. Descifrar la estructura en un *SDC* requiere, por tanto, determinar y caracterizar los elementos y sus relaciones, a través de indicadores; así como de la aplicación de modelos matemáticos, que pueden dar razón sobre su condición jerárquica, holárquica o heterárquica, por ejemplo.

La condición deseable de llegar a estructuras "más estables" en el *SDC*, para periodos determinados de tiempo, será el producto del alcance de ciertos objetivos globales, como resultado del dinamismo del proceso auto-organizante. La estabilidad de una estructura dependerá, básicamente, del mantenimiento de la misma en una zona de viabilidad ($Z_v$), es decir, de la condición homeostática que le proporciona la auto-organización. No obstante, la tendencia auto-organizante de mantener tal estabilidad puede conllevar a que, en caso de la superación de los umbrales máximos y mínimos en una determinada $Z_v$, se dé un cambio importante en la estructura, como respuesta evolutiva del *SDC*. De allí se generará un nuevo estado inicial para la estructura, que buscará llegar nuevamente a su estado estable, y su desempeño se dará en una nueva $Z_v$. A este respecto vale la pena tener presente la consideración de (Kaufmann, 1993), quien asevera que los cambios de estado dentro de la auto-organización, al no ser graduales y constantes, se dan por la superación de umbrales de complejidad que llevan a nuevas formas estables.

Lo expuesto anteriormente toma sentido al considerar que, hipotéticamente en un *SDC*, pueden existir tantas estructuras como tantas posibles combinaciones de elementos e interacciones puedan haber. En términos prácticos, esta situación puede ser observada como la variación dinámica en una intrincada red de elementos que se relacionan a partir de sus interacciones. En consecuencia, se puede estimar que un *SDC* puede ser descrito, en términos matemáticos, sobre la base de un grafo o red dinámica, cuyos aspectos formales se explican a continuación (Fernández et al., 2011).

## 2.2 Los *SDC* cómo Redes Computacionales de Agentes.

Los siguientes apartados se basan en lo expresado en (Fernández et al., 2012a, 2012b, 2011, 2010b).

Como se ha venido observando, la consolidación del *SDC* como un todo se puede establecer a través de las interacciones relevantes entre sus elementos que dan como resultado propiedades cómo la auto-organización, emergencia, homeostasis y autopoiesis. En este aspecto, los *SDC* pueden ser especificados cómo un grafo o red de interacciones, de manera que se puede representar en una estructura matemática formal que representa a los elementos y sus relaciones. Las redes, particularmente, se han convertido en una herramienta central en el estudio de sistemas complejos (Gershenson and Prokopenko, 2011), donde los nodos son considerados como "agentes".

La definición básica de agente que adoptamos corresponde con la de Gershenson (2009), referida a una entidad que está situada en un entorno y que puede afectarse o afectar lo que en él sucede. De acuerdo con su tipo, un agente puede generar un conocimiento apropiado del sistema por la creación de una representación interna de los componentes del sistema externo. Cuando los agentes interactúan dinámicamente para alcanzar una función o comportamiento global, el sistema que los contiene puede ser descrito cómo auto-organizante (Gershenson, 2006). Estas definición general se sintoniza con lo requerido para la construcción de sistemas multi-agentes de acuerdo con lo expresado por Wooldridge and Jennings (1995).

Tanto la red en su globalidad, como en sus componentes locales, vistos cómo *agentes*, pueden ser especificados como *SDs* tradicionales. Es decir con cambios y/o evolución de estados en el tiempo, a partir de reglas que pueden ser expresadas funcionalmente. Con una red computacional de agentes, se pueden lograr descripciones de *SDC* más adecuadas, y se tiene la ventaja de poder analizar la información nueva producida por las interacciones entre agentes.

Para la especificación de una red computacional de agentes, se requiere integrar diferentes aspectos de la teoría de los *SDC*, grafos y agentes, como se describirá en el siguiente apartado.

### 2.2.1 Aportes de la Teoría de *SD*s

La teoría de modelado y simulación (Zeigler et al. 2000), permite considerar los elementos básicos para la especificación de la red computacional de agentes (*SDC*-Ag), como un sistema dinámico tradicional (*SD*). Para ello se tendrá en cuenta la siguiente estructura (ver Ecuación 2.1):

$$SD = (X, Y, \Omega, Q, F, F, T) \tag{2.1}$$

Donde:

$SD$: Sistema Dinámico

$X$: Conjunto de Entradas.

$Y$: Conjunto de Salidas.

$\Omega$: Conjunto total de entradas posibles.

$Q$: Conjunto de Estados.

$F: (Q \times \Omega \rightarrow Q)$: Función de Transición entre estados, que actualiza el estado.

$Fs: (Q \times \Omega \rightarrow Y)$: Función de Salida.

$T$: Posibles valores para el tiempo.

En esta estructura, $X, \Omega$ y $Y$ definen las entradas y salidas del *SD*, mientras que $Q$ y $F$ representan la estructura interna y el comportamiento. La función de transición de estados $F$ debe garantizar que siempre se puede determinar el próximo estado del *SD*, a partir del estado actual.

La representación gráfica de los elementos básicos en una red *SDC* -Ag, desde la perspectiva de un *SD* acoplado como lo propone Zeigler et al. (2000), se observa en la figura 2-2.

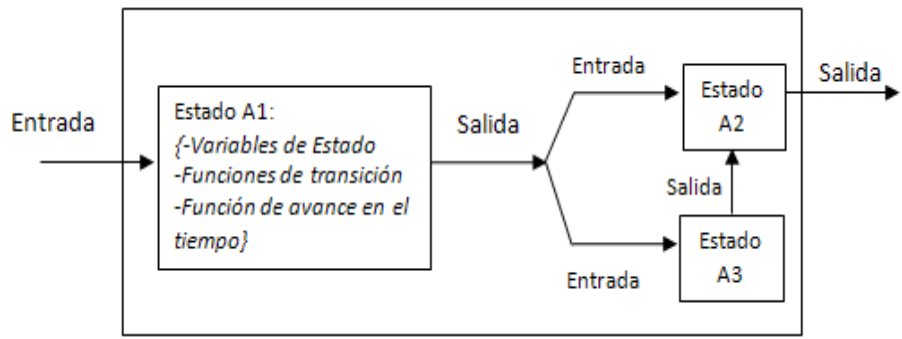

**Figura 2-2 Representación Simplificada de un *SDC* -*Ag* sobre la Base de la Teoría de Modelado y Simulación de Zeigler et al. (2000). Las designaciones A1, A2, A3 se refieren a Agentes.**

## 2.2.2  Aportes de la Teoría de agentes y de los Sistemas Multi-agentes-*SMA*

La teoría de agentes (Desalles et al., 2008; Ferber and Gutknecht, 1998; Ferber et al., 2004; Jennings and Wooldridge, 1995; Müller and Et Al., 1997; Wooldridge et al., 1996), brinda elementos de interés que han permito establecer una de las principales áreas de la inteligencia artificial distribuida, como lo son los *SMA*.

Desde una base general, un agente puede describirse como una entidad que actúa sobre su ambiente o entorno (Gershenson, 2007). Sobre esta base, y en términos de un *SDC*, se estima que los *Ag* son entidades que perciben su ambiente (entradas) y actúan (cambio de estado) en concordancia, deliberando sobre las posibilidades de sus acciones (salidas). Igualmente, se ha considerado que los *Ag*s pueden tener rasgos inspirados en el comportamiento humano como: autonomía, reactividad, iniciativa, habilidad social. Estos rasgos le permiten al *Ag* compartir el entorno o ambiente, recursos y conocimiento, además de comunicarse y coordinar sus actividades con otros (Aguilar, 2014; Aguilar et al., 2012).

Desde la perspectiva de la teoría de *Ag*s los *SMA*, como sistemas compuestos por diversos *Ag*s, presentan características como la modularidad, redundancia, descentralización, comportamiento emergente, y especialmente, coordinación. Esta última, basada en la compartimentalización de información y resultados, la optimización de acciones y recursos, y el trabajo conjunto independiente de los objetivos individuales. Los mecanismos propios para lograr una coordinación efectiva se sustentan en la sincronización de acciones, vista como la simultaneidad de las mismas; en la planificación, es decir en la subdivisión y distribución de tareas; y en la coordinación reactiva, a partir de la definición de marcas o reglas de comportamiento reactivo general. Bases como éstas, permiten el desarrollo de *SMA* auto-organizantes emergentes cuyo sustento son mecanismos que consideran las interacciones directas, la estigmergia, el refuerzo, la cooperación y la definición de arquitecturas genéricas (Perozo et al., 2008).

Operacionalmente, a cada *Ag* le puede ser asignado uno o más roles para la consecución de los objetivos del *SDC*, y ser responsable tanto de coordinar sus interacciones con otros *Ag*s, como de ejecutar las actividades bajo su cargo. En este aspecto, un *Ag* inteligente es un "tomador de decisiones interactivo", capaz de alcanzar de manera pro-activa sus objetivos, mientras que se adapta a un ambiente dinámico. De allí que los *SMA* son vistos como sistemas computacionales auto-organizantes con un grado apropiado de autonomía, dada la inexistencia de un control central (Aguilar, 2014).

### 2.2.3 Aportes de las Teorías de Grafos y Redes

La teoría de redes se remonta al siglo XVIII en la antigua ciudad alemana Königsberg (hoy Kaliningrado, ciudad rusa), donde se planteó un interesante problema matemático, hoy conocido como el de "los puentes de Königsberg". El planteamiento de éste problema consistía en hallar la forma de cruzar los 6 puentes, en un solo sentido, y sin repetir el tránsito por alguno de ellos. El problema fue finalmente resuelto en 1736 por el suizo Leonhard Paul Euler (Kruja et al., 2002), a través de la representación de las rutas posibles como enlaces o ligas, y los puntos de conexión entre ellas como nodos. Este hecho es frecuentemente citado como el nacimiento de la teoría de grafos, actualmente una de las formas más comunes de representación de las redes (Hedrih, 2007).

Sí bien el modelado *SMA* ha constituido una herramienta computacional adecuada para lograr explicaciones o predicciones sobre fenómenos específicos en *SDC*, existe dificultad en extraer algunos parámetros estadísticos que den razón de su comportamiento global. La teoría de redes, como extensión de la teoría de grafos, cuenta con las posibilidades que no tienen los *SMA*, tales como la representación del flujo de información y las interacciones topológica entre agentes-*Ag*. Entre los parámetros topológicos de interés se hallan el grado (*Gd* en ecuación 2.2) y el coeficiente de agrupación (*Ca* en ecuación 2), las cuales pueden ser determinados considerando su carácter dinámico (Habiba and Berger-Wolf, n.d.). *Gd* se entiende como el cambio de vecindad $V(i_t)$ de un nodo $N_i$ en el tiempo $t$. *Ca* de un nodo $i$ es la fracción de sus vecinos, quienes han sido vecinos entre ellos mismos en cualquier tiempo previo $t$ (ecuación 2-3).

$$Gd_T(i) = \sum_{1<t<T} \frac{|V(i_{t-1}) - V(i)|}{|V(i_{t-1}) \cup V(i_t)|} \times V(i_t) \tag{2.2}$$

$$Ca_T(i) = \sum_{i \in V(i_T)} \frac{|\cup_{1<t<T} V(i_t) \cap V(i_T)|}{V(i_T)} \tag{2.3}$$

La importancia de estos parámetros se halla en que, a partir de ellos, se podrían determinar patrones resultantes, o no, de procesos de auto-organización en la red *SDC*. Adicionalmente, existe la posibilidad de determinar *estados* subyacente para la red *SDC*, que definen su "estado de equilibrio" como resultado de los procesos homeostáticos. Esto se puede lograr a través de la aplicación de algoritmos como el de Kamada-Kawai (Kamada and Kawai, 1989).

Dado que desde la teoría de redes, los nodos pueden verse como estructuras excesivamente simplificadas, la hibridación de redes con nodos tipo *Ag* puede ser una concepción de gran importancia. De esta manera, se hace posible ver un *SMA* como un conjunto de nodos y aristas interactuantes que conforman una red computacional, es decir una red *SDC-Ag*.

En términos generales, una red *SDC-Ag*, se sintoniza con la estructura definida por (Gershenson, 2010) para una red computacional $C(N, K, a, f)$, que la define como un conjunto de nodos $N$, ligados por un conjunto de aristas $K$, usadas por un algoritmo a para computar una función $f$. A nivel operativo, $N$ y $K$ tienen variables internas que determinan su estado, y funciones $f$ que determinan sus cambios de estado, respectivamente. Se resalta que, esta generalización ha resultado de gran utilidad para comparar las arquitecturas de redes neuronales, colonias de hormiga y partículas colectivas; al tiempo que, puede ser un punto de partida y alternativa interesante para la representación, igualmente, genérica, de un *SMA*.

No obstante, para describir completamente la dinámica de un *SMA* como red computacional, se requiere ahondar en los detalles de cada componente de la red y sobre su dinámica. Por ejemplo, el flujo de información en una red da como resultado la evolución de los cambios de estado en los nodos (Smith et al., 2008). El comportamiento global de esta dinámica depende de varios factores como el tipo de nodo, su modelo cognitivo, sus funciones de transición interna, el tipo de información que comparte, entre otros. Desde esta perspectiva, se haría posible descifrar la trama de interacciones entre *Ag*s, y profundizar en la forma autónoma como se logran objetivos globales a partir de reglas locales, que podrían evolucionar de simples a complejas (Aguilar et al., 2007).

### 2.2.4 Aspectos Formales de la Red *SDC*-Ag.

A continuación se muestran los desarrollos logrados hasta el momento, que serán la base para llevar, en el corto y mediano plazo, al plano del modelado los elementos conceptuales hasta ahora expuestos.

- La forma general de la red *SDC*-Ag

La red *SDC-Ag* se define, básicamente, como un grafo $(N, K)$. Cada nodo $i \in N$ representa un elemento $N_i$, y cada arista $(i, j) \in K$ representa flujo de información entre $N_i, N_j$. Los vecinos son elementos $(N_i, N_j)$ unidos por un enlace de información $(i, j) \in K$. El grafo $(N, K)$ determina la estructura del *SDC*-Ag.

La conectividad de la red *SDC* se puede representar en un nivel básico por una matriz de adyacencia $N \times N$: $M(i, j) = \begin{cases} 1, i \to j \\ 0, i \not\to j \end{cases}$. Se dice que $M$ contiene la topología de la red, y puede contener el "peso" del acoplamiento $i, j$, tal que $M_{ij} \in [0,1]$. No obstante, en una formalización más general, cada elemento de la matriz $M$ puede ser más complejo, conteniendo vectores, etiquetas, y/o funciones.

- El sentido dinámico de la red *SDC*-Ag

Se puede estimar que la dinámica de una red *SDC-Ag* estará dada por $SD = (Q^n\ F)$; Donde $Q^n$ es un conjunto no vacío, compuesto en el caso de sistemas discretos de un número finito de estados $q \in Q^n$. $Q^n$ también es llamada espacio de estados, o espacio de todas las configuraciones posibles. El exponente $n \in \mathbb{N}$ hace referencia al número de total de elementos en la red *SDC*. En el caso de sistemas continuos, la dinámica se desenvuelve en un espacio de estados.

Por su parte, $F$ es una función de transición, o de activación global, tal que $F: Q^n \to Q^n$. Igualmente, $F$ corresponde a una familia de funciones, denominadas funciones de activación local de la red ($f_i$), tal que $F = (f_1, f_2, \ldots, f_n)$. Igualmente, $f_i: q^n \to q^n$ dado que $F(q) = (f_1(q), \ldots, f_n(q)), \forall q \in Q^n$. $F$ puede determinar a un sistema compuesto por funciones puramente locales.

Se destaca que cada estado $q \in Q^n$ hace parte de una configuración global de la red ($C$), entendida como la descripción completa de la situación en que se encuentra el *SDC* en un momento dado. Así, $C$ estará representada por su estado actual $q_t \in Q$, y un evento $e$ que genera la transición $F$, donde $e \in \Sigma$. $\Sigma$, por su parte, será el conjunto de todos los eventos que generan transición. Entonces, a partir de un evento $e$ se dará una $F(q(t), e)$, que producirá la configuración global en el tiempo siguiente $t + 1$, ($q(t + 1)$). Dependiendo del caso, el evento $e$ puede corresponder con el punto crítico ($r_c$) del parámetro $r$ mencionado anteriormente.

Consecuente con lo expuesto, para pasar de una configuración $c_i$ a $c_j$, cada *Ni* tendrá su función de transición $f_i$, tal que $F(q_i \to q_i(t + 1)) = f_1(q1_i), \ldots, f_n(q(t + 1)) = q_i \ldots q_i(t + 1)$. Para una configuración global $c \in C$ en el tiempo t de la red ($q(t)$), $F(q(t,e))$ dará la configuración global siguiente en el tiempo $t + 1$, ($c(t + 1)$). Un sistema donde todo estado tiene un sólo sucesor es determinista, mientras que si hay más de un sucesor posible para por lo menos un estado, habrá cierto grado de no determinismo.

- El vecindario y el entorno en la red *SDC*-Ag.

En cuanto a $N_i$, se tiene para cada uno de ellos un $q_i$ determinado en el tiempo $t$, es decir $q_i(t)$. Por poseer $N_i$ un conjunto de vecinos $V_i$, la evolución del *SDC* estará descrito por la regla $q_i(t + 1) = F(q_i(t); V_i)$, donde $i = 1, \ldots, n$. Esto es debido a la influencia que pueden tener los nodos vecinos entre sí.

De forma complementaria, el estado de la red *SDC* en un tiempo $t$, puede ser representado por un vector de estado, tal que $\vec{q}(t) = (q_1(t), \ldots, q_n(t))$. Así, la forma vectorial de la red dinámica sería $\vec{q}(t + 1) = \vec{F}(\vec{q}(t)) + V\vec{q}(t) + Ab\vec{q}(t))$, donde $V\vec{q}(t)$ es el estado de los vecinos y $\overrightarrow{Abq}(t)$ corresponde al ambiente o entorno.

Se aclara que los componentes de $\vec{Abq}(t)$, pueden ser exógenos o endógenos a los *Ni*. Como también, pueden darse de manera conjunta o separada. Al mismo tiempo, pueden tener efecto sobre la totalidad de los $N_i$ (incidencia global) o sobre $N_i$ determinados (incidencia parcial). La intensidad relativa de la incidencia, o la probabilidad de interacción, de $\vec{Abq}(t)$, de manera totalizada, sobre determinado $N_i$, estaría dada por un parámetro $B \in [0,1]$ (González-Avella *et al.*, 2007). Se puede asumir que la influencia de $B$ sobre los $N_i i$ será uniforme, más no así la respuesta que los $N_i$ den al mismo. Esto será relativo a la capacidad homeostática de cada $N_i$ para responder a influencias bajas, intermedias o altas de $\vec{Abq}(t)$.

### 2.2.5  La Especificación de Nodo Tipo *Ag* y *SMA*

Desde la perspectiva de (Zeigler et al. 2000) se puede afirmar que un nodo $N_i$, cuando representa a un proceso tendrá reglas de comportamiento internas y externas. Las primeras estarán definidas por sus variables de estado, funciones de transición, y funciones de avance en el tiempo; y las segundas estarán dadas por el conjunto de salidas. De la misma manera, se debe caracterizar la naturaleza discreta o continua de los nodos, lo que definirá su forma particular de especificación.

Por ejemplo, en el caso de sistemas naturales como los socio-ecológicos, representados como una red de agentes, los procesos como la contaminación o el cambio climático pueden ser representados como nodos. La dinámica de estos nodos puede ser descrita a través de sistemas de ecuaciones diferenciales de primer orden (*SEDO*), dada su naturaleza continua. Para el caso de procesos, como los de manufactura industrial, como los que se dan en una fábrica de ensamblado, cada operación puede ser representada como un nodo. No obstante, la dinámica será descrita a través de un sistema de eventos discretos *SED* (*DEVS* en inglés), o de ecuaciones en diferencia (*EED*). Nosotros consideraremos a las *EED* dentro de las *SED*, ya que las *EED* están embebidas en los *SED* según (Zeigler et al. 2000). Ahora bien, en este modelo también puede haber nodos tipos agentes ($N_{Ag}$) que tienen una estructura dinámica dada por la tupla 2.4:

$$N_{Ag} = < X, Y, Q_{Ag}, f_{Ag-int}, f_{Ag-ext}, f_s, t_a, W, \Gamma, \wedge, Op > \tag{2.4}$$

Donde :
- $N_{Ag}$: Nodo tipo *Ag*
- $X$: conjunto de eventos externos de entrada al *Ag*. Pueden corresponder a información proveniente desde el *Ab* ($I_{Ab}$) o desde otros *Ag* ($I_{Ag}$). En algunos casos, el ambiente puede ser modelado también como un *Ag*.
- $Q_{Ag}$: conjunto de estados del $Ag_i : Q_{Ag} \in Q^n$
- $Y$: conjunto de eventos de salida desde el *Ag*

- $f_{Ag-int}: Q_{Ag} \rightarrow Q_{Ag}$ función de transición interna por la que el *Ag* cambia de estado. Determina el tiempo de vida de un estado $q$.
- $f_{Ag-ext}: Q_{Ag} \times X \rightarrow Q_{Ag}$ función de transición externa, donde: $Q_{Ag} = \{(q,t|q \in QAg, 0 \leq t \leq t_a(q)\}$ es el conjunto total de estados, $t$ es el tiempo transcurrido desde la última transición. Especifica la respuesta del *Ag* a los eventos de entrada.
- $f_s = Q_{Ag} \rightarrow Y$: función de salida. Es una aplicación de $Q_{Ag}$ dentro del conjunto de eventos de salida $Y$. Se activa cuando el tiempo que se ha permanecido en un estado dado es igual a su tiempo de vida. Consecuentemente $Y$ es definida solo por estados activos, p.e. $\forall q : ta(q) \neq \infty$.
- $t_a: Q_{Ag} \rightarrow \Re + 0, \infty$ o función de avance en el tiempo, donde $\Re + 0, \infty$ es el conjunto de valores positivos reales entre 0 y $\infty$.
- $C_{Ag}: f_{Ag-int} \cup f_{Ag-ext}$ comportamiento del *Ag*
- $W$ es el conjunto de los estados del ambiente que rodea a los *Ag*s, representado por las variables de estado del *SDC* ($W = \{wi, ..., wn\}$, donde $w$: es variable de estado). Estas variables pueden ser útiles en la computación de las funciones de transición.
- $\Gamma$ es el conjunto de influencias del *Ag* para generar nuevos eventos. Corresponden con acciones propias con las que el *Ag* intenta modificar el curso de los eventos que ocurrirán.
- $\Lambda$ descripción del sistema a través de las leyes o reglas de cambio del *SDC*.
- $Op$ es un conjunto de operadores del *Ag* para generar influencias.

En razón que la red *SDC* puede contener más de un $N_{Ag}$, se conformará entonces un *SMA*. En este aspecto, un *SMA* se definirá como un sistema informático compuesto por un grupo de *Ag*s que interactúan entre sí. La interacción dará como resultado respuestas que irán más allá de las acciones individuales de cada $N_{Ag}$.

Desde la teoría multi-agentes de Ferber y Müller (1996) y su extensión por Dávila et al (2005), se puede estimar que la dinámica del *Ab* estará dada por una *tupla* de la forma $< W, \Gamma, Op, Acción, Reacción >$. Donde *Acción*, será el intento del *Ag* por influenciar a la red *SDC-Ag* tal que: $Op \times W \times \Gamma \rightarrow \Gamma$; y *Reacción* del *Ab* ante esas influencias será $\Lambda \times W \times \Gamma \rightarrow W$. Al mismo tiempo, se tendrá que la evolución del *SDC-Ag* será una función infinita recursiva tal que $Evolución: W \times \Gamma \rightarrow \tau$, donde $\tau$ será una función que no retorna valor o que retorna valores en un dominio de errores.

Cabe resaltar que en la formalización propuesta, $\Lambda, W, \Gamma$ y $Op$ no se explicitan en cuanto a sus componentes, dado que dependerían del *SDC-Ag* a modelar.

A partir de lo anterior, el *SMA* será modelado como un conjunto de *Ag*s que interacionan entre ellos y su *Ab*. La función global del *SDC-Ag* permitirá relacionar el cambio en las conexiones de los *Ag*s a partir de las funciones locales en los *Ag*s.

## 2.3 Síntesis

Hemos revisado la descripción y especificación de un sistema dinámico sobre la base de un conjunto mayor de propiedades, más allá de la auto-organización y emergencia. La definición obtenida incluye las propiedades de complejidad requeridos para descifrar cómo un comportamiento global y coherente, puede surgir desde las actividades locales. Con estas propiedades, es posible describir un gran conjunto de sistemas abstraídos del cosmos, de la biosfera, así como sistemas construidos por nosotros.

Los elementos conceptuales y formales desarrollados sugieren un enfoque ampliado para el estudio de los *SDC*, y de sus propiedades emergentes de auto-organización, homeostasis y autopoiesis. Estas propiedades son producto de las interacciones locales que se expresan como una conducta global y sintética observable de todo el sistema.

Como respuesta a la tarea de trasladar a aspectos formales las características emergentes mencionadas, se tiene que la teoría de grafos, y su instanciación en la teoría de redes, brindan un soporte adecuado de interpretación. A partir de ellas, los nodos pueden ser observados como agentes que pueden calcular funciones locales, las aristas pueden describir las interacciones, y una función global puede regular las interacciones entre agentes para alcanzar un estado global. El estado global, consecuentemente, se alcanzaría con la auto-organización como método, la cual se soportará en mecanismos de auto-regulación y auto-mantenimiento.

De igual manera, la emergencia del sistema tendrá como base la cuantificación de la auto-organización y los patrones, situación que unifica muchas de las visiones hasta ahora analizadas en el campo de la complejidad. En este sentido, el enfoque y las formalizaciones aquí propuestas, enriquecen las aproximaciones hasta ahora desarrolladas para el estudio de los sistemas emergentes como redes computacionales de agentes.

Trabajos futuros plantean la necesidad de ejemplificar los desarrollos aquí logrados en diversos casos de estudio, de manera a clarificar su utilidad. Al mismo tiempo, dada la complejidad del modelado, será apropiado desarrollar una herramienta computacional que facilite su uso en dichos casos de estudios.

# CAPITULO 3: FORMALISMOS MATEMÁTICOS PARA LA MEDICIÓN DE LA EMERGENCIA, AUTO-ORGANIZACIÓN, COMPLEJIDAD, HOMEOSTASIS Y AUTOPOIESIS EN SISTEMAS DINÁMICOS.

## Resumen

*Habiendo definido un SDC con rasgos de complejidad (capitulo 1) y habiéndolo especificado como una red computacional de agentes (capitulo 2), surge la necesidad de formalizar las definiciones relativas a la emergencia, auto-organización, complejidad, homeostasis y autopoiesis. El objetivo del establecimiento de estos formalismos, es permitir la valoración de estas propiedades. El fundamento para todos los formalismos fue la teoría de la información, la cual permitió generar definiciones claras, sencillas y consistentes, que pueden relacionar los procesos relativos a la complejidad en SDC. Estos formalismos son de utilidad para descripciones, a diferentes escalas, en el mundo físico y en diversos campos como la biología, la ecología, la sociología, entre muchos otros, como se verá en los capítulos 4 y 5.*

## 3.1 Teoría de la Información cómo Fundamento

Creada por Claude Shannon (Shannon, 1948), en el contexto de las telecomunicaciones, la Teoría de la Información analizó la posibilidad de reconstruir los datos trasmitidos, a través de un canal con ruido. En este modelo, la *información* es representada por una cadena $X = x_0, x_1 ... x_n$ donde cada $x_i$ es un símbolo desde un conjunto finito de símbolos $A$, llamado *alfabeto*. Cada símbolo en el alfabeto $A$, tiene una probabilidad de ocurrencia $P(x)$. En consecuencia, símbolos que sean muy comunes tendrán alta $P(x)$, mientras que símbolos con baja frecuencia de aparición tendrán baja $P(x)$.

Shannon estuvo interesado en hallar una función para medir que tanta información era *producida* por un proceso. Citándolo textualmente Shannon (1948)[3], argumentó:

*…Suponga que tenemos un conjunto de posibles eventos cuyas probabilidades de ocurrencia son $p_1, p_2,…,p_n$ . Estas probabilidades son conocidas y es todo lo que nosotros conocemos acerca del evento que podría ocurrir. ¿Podemos encontrar una medida de qué tanta "oportunidad" está involucrada en la selección del evento, o cuanta incertidumbre tenemos de la respuesta o salida? Si existe tal medida, se puede decir que sería razonable que $(p_1, p_2,…,p_n)$ cumpliera con las siguientes propiedades:*

- *Continuidad: I debería ser continua en cada $p_i$*
- *Monotonía: Sí todas las $p_i$ son iguales, $p_i = 1/n$, entonces I debe ser una función de incremento monotónico de n. Con n eventos de igual probabilidad hay más opción, o incertidumbre, que cuando hay eventos de mayor probabilidad.*
- *Recursión: Sí una opción se divide en dos, y se consideran opciones sucesivas, la I original debe ser la suma ponderada de los valores individuales de I…*

Con estos pocos *axiomas,* Shannon demostró que la única función $I$ que los satisface, tiene la forma:

$$I = -K \sum_{i=1}^{n} p_i \log p_i \qquad (3.1)$$

Donde $K$ es una constante positiva

La ecuación 3.1 se representa de manera gráfica en la figura 3-1. De allí podemos extraer las siguientes consideraciones para cadenas binarias: (i) si la probabilidad de recibir unos es máxima $p(1) = 1$, y la probabilidad de recibir ceros es mínima $p(0) = 0$, la información es mínima. Esto sucede porque conocemos de antemano que el futuro valor de $x$ será 1. La información (I) será 0 debido a que futuros valores de $x$ no adicionan nada nuevo, es decir, los valores son conocidos de manera anticipada. Ahora, si no conocemos acerca del futuro valor de $x$, como con la probabilidad para cara o sello de una moneda $p(0) = p(1) = 0.5$, entonces la información será máxima ($I = 1$), debido a que una futura observación nos traerá información relevante que será independiente de los valores previos. Desde esta condición, la información también ha sido vista como una medida de incertidumbre (Gershenson, 2013a). Sí hay absoluta certeza que el futuro de x será cero $(P(0) = 1)$ o uno $(P(1) = 1)$, entonces la información recibida será cero. Si no existe certidumbre debido a la distribución de probabilidad ($(P(0) = P(1) = 0.5)$), la información recibida será máxima (Fig. 3-1).

---

[3] Se reemplaza **H** de Shannon por **I** de información

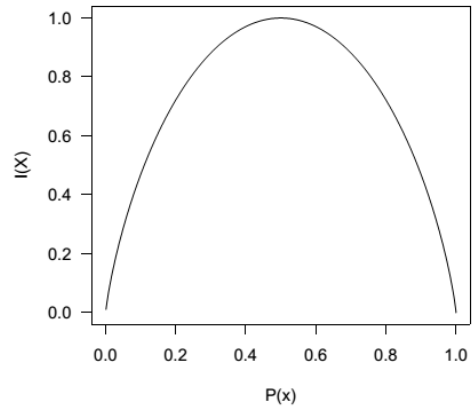

**Figura 3-1 Información de Shannon a Diferentes Probabilidades para una Cadena Binaria** (Gershenson and Fernández, 2012a)

Por ejemplo, si tenemos una cadena '0001000100010001…', podemos estimar que $P(0) = 0.75$ y P(1) = 0.25, entonces $I = -(0.75 * \log 0.75 + 0.25 * log\, 0.25$ ). Sí usamos $K = 1$ y logaritmo base 2, entonces $I \approx 0.811$. Sí transformáramos a escala porcentual, diríamos que el próximo valor que venga nos traerá un 81.1% de información, pues presumimos que tener un cero tendría mayor probabilidad. No obstante, puede llegar un uno.

Shannon usó $H$ para describir información (nosotros usamos $I$), debido a que él cuando desarrolló la teoría estaba pensando en el teorema de Boltzmann. El llamó a la ecuación 3.1, la entropía de un conjunto de probabilidades $p1, p2, …, pn$. En palabras modernas, $I$ es una función de una variable aleatoria $X$.

La unidad de información es el bit (binary digit). Un bit representa la información ganada cuando una variable binaria aleatoria se vuelve conocida. Sin embargo, desde la ecuación 3.1 es una suma de probabilidades, por lo que resulta ser una medida adimensional. Más detalles sobre la teoría de la información, pueden ser encontrados en (Ash, 2000, 1965).

En años recientes, la teoría de la información ($TI$) ha sido relacionada con la complejidad, auto-organización y emergencia (Prokopenko, 2009b). De ello se destaca que la teoría de la información presenta, en el contexto de la ciencia de los sistemas complejos-$CSC$, razones importantes cómo:

- Un considerable cuerpo de trabajo en $CSC$ ha sido desarrollado dentro de la $TI$, cómo pioneros tenemos al instituto Santa Fé, el cual ha direccionado la investigación de muchos de ellos.
- La $TI$ brinda definiciones que pueden ser formuladas matemáticamente

- La *TI* brinda herramientas computacionales disponibles al instante; un número de medidas puede ser actualmente computadas.

## 3.2 Medidas

Las medidas presentadas en este capítulo corresponden con las publicadas recientemente en Gershenson and Fernández(2012), y posteriores más refinadas y basadas en axiomas explicitas en Fernández et al. (2014b). El beneficio de usar axiomas está en que permite centrar la discusión en las presunciones o propiedades que estas medidas pueden tener, más que en la medida en sí misma. Adicionalmente, se incluyen medidas adicionales desde los trabajos de Fernández et al. (2012a, 2011, 2010a), que ahondan en las características de la homeostasis y la autopoiesis.

### 3.2.1 Emergencia

Hemos mencionado que la emergencia se refiere a las propiedades de un fenómeno que están presentes en una escala y no en otra. Las escalas pueden ser temporales, espaciales o numéricas.

Si describimos un fenómeno en términos de información, para tener *nueva* información, la información *vieja* debe haber sido transformada. Esta transformación puede ser *dinámica*, *estática*, *activa*, o *estimérgica* (Gershenson, 2012a). Por ejemplo, nueva información se produce cuando un sistema dinámico cambia su comportamiento, pero también cuando una descripción del sistema cambia. Concerniente con el primer caso, se han propuesto enfoques de medición de la diferencia pasada y futura (p.e. Shalizi & Crutchfield (2001)). Podemos llamar esta emergencia, *emergencia dinámica*. Concerniente con el segundo caso, los enfoques de medición se dan entre las escalas que han sido usadas (Holzer and de Meer, 2011; Shalizi, 2001), podemos llamar esta emergencia, *emergencia de escala*.

Aun cuando existan diferencias entre las emergencias *dinámicas* y de *escalas*, ambas pueden ser vistas cómo la producción de información nueva. En el primer caso, la dinámica produce nueva información. En el segundo caso, el cambio de descripción produce nueva información. Entonces, una medida de emergencia basada en información $E$ incluiría tanto la emergencia dinámica cómo la de escala. Si retomamos, Shannon propuso una cantidad que medía qué tanta información era producida por un proceso. En consecuencia, podemos decir que la emergencia es la misma información de Shannon $I$ (Ecuación 3.2.). En lo sucesivo, consideraremos a la emergencia de un proceso $E$, cómo la información $I$, y usaremos el logaritmo en base 2.

$$E = I \tag{3.2}$$

La idea intuitiva de emergencia cumple con las tres nociones básicas (axiomas) que Shannon usó para derivar la $I$ (Shannon $H$). Para el axioma de continuidad, se espera que una medida no tenga grandes saltos cuando se presentan pequeños cambios. El segundo axioma es más difícil de mostrar. Este establece que si consideramos una función auxiliar $i$, la cual es función de $I$, cuando existan $n$ eventos con la misma probabilidad $1/n$, entonces la función $i$ tendrá un incremento monotónico. Si tenemos la misma configuración para la emergencia, entonces podríamos pensar que el proceso tendría igual variabilidad en cualquiera de los $n$ estados disponibles. Si algo pasa y ahora el proceso puede estar en $n + k$ estados, igualmente probables, podemos decir que el proceso ha tenido emergencia, dado que ahora necesitamos más información para conocer en cual estado del proceso está. Para el tercer axioma, necesitamos encontrar una forma para resolver cómo se puede dividir el proceso. Recordemos que la tercera propiedad requerida para la $I$ de Shannon, es que si una opción puede ser dividida en dos diferentes, la $I$ original debe ser promedio de las otras dos $I$. En un proceso, podemos pensar en las opciones cómo una fracción del proceso que actualmente observamos. Para este propósito, podemos hacer una partición del dominio. En nuestro caso, podemos tomar dos subconjuntos cuya intersección es el conjunto nulo y cuya unión es el conjunto original completo. Después de ello, computamos la función $I$ para cada uno. Dado que observamos dos diferentes partes de un proceso y de cada observación tenemos el promedio de información requerida para describir el proceso (parcial), entonces toma sentido tomar el promedio de ambos cuando se observa el proceso total.

Al estar $E$ basada en $I$, tenemos que es una medida probabilística. $E = 1$ significa que cuando cualquier variable binaria se vuelve conocida, un bit de información emerge. Sí $E = 0$, entonces no emergerá nueva información, incluso variables aleatorias se vuelven "conocidas" (se conocen de manera anticipada, de antemano). Enfatizamos que la emergencia puede tener lugar en un nivel de observación de un fenómeno o en un nivel de descripción del fenómeno. Uno u otro, pueden producir nueva información.

- Múltiples Escalas y Normalización.

Para medir la emergencia, auto-organización, complejidad y homeostasis a múltiples escalas numéricas, las cadenas binarias pueden ser convertidas a diferentes bases. Si tomamos dos bits, tendremos cadenas de base 4. Si tomamos tres bits tendremos cadenas de base 8. Tomando $b$ bits tendremos cadenas de base $2^b$.

Cuando Shannon definió la ecuación 3.1, él incluyó una constante positiva $K$. Este hecho es importante debido a que podemos cambiar el valor de $K$ para normalizar la medida en el intervalo [0,1]. El valor de $K$ puede depender de la longitud del alfabeto finito $A$ que usemos. En el caso particular de las redes Booleanas, que tienen un alfabeto $A = \{0, 1\}$, su longitud es $|A| = 2$. Entonces, el valor de $K = 1$ normalizará la medida en el intervalo [0,1]. Debido a la relevancia de la notación binaria en ciencias computacionales y otras aplicaciones, en este trabajo se usará a menudo el alfabeto booleano. Sin embargo, podemos computar la entropía para alfabetos de diferentes longitudes. Para ello consideramos la ecuación 3.3

$$K = 1/log_2 b \qquad (3.3)$$

Donde $b$ es la longitud del alfabeto usado. En esta forma $E$ y las medidas derivadas de ella son normalizados, teniendo un máximo de 1 y un mínimo de 0. Por ejemplo, consideremos la cadena en base 4 '0133013301330133…'. Podemos estimar que $P(0) = P(1) = 0.25$, $P(2) = 0$, y $P(3) = 0.5$. Siguiendo la ecuación (3.1), tenemos que $I = -K((0.25 \times log\ 0.25) + (0.25 \times log\ 0.25) + (0.5 \times log\ 0.5))$, dado que $b = 4$, entonces $K = 1/log_2 4$, y obtenemos una $I = 0.75$ normalizada.

### 3.2.2 Auto-organización

La auto-organización ha sido asociada, correlacionada, con el *incremento* de orden. Por ejemplo, con la *reducción* de entropía (Gershenson and Heylighen, 2003). Si la emergencia implica un incremento de información, la cual es análoga a la entropía y al desorden, la auto-organización estará anti-correlacionada con la emergencia.

Una medida de auto-organización $S$ (por self-organization en inglés), será una función $S: \sum \to \mathbb{R}$ D donde $\sum = A^{\mathbb{N}}$, con las siguientes propiedades:

a. El rango de $S$ está en el intervalo de los reales [0,1]
b. $S(X) = 1$ sí y sólo sí $X$ es determinística. P.e. conocemos de antemano el valor del proceso.
c. $S(X) = 0$ sí y sólo sí $X$ tiene una distribución uniforme. P.e. cualquier estado de un proceso tiene la misma probabilidad.
d. $S(X)$ tiene una correlación negativa con la emergencia E

Proponemos cómo medida:

$$S = 1 - I = 1 - E \qquad (3.4)$$

Es sencillo comprobar que esta función cumple con los axiomas declarados. Sin embargo, no es la única formulación posible. No obstante, es la única función (lineal) ajustada que satisface los axiomas. Por simplicidad, proponemos usar esta expresión cómo medida de la auto-organización.

$S = 1$ significa que existe orden máximo. Es decir, no se produce nueva información ($I = E = 0$). En el otro extremo, cuando $S = 0$, no existe orden en absoluto. Es decir, cuando ninguna variable aleatoria se vuelve conocida se produce/emerge información.

Cuando $S = 1$, el orden es máximo, la dinámica no produce nueva información, entonces el futuro, desde el pasado, es completamente conocido. De otra parte, cuando $S = 0$ el orden es mínimo, la información pasada no nos dice nada acerca de la información futura.

Nótese que la ecuación 3.4 no hace distinción entre si el orden es producido por el (mismo) sistema o por su ambiente. Entonces, $S$ podría tener un gran valor en sistemas con alta organización, independientemente sí es un producto de las interacciones locales o impuesta externamente. Esta distinción puede hacerse fácilmente por la descripción en detalle del comportamiento de los sistemas, pero es difícil e innecesario de capturar con una medida de probabilidad general cómo $S$. Cómo analogía, podemos medir la temperatura de una substancia, pero la temperatura no nos diferencia (y no es necesaria para diferenciar) entre las substancias que están calientes o frías desde el exterior, y substancias cuya temperatura es principalmente dependiente de sus reacciones químicas internas.

Resaltamos que con esta medida de auto-organización los patrones y/o dinámicas estáticas pueden ser valorados adecuadamente por su alta S y baja entropía. Por ejemplo, los sistemas vivos reducen su entropía termodinámica para mantenerse, lo cual ha sido descrito con el concepto de auto-organización (Kauffman, 1993).

### 3.2.3 Complejidad

Sobre la base de Lopez-Ruiz et al. (1995), podemos definir complejidad $C$, cómo el balance entre el cambio (caos) y estabilidad (orden). Dado que hemos asociado tales aspectos a las medidas de emergencia y auto-organización, la complejidad es una función $C: \Sigma \to \mathbb{R}$, que tiene las siguientes propiedades:

- Su rango son los reales en el intervalo [0,1]
- $C = 1$ sí y sólo sí $S = E$
- $C = 0$ sí y sólo sí $S = 0$ y $E = 0$

Es natural considerar que el producto de $S$ y $E$ satisface los dos últimos requerimientos. Por lo tanto, proponemos:

$$C = 4 \times E \times S \tag{3.5}$$

Donde 4 es adicionada cómo una constante de normalización para cumplir con el primer axioma, tal que $C$ pertenecerá al intervalo [0,1].

$C$ puede ser representada en términos de $I$ cómo:

$$C = 4 \times I \times (1 - I) \tag{3.6}$$

La Figura 3-2 muestra las medidas propuestas para diferentes valores de $P(x)$. Allí es notorio que $E$ es máxima cuando $P(x) = 0.5$ y mínima cuando $P(x) = 0$ or $P(x) = 1$. Lo opuesto es para $S$, esto es mínima cuando $P(x) = 0.5$ y máxima cuando $P(x) = 0$ o $P(x) = 1$. $C$ es máxima cuando $E = S = 0.5$, lo cual ocurre cuando $P(x) \approx 0.11$ or $P(x) \approx 0.89$. La información de Shannon puede ser vista como un balance de ceros y unos (máxima cuando $P(0) = P(1) = 0.5$), en tanto que $C$ es vista como un balance de $E$ y $S$ (máxima cuando $E = S = 0.5$).

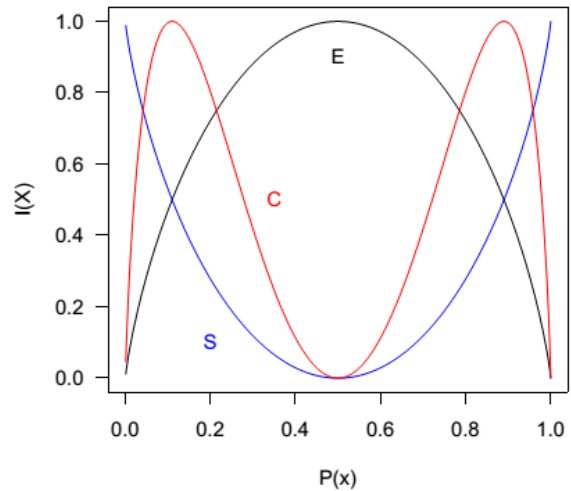

**Figura 3-2 Emergencia, Auto-organización y Complejidad vs Probabilidad en una Cadena Binaria**
(Gershenson and Fernández, 2012a)

Cabe destacar que históricamente, sí bien Shannon definió la información cómo entropía (Equivalente a nuestra $E$), Wiener y von Bertalanfy definieron información como el opuesto, cómo *negentropía* (Equivalente a nuestra $S$). En este sentido, nuestra medida de complejidad $C$, reconcilia estos dos puntos de vista al ser producto del balance entre el orden y el caos. Desde esta perspectiva, las ventajas de ésta medida son importantes, pues pueden caracterizar sistemas dinámicos con máxima complejidad en regiones complejas, cómo las redes booleanas y sistemas vivos que requieren un balance de estabilidad ($S$) y adaptabilidad ($E$). Casos que ilustran este tipo de situaciones se observarán en el capítulo de aplicaciones.

### 3.2.4 Homeostasis

- Homeostasis del Sistema Total (Homeostasis de los Estados del Sistema en el Tiempo)

Las anteriores medidas ($E, S$ y $C$) estudian cómo variables individuales cambian en el tiempo. Para calcular las medidas para un sistema total, podemos graficar el histograma, o simplemente promediar las medidas para todas las variables del sistema. Para la homeostasis $Hm$, estamos interesados en cómo todas las variables del sistema cambian, o no, en el tiempo. Se resalta que $E, S$ y $C$ se calculan desde las series de tiempo de las variables halladas en las columnas, mientras $Hm$ se enfoca en los estados del sistema que cambia en el tiempo, y cuya información se halla en las filas.

Sí el sistema ha tenido alta homeostasis, se espera que sus estados no cambien mucho en el tiempo. La homeostasis es una función $H: \sum \times \sum \to \mathbb{R}$ que tiene las siguientes propiedades:

- Su rango es el intervalo real [0,1]
- $H = 1$, Sí y sólo sí, para los estados $q$ y $q'$, $q = q'$, esto es: no hay cambio en el tiempo.
- $H = 0$, Sí y sólo sí, $\forall_i$, $q_i \neq q'_i$, esto es: todas las variables en el sistema cambiaron

Una función útil para la comparación de cadenas de igual longitud es la distancia de Hamming, considerada también en la teoría de la información (Hamming, 1950). La distancia de Hamming $d$ (ecuación 3.7) mide el porcentaje de símbolos diferentes en dos cadenas $X$ y $X'$. Para cadenas binarias, puede ser calculada con la función $XOR$ ($\oplus$). Su normalización lleva al intervalo [0,1]:

$$d(X, X') = \frac{\sum_{i \in \{0,\ldots,|X|\}} x_i \oplus x'_i}{|X|} \qquad (3.7)$$

$d$ mide la fracción de símbolos diferentes entre $X$ y $X'$. Para casos booleanos $d = 0 \Leftrightarrow X = \neg X'$, mientras $X$ y $X'$ están no correlacionadas $\Leftrightarrow d \approx 0.5$.

Nosotros podemos usar el inverso de $d$ para definir $h$:

$$h(X^t, X^{t+1}) = 1 - d(X^t, X^{t+1}) \qquad (3.8)$$

La ecuación 3.8 de homeostasis, claramente cumple con las propiedades deseadas para la homeostasis entre dos estados. Para medir la homeostasis de un sistema en el tiempo, podemos generalizarlo como se muestra en la ecuación 3.9:

$$H = \frac{1}{m-1} \sum_{t=0}^{m-1} h(X^t, X^{t+1}) \qquad (3.9)$$

Donde $m$ es el número total de sucesos de tiempo que serán evaluados. $H$ será el promedio de las diferentes $h$ desde $t = 0$ a $t = m - 1$. Cómo las medidas previas basadas en $I$, $H$ es una medida adimensional.

Cuando $H$ es medida a grandes escalas, captura dinámicas periódicas. Por ejemplo, sea nuestro sistema con $n = 2$ variables y un ciclo de periodo 2: $11 \to 00 \to 11$. $H$ para base 2 será mínima, dado que en todos los pasos de tiempo todas las variables cambian, los unos se cambian a ceros o los ceros a unos. Sin embargo, si la medida de $H$ se hace en base 4 tomamos dos eventos de tiempo binarios. Así, en base 4 el atractor será $22 \rightarrow 22$ y la $H = 1$. Lo mismo aplica para bases grandes. Ejemplos de éste funcionamiento se demostrará en el capítulo de aplicaciones.

- Homeostasis para Una Variable del Sistema ($HmV$)

Cómo se mencionó en el capítulo 1 numeral 1.3.4 (definición de homeostasis), se ha establecido que las variables esenciales de un sistema se mantienen en una zona de viabilidad. Esta zona, está definida por los límites inferiores y superiores (rango) en los que se mueve la variable (Ross Ashby, 1960b). Con esto en mente, desarrollamos una medida que valorara la dispersión de los estados de una variable (desviación), calculada a partir de los valores de probabilidad de tales estados. Se considera que a mayor dispersión- desviación entre estados, el sistema mostrará menor regularidad y menor será la homeostasis.

Para complementar la medida de $Hmv$ tiene en cuenta la equitabilidad entre estados (evenness, en inglés; relativo a uniformidad, regularidad). Más equitabilidad es más homeostasis, dado que las trayectorias del sistema serán más regulares (mayor auto-organización).

En el anterior contexto, definimos la homeostasis de una variable $HmV$ como una función de la dispersión entre las probabilidades de sus estados, representada en su desviación estándar ($\sigma P_i$) y el inverso de la equitabilidad (Buzas and Gibson's evenness). El cálculo tiene en cuenta la transformación de la variable según la escala elegida. La ecuación 3.10 muestra la relación planteada.

$$HmV = \frac{\sigma P_i}{Evenness} = \frac{\sigma P_i}{e^I/Cs} \qquad (3.10)$$

Donde:

$I$: Información de Shannon: $I = -\sum Pi * LogPi$

$Cs$: Número de casos a partir de los que se calculan las probabilidades.

Para efectos de normalización en el rango [0,1] se obtiene de la siguiente manera:

$$HmV_N = 1 - \frac{1}{Hm + 1} \qquad (3.11)$$

- Mecanismos Homeostáticos en Redes Computacionales de Agentes

A través de diferentes mecanismos homeostáticos, un sistema multi-agentes puede hacer frente a los cambios, influencias y/o perturbaciones de su ambiente interno o endógeno, y/o externo o exógeno $Ab$. Estos mecanismos de auto-regulación promueven la estabilidad y flexibilidad del sistema, en su paso por los múltiples estados de su ciclo adaptativo, sin que se dé su destrucción.

La base del mecanismo homeostático está en el desenvolvimiento de la red en una zona de viabilidad ($Zv$), definida por los rangos de viabilidad ($RX$), de nodos y aristas, a los factores $Ab$. Como consecuencia de la búsqueda de equilibrio ante la influencia de $Ab$, la homeostasis puede generar cambios semipermanentes en la red, como respuesta a los cambios semipermanentes de $Ab$.

En una red computacional de agentes, representados por nodos, cómo se explicó en el capítulo 2, los procesos de auto-regulación se pueden dar de manera complementaria, tanto de forma activa como pasiva. Para el primer caso se parte que la función global $F$ es una función de transición, tal que $F: Q \to Q$, donde $Q$ es un conjunto no vacío compuesto de un número finito de estados $q_i \in Q$. Al mismo tiempo, $F$ corresponde a una familia de funciones, denominadas funciones de activación local de la red ($f_i$), tal que $F = (f_1, f_2, \ldots, f_n)$. Por lo tanto, la auto-regulación activa se puede dar, acorde a la forma en que se den las diferentes combinaciones de los $f_i$. Así, al tener a "$o$" como operador, tal que $F = f_1 o\ f_2 o, \ldots, f_n$, será evidente que $F = f_1 o\ f_2 \neq F = f_2 o\ f_1$. Lo anterior permite observar que según se combinen las funciones parciales, se generará un cambio en la función global. Esta condición será un mecanismo básico que le dará a la red mayor adaptabilidad ante los cambios del entorno $Ab$.

La auto-regulación pasiva, por su parte, estará definida por situaciones como los rangos de viabilidad $RX$ de la red a la influencia de un factor ambiental $Ab$, como a continuación se aprecia:

- Ante la influencia $Ab$, la red se desenvuelve en una zona de viabilidad $Zv$.
- Tal $Zv$ está en correspondencia con los valores máximos ($Xmax$) y mínimos ($Xmin$) de un factor $i$ de $Ab$. Este hecho define la respuesta (*tolerancia*) de la red al factor $i$ de $Ab$, acorde con sus rangos de viabilidad ($RX_i$). En consecuencia, para cada factor $i$ de $Ab$, un nodo $j$ de la red tendrá un rango de viabilidad al factor $i$, resultante de la diferencia de los valores de extremos, tal que $RX_i^j = Xmax - Xmin$.
- Para el caso del rango de viabilidad en las aristas, que considerará igualmente los valores máximos y mínimos ante el factor $i$ de $Ab$, su notación estará dada por $RX'^j_i$.
- La respuesta global $RX_i$ se dará por el promedio de $RX_i^j$ y $RX'^j_i$.
- Se estima que la respuesta de la red o tolerancia ($T$) a un factor $i\ Ab$, puede coincidir, con una función gaussiana con un valor promedio ($\mu_i$) de su rango $RX_i$ (Fig. 3-3). La escogencia de la función normal, se debe a que ella coincide con la respuesta de muchos sistemas complejos, como los sistemas socio-ecológicos. No obstante, en caso de respuestas diferentes se pueden usar otros tipos de distribuciones.

- El valor de tolerancia para un determinado valor específico $X_i \in RX_i$ del factor $Ab_i$, deberá considerar como parámetros para su cálculo la desviación estándar $\sigma_i$, y el promedio $\mu_i$. La función también cuenta con las constantes $\pi$ y $e$, y su formulación se expresa en la ecuación 3.12.

- $$T(A_i) = \frac{1}{\sigma_i \sqrt{2\pi}} e^{\frac{-(X_i-\mu_i)^2}{2\sigma_i^2}} \tag{3.12}$$

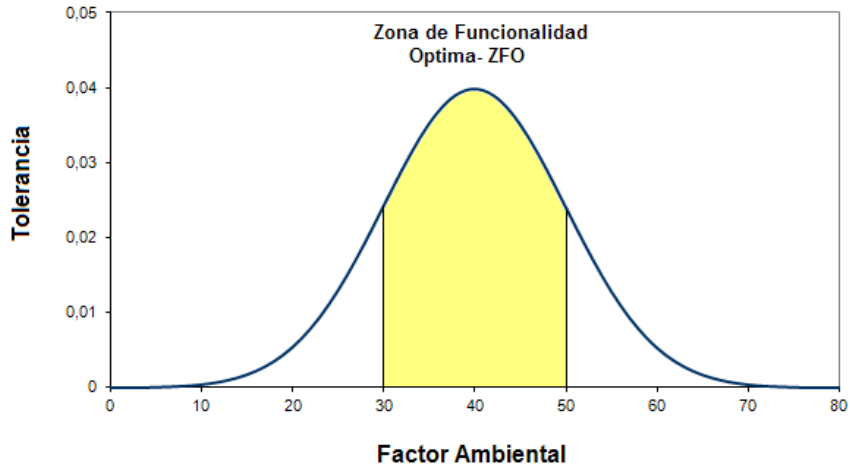

**Figura 3-3 Zona de Viabilidad *Zv* y Zona de Funcionamiento Óptimo (*ZFO*) de un factor $Ab_i$ o $B_i$ para una red computacional. Los valores del eje *x* dependen del dominio del factor *i*.** (Fernández et al., 2012b)

- Se destaca que dentro de la zona de viabilidad $Zv$, definida por la función de tolerancia $T(Ab)$, existe una zona de funcionalidad óptima $ZFO$, que se corresponde con una zona cercana al promedio, de cada factor $\mu_i$ del factor $i$ de $Ab$. Estadísticamente, esta área se encuentra entre $\mu_i \mp \sigma_i$.

- En general, los niveles más altos de tolerancia serán aquellos en los cuales el nivel de satisfacción de un nodo o agente es máximo. Ello sucede cuando el agente alcanza un objetivo particular, y al mismo tiempo se minimiza la fricción[4] con los demás agentes. Los mecanismos de auto-regulación como el descrito, son los que soportan el proceso auto-organizativo que alcanza niveles más altos de tolerancia y satisfacción, y mínimos de fricción. Lo anterior, coincide lo que Gershenson (2010) ha definido como perfil sigma.

---

[4] interacciones negativas que no favorecen el alcance del objetivo global del sistema.

- En caso que los nodos y/o aristas tengan tolerancias individuales heterogéneas, la tolerancia global de la red será producto del acople de las tolerancias individuales. Esta condición se muestra en la función de tolerancia global desde las funciones locales de los nodos y aristas, representada en la figura 3-4.

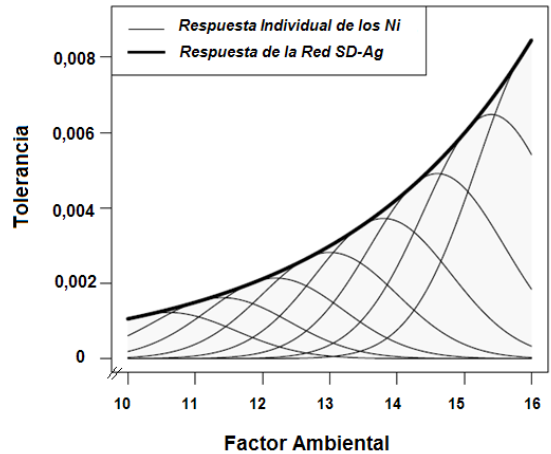

**Figura 3-4 Detalle de la Respuesta Global y Local en la Red *SD-Ag*. Los valores de los ejes son simplemente demostrativos** (Fernández et al., 2012b)

- Consecuente con lo anterior, se puede considerar que un factor $i$ de $Ab$ puede tener efecto sobre la totalidad de los nodos $Ni$ (incidencia global) o sobre determinados $Ni$ (incidencia parcial). No obstante, si se asumiera en algún caso que la influencia de $Ab_i$ sobre los $Ni$ es uniforme, no se podrá asumir que su respuesta sea igual. La respuesta diferencial de los $Ni$, será relativa a la capacidad de auto-regulación de cada uno de ellos para responder a las influencias de $Ab_i$.

- Si el funcionamiento de la red, nodos y/o aristas individuales se diese por fuera de la zona de viabilidad-$Zv$, ellos operarán en zonas definidas como críticas ($Zc$). Estar en $Zc$, puede generar su fallo extremo o degradación. En términos de la autopoiesis, esto es una limitación para el auto-mantenimiento pues se da la afectación de la estructura y función en la red. No obstante, al hallarse dentro de $Zc$, la red deberá compensar este efecto tanto en un sentido homeostático (de auto-regulación) como autopoiético (de auto-mantenimiento), con el fin de seguir operando. El mecanismo básico de compensación para ello se presenta en la siguiente sección.

▪ Mecanismo Homeostático General de Compensación ante Factores del Ambiente $Ab$

La siguiente es la secuencia de compensación en la red.

- La compensación será parte del mecanismo homeostático, y consistirá en la escogencia de una *acción* $y \in Y$ apropiada de respuesta, ante una influencia $Ab_i$. Para ello, se contará con la ejecución de una función objetivo ($f_{ob}$) que restablecerá el sistema al llevarlo nuevamente a la zona de viabilidad $Zv$, tal que $y(i, f_{ob}, Xmax, Xmin)$. La ejecución de $f_{ob}$ tendrá como parámetros: (i) que la acción relacionada controle efectivamente el desbalance, y (ii) que no se cause desequilibrio en los demás componentes de la red. Por lo tanto, $f_{ob}$ causará el mínimo trauma en la red y compensará efectivamente la influencia.

- La anterior visión coincide, nuevamente, con la maximización de la satisfacción y la reducción de la fricción en el sistema, como lo propone Gershenson(2010). Se puede estimar que $f_{ob}$ podría ser parte del algoritmo que modifica la red computacional.

- La formalización del mecanismo homeostático de compensación es extensible a la autopoiesis. Para tal fin, existirá una función $f_{ob}$ involucrada en la producción (síntesis-degradación) de los elementos de la red. Esto, debido a que la autopoiesis puede ser observada como un tipo particular de homeostasis.

- Otros indicadores del proceso homeostático

Hasta ahora la estabilidad y flexibilidad en la red computacional de agentes se han soportado en la función de tolerancia $T(Ab_i)$, que involucra al rango ($RX$), y en los mecanismos de auto-regulación pasiva y activa. Sin embargo, es posible aportar otros elementos que enriquecen la caracterización de esta misma estabilidad, e incluso determinar la "*fragilidad*" de la red. Como base se tienen los conceptos expresados por (Pimm, 1991) acerca de la resiliencia, persistencia y resistencia, en sistemas ecológicos. A partir de estas definiciones, se establecen nuevos formalismos para su representación matemática que pueden hacer posible la integración de todos ellos, en una medida extendida de homeostasis.

- **Resiliencia ($Rs_i$):** tasa (velocidad) a la que se retorna al punto medio $\mu_i$ en la zona de viabilidad $Zv$. Esto puede ser definido como $Rs_i := |Di|/t$, donde $|Di|$ o *distancia* es el valor absoluto de la diferencia entre el valor $X_i$ del factor $i$ de $Ab$ y su promedio $\mu_i$, tal que $Di = |X_i - \mu_i|$. Entre tanto, $t$ es el tiempo que demora en retornar al valor $\mu_i$. $Rs_i$ puede ser extendido a la tasa de retorno a un punto determinado de la zona de funcionalidad óptima-$ZFO$. Mayores valores de $Rs_i$ indican mayor capacidad homeostática.

- **Persistencia ($Pe_i$):** es la medida del tiempo que se pasa en el punto medio $\mu_i$, o en $ZFO$. $Pe_i = t_r$, donde $t_r$ es el tiempo de residencia en esos valores anteriores. Valores grandes de $Pe_i$ indican más capacidad homeostática.

Los dos indicadores anteriores ($Rs_i, Pe_i$) tienen un comportamiento opuesto en el tiempo (Fig. 3-5), dado que a mayores tiempos de retorno la resiliencia se hace menor, y en consecuencia, la capacidad homeostática. Entre tanto, al aumentar el tiempo de mantenimiento en el punto medio, la persistencia es mayor, y así mismo, la capacidad homeostática de la red.

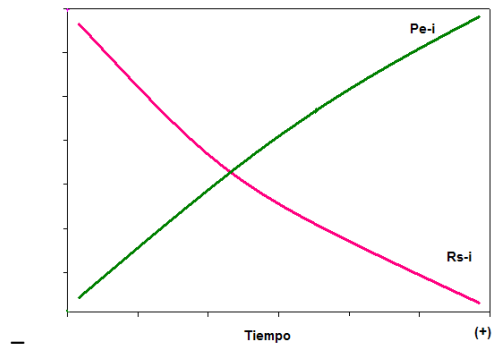

**Figura 3-5 Comportamiento de la resiliencia ($Rs_i$) y la persistencia ($Pe_i$):) en el tiempo de retorno y el tiempo de estancia en el punto medio respectivamente, en constraste con la capacidad homeostática** (Fernández et al., 2012b).

- **Resistencia Máxima ($Re - Max_i$)**: es la medida del *grado* de desplazamiento máximo con referencia al $|\mu_i|$, tal que el grado de desplazamiento máximo se corresponde con el valor absoluto de $RX_i$. Es decir, $Re - Max_i = |RX_i|/|\mu_i|$. Mayores valores de $Re - Max_i$ indican mayor capacidad homeostática. Cabe notar que, de la misma forma que se obtiene resistencia máxima se puede obtener otras variantes como la Resistencia $ZFO$, o la resistencia instantánea, acorde con los valores de $X_i$ que se tengan en cuenta.

- **Vulnerabilidad ($Vu_i$):** la susceptibilidad de la red ante el factor *i* de *Ab* estará definida por una función inversa con el rango de viabilidad $RX_i$, tal que $Vu_i$=1/(1+RX$_i$), donde $Vu_i \in [0,1]$. Mayor $Vu_i$ indica menor capacidad homeostática.

El comportamiento de la resistencia máxima y la vulnerabilidad a medida que se incrementa el rango, se puede observar en la figura 3-6. En ella es notorio que la resistencia máxima se incrementa con el rango, y así mismo se transfiere mayor capacidad homeostática a la red. En tanto, la vulnerabilidad se hace menor a medida que el rango se incrementa, y por ende, la capacidad homeostática aumenta.

Dado que el proceso de auto-regulación es un proceso dinámico, los anteriores indicadores deben ser observados en el tiempo para determinar su cambio.

A manera de síntesis, se puede decir que la capacidad de auto-regulación u homeostasis de la red multi-agentes ($H_{SD-Ag}$) a un factor $i$ dado de *Ab*, estará dado por la caracterización de los indicadores. Así, la $H_{SD-Ag}$ de la red será una función $g$ de la tolerancia, resiliencia, persistencia, resistencia máxima, y vulnerabilidad, tal que $H_{SD-Ag} = g(T(Ab_i), Rs_i, Pe_i, Re-Max_i, Vu_i)$. Al normalizar por cualquier procedimiento $Rs_i, Pe_i, Re-Max_{i_i}$ a una escala $[0,1]$, la función $g$ de $H_{SD-Ag}$ podrá calcular el promedio de los mismos.

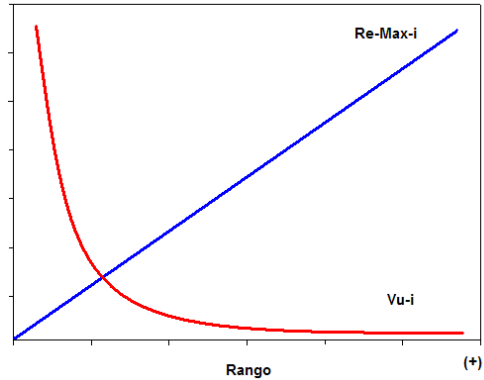

**Figura 3-6 Comportamiento de la resistencia máxima ($Re-Max_i$) y la vulnerabilidad ($Vu_i$) en cuanto al incremento del rango de viabilidad ($RX_i$)** (Fernández et al., 2012b).

### 3.2.5 Autopoiesis

- Autopoiesis cómo expresión de la Autonomía del Sistema

Los siguientes aspectos formales de la autopoiesis se basan en la consideración que los sistemas adaptativos requieren alta complejidad o $C$ para enfrentar los cambios ambientales y mantener su integridad (Kaufmann, 1993; Langton, 1990). Desde esta perspectiva, fue posible definir una expresión de autopoiesis desde la autonomía del sistema. A continuación presentamos las consideraciones axiomáticas para ello.

Sea $\bar{X}$ las trayectorias de las variables de un sistema y $\bar{Y}$ las trayectorias de las variables del ambiente del sistema. Una medida de autopoiesis $A: \sum \times \sum \to \mathbb{R}$ tiene las siguientes propiedades:

- $A \geq 0$
- $A$ deberá reflejar la independencia de $\bar{X}$ y $\bar{Y}$, lo que implica:
    - $A > A' \Leftrightarrow \bar{X}$ produce más de su propia información que $\bar{X}'$ para un $\bar{Y}$ dado.
    - $A > A' \Leftrightarrow \bar{X}$ produce más de su propia información en $\bar{Y}$ que en $\bar{Y}'$
    - $A = A' \Leftrightarrow \bar{X}$ produce mucho más de su propia información que $\bar{X}'$ para un $\bar{Y}$ dado.
    - $A = A' \Leftrightarrow \bar{X}$ produce mucha más de su propia información en $\bar{Y}$ que en $\bar{Y}'$
    - $A = 0$ si toda la información en $\bar{X}$ es producida por $\bar{Y}$.

Siguiendo la clasificación de tipos de transformación de información propuesta por Gershenson (2012b), las transformaciones dinámicas y estáticas son internas (un sistema produce más de su propia información), mientras que las transformaciones activas y estimérgicas son externas (información producida por otro sistema).

Desde la alta complejidad requerida por un sistema para afrontar el ambiente, tenemos que sí el sistema $\bar{X}$ tiene alta $E$, entonces podría pasar que no esté apto para producir los mismos patrones para diferentes $\bar{Y}$. Por otra parte, si se tiene alta $S$, $\bar{X}$ podría no estar apta o ser capaz para adaptarse a los cambios en $\bar{Y}$.

Por lo tanto, proponemos la autopoiesis como una relación de las complejidades del sistema $C(\overline{X})$ y su ambiente $C(\bar{Y})$, acorde con la ecuación 3-13

$$A = \frac{C(\overline{X})}{C(\bar{Y})} \tag{3.13}$$

Sí $C(\overline{X}) = 0$, entonces $\bar{X}$ es estática ($E(\bar{X}) = 0$) o seudoaleatoria ($S(\bar{X}) = 0$). Esto implica que cualquier patrón (complejidad) que pueda ser observado en cualquier $\bar{X}$, debe venir de $\bar{Y}$. Este caso da una mínima $A$. Por otra parte, sí $C(\bar{Y}) = 0$, esto implica que cualquier patrón en $\bar{X}$, sí lo hay, debe venir de sí mismo. Este caso da una $A = \infty$. Un caso particular ocurre sí $C(\bar{X}) = 0$ y $C(\bar{Y}) = 0$, $A$ resulta indefinida. Pero cómo podemos decir algo sobre la autopoiesis si comparamos dos sistemas los cuales son sín variación ($S = 1$) o seudoaleatorios ($E = 1$), este debe ser un caso indefinido. El resto de las propiedades son evidentemente cumplidas por la ecuación 3.13. Se aclara, que esto no es ciertamente la única función que cumple con los axiomas deseados. La exploración de alternativas requiere más estudios.

Dado que $A$ representa la proporción de las probabilidades, esta es una medida adimensional. $A \in [0, \infty)$, sin embargo ella puede ser mapeada a [0,1] usando una función cómo $f(A) = \frac{A}{1+A}$. No obstante, nosotros no normalizamos $A$ debido a que es útil distinguir cuando $A > 1$ y $A < 1$. El primer caso corresponde cuando el sistema es más autónomo y adaptable que su ambiente, y el segundo cuando el sistema, por su menor complejidad, tiene menores posibilidades de adaptación que su medio, hecho que le confiere menor autonomía. Esto se podrá ver claramente en el capítulo de aplicaciones.

- Procesos Autopoiéticos de Auto-Mantenimiento en Redes Computacionales de Agentes

Desde el punto de vista del auto-mantenimiento, *A* es la responsable de los procesos de auto-regulación de la síntesis[5] y degradación[6] de los agentes y sus interacciones (Fernández et al., 2010a). *A* se soporta en mecanismos homeostáticos específicos, enfocados en la constitución de los elementos de la red computacional, y promueve la "producción del sistema" y la "preservación de la estructura" (Iba, 2011). Por medio de *A* se establece la capacidad de la red para desarrollar, mantener, producir y restablecer su unidad e identidad en un nivel dado. Particularmente, este proceso está basado en la hetero y autoreferenciación[7], hecho que involucra un grado determinado de cognición. La cognición, según Gershenson (2010), se refiere al conocimiento que el sistema tiene de cómo actuar en su ambiente.

Se puede estimar que en una red de agentes, *A* se corresponde con una función *f* de auto-mantenimiento ($f_a$) que hace parte de la combinación de las funciones de activación local $f_i$ que regula el "desgaste" natural de la red. El desgaste, en el sentido homeostático, es visto como una variante de un factor *i* de *Ab*. En consecuencia, el mecanismo de ejecución de la autopoiesis corresponde con el mecanismo general de auto-regulación, el cual puede estar inmerso en el algoritmo *a* de la red computacional.

El auto-mantenimiento, referido a *A*, puede ser medible a través del grado de "auto-mantenimiento" de la red en la constitución de nodos *N* y aristas *K*. La medida del auto-mantenimiento autopoiético puede ser expresada a través de un balance de masas (Holzbecher, 2007). El balance considera la dinámica de la constitución de los *N*, a partir de la diferencia de sus tasas de síntesis y de degradación. Para un *N* de la red, la dinámica de su auto-mantenimiento será igual a su síntesis (*Si*), o producción local, menos su degradación local (*Dg*), de tal manera que:

$$dN/dt = dSi/dt - dDg/dt \qquad (3.14)$$

La síntesis de los nodos *N* estará dada por $Si = \gamma N$, donde γ será el coeficiente o la razón promedio de síntesis de *N*. La degradación o perdida de *N*, será proporcional a $D = -\lambda N^d$, donde $\lambda$ corresponde a la constante de degradación promedio de *N*. El exponente *d*, es el orden de la degradación (Holzbecher, 2007). Así, la ecuación 3.14 toma la forma:

---

[5] La síntesis autopoiético se refiere al proceso de constitución y mantenimiento tanto de los *N*, como de la estructura de que depende la constitución del elemento. Esto se logra a partir del establecimiento de una red de procesos de producción (Luhmann 1995).
[6] La degradación, por su parte, se refiere a la perdida de integridad estructural o unidad de los *N*, y en consecuencia de la estructura que los soporta y de las condiciones para la reproducción (Niklas Luhmann, 1995).
[7] *Referida a relación, semejanza o dependencia que como producto de la cognición el elemento hace sobre el exterior (hetero-referencia) y sobre sí mismo (auto-referencia).*

$$dN/dt = \gamma N - \lambda N^d \tag{3.15}$$

Cuando $d$ es de orden 1, la diferencia de los promedios de las tasas de *Síntesis* y *Degradación* en una red es la siguiente:

$$\frac{dN}{dt} = \gamma N - \lambda N^1 = N(\gamma - \lambda) = Nrp \tag{3.16}$$

Ahora podemos definir la tasa de reparabilidad de la red ($rp$). Acorde con los valores que $rp$ puede tomar, se tiene que sí $rp > 0$, la capacidad auto-regenerativa se incrementará y habrá auto-crecimiento de la red; sí $rp = 0$, la capacidad de regeneración se mantiene estable. Sí $rp < 0$ la capacidad decrecerá y el sistema evidenciará un proceso de desgaste o degradación.

El anterior razonamiento puede ser instanciado en la síntesis-degradación de las aristas, de forma que para una arista $K(i,j)$, con sus correspondientes tasas de síntesis ($\gamma'$) y degradación ($\lambda'$), la ecuación correspondiente será:

$$\frac{dK}{dt} = \gamma' K - \lambda' K^1 = K(\gamma' - \lambda') = Krp' \tag{3.17}$$

El promedio de $rp$ y $rp'$ dará la tasa de reparabilidad de la red-$R$.

Cabe anotar que si bien todos los sistemas autopoiéticos son homeostáticos, todos los sistemas homeostáticos pueden no ser autopoiéticos. Razón por la cual, para los que sí lo son, la formulación de homeostasis incluirá las tasas de reparabilidad $rp$ y $rp'$.

## 3.3 Síntesis

A partir de la proposición de diversos axiomas, en este capítulo se han propuesto medidas para la emergencia, auto-organización, complejidad, homeostasis y autopoiesis, basadas principalmente, en la teoría de la información. Estas medidas son de tipo simple y general, y básicamente representan de forma adecuada lo que miden. El beneficio potencial de estas medidas y aplicaciones son múltiples. Aún si mejores medidas son halladas, las nuestras alcanzan el objetivo de caracterizar de forma general la complejidad, emergencia, auto-organización, homeostasis y autopoiesis.

Adicionalmente, se profundiza en aspectos importantes para la homeostasis y la autopoiesis. Para la primera, el resultado es el poder medir, no sólo cómo el sistema cambia o no su equilibrio en el tiempo, sino cómo varia a través de sus cambios de estado. Se complementa la homeostasis, con la descripción de sus procesos activos y pasivos, fundamentales para el entendimiento como mecanismo de auto-regulación, y se brinda un conjunto de indicadores interesantes que la describen detalladamente. Para la autopoiesis, desde la perspectiva de redes computacionales de agentes, se profundiza en su significado como proceso de auto-mantenimiento, y se brinda un formalismo.

Cabe destacar que a nivel formal, son carentes las expresiones de Homeostasis, de manera que los formalismos para ella desarrollados, son de especial interés. En este mismo sentido se puede opinar de la autopoiesis, desde la perspectiva de la autonomía del sistema respecto de su ambiente.

La importancia de desarrollar una medida de complejidad, está en que ella debe estar disponible para permitirnos responder preguntas como: ¿Es un desierto más o menos complejos que una paramo? ¿Cuál es la complejidad de diferentes brotes de influenza? ¿Cuáles organismos son más complejos: predadores o presas; parásitos u hospederos; individuos o sociedades? ¿Cuál es la complejidad de situaciones tan disimiles cómo diferentes como los géneros musicales y los diferentes regímenes de tráfico? ¿Cuál es la complejidad requerida de una compañía para enfrentar la complejidad de un mercado? ¿Cuándo un sistema es más o menos complejo? Dada la dificultad existente en la noción de complejidad, y más aún en su medición, el desarrollo de medidas generales y simples de ella, es una tarea de gran importancia y utilidad.

Como se dijo antes, la conveniencia de tener una medida de auto-organización que capture la naturaleza de la dinámica local en una escala global, es especialmente relevante en el campo de la auto-organización guiada u orientada (*AOG*) (Ay et al., 2012; Prokopenko, 2009b).

Para el caso de la autopoiesis, desde las tasas de síntesis y reparabilidad, futuros trabajos pueden probar su utilidad al aplicarla a modelos dinámicos de autopoiesis. Para ello trabajos como el de Bourgine & Stewart (2004) pueden dar bases para la especificación de las tasas.

# Capítulo 4: APLICACIONES EN SISTEMAS DISCRETOS Y SISTEMAS URBANOS

## Resumen

*Definidos los aspectos teóricos, conceptuales y formales de la complejidad en SDC con múltiples agentes, en este capítulo desarrollamos experimentos y aplicaciones computacionales, para mostrar el significado y la utilidad de nuestras medidas de complejidad, auto-organización, emergencia, homeostasis y autopoiesis. El primer casos de estudio corresponde a redes booleanas y autómatas celulares elementales, los cuales han sido extensamente utilizados en la literatura científica como modelos clásicos para representar dinámicas ordenadas, críticas y caóticas, al igual que para detectar cambios de fase. El segundo caso corresponde a la caracterización de dos tipos de modelos de tráfico urbano, uno con coordinación tradicional y otro con un método auto-organizante y fronteras no orientables, que demuestra la utilidad de nuestras medidas en la identificación y descripción de diferentes fases dinámicas del flujo y de la velocidad vehicular. Los resultados muestran que los formalismos desarrollados, efectivamente, miden lo que representan, y describen regímenes desde ordenados hasta caóticos, pasando por los regímenes complejos. Dado que todo puede ser descrito y medido en términos de información, con nuestras medidas se puede caracterizar las diferentes fases de un SDC.*

## 4.1 Sistemas Discretos

### 4.1.1 Las Redes Booleanas Aleatorias-*RBA*

Las redes booleanas aleatorias (*RBA*) son modelos computacionales abstractos, originalmente propuestos para el estudio de redes genéticas de regulación (Kauffman, 1969; Kaufmann, 1993). No obstante, han sido de utilidad para abordar el conocimiento de diversos casos, como en sistemas auto-organizantes, en los que se estudia cómo cambios en los nodos y las conexiones afectan la dinámica global de la red. Igualmente, las *RBA* han sido de utilidad para estudiar la auto-organización guiada, la cual consiste en que, dependiendo de la configuración y régimen del sistema dinámico, se pueden realizar diferentes intervenciones para llevarlo a un punto de mayor o menor auto-organización (Aldana-González et al., 2003; Gershenson, 2012a, 2004a).

Básicamente, una *RBA* está formada por *N* nodos unidos por *K* conexiones. Cada nodo tiene un estado booleano, como por ejemplo cero o uno, vivo o muerto, perdió o ganó, etc. El estado futuro de cada nodo está determinado por los estados actuales de los demás nodos, con los que tiene conexión. Su estructura se determina por la red de los nodos interactuantes. Su funcionamiento se basa en tablas de búsqueda que especifican cómo es la actualización del estado de cada nodo, en dependencia con el estado de sus entradas. La conectividad y la actualización de los estados de los nodos se generan aleatoriamente al iniciar la red, pero permanece fija la conectividad en cuanto inicia la dinámica.

Las *RBA* clásicas se actualizan sincrónicamente (Gershenson, 2004b), lo que define una dinámica determinística. Cómo su espacio de estados es finito ($2^N$ estados), tarde o temprano, uno o más estados se repetirán. Estos estados pueden ser definidos como atractores. Los atractores alcanzados, pueden ser puntuales (un estado único), o cíclicos de diferentes longitudes.

En dependencia con sus propiedades estructurales o funcionales, se han definido para las *RBA* tres tipos de regímenes, que han sido estudiados de forma extensa (Gershenson, 2004a). (i) Ordenado: muchos nodos son estáticos, lo que da robustez a la *RBA* a perturbaciones. (ii) Caótico: muchos nodos son cambiantes, lo que genera *RBA* frágiles a perturbaciones. (iii) Crítico: algunos nodos cambian, lo que le da a la *RBA* un potencial de adaptabilidad por el balance de los regímenes caóticos con ordenados. Dinámicas críticas están relacionadas con alta complejidad.

Se puede decir que el régimen crítico es el balance entre la robustez del régimen ordenado y la variabilidad del régimen caótico. Desde los 90s, se ha argumentado que la computación y la vida requieren de este balance para computar y adaptarse (Kaufmann, 1993; Langton, 1990).

### 4.1.2  Midiendo la Complejidad en Redes Booleanas con Diferentes Regímenes y Grados de Conectividad.

A continuación, se aplican las medidas propuesta en el capítulo 3 al caso de estudio de las redes booleanas con diferentes regímenes. Los resultados se obtuvieron desde el promedio de 1000 *RBA*, después de 1000 iteraciones, que iniciaron en estado aleatorio y que dieron como resultado la información de entrada. Seguidamente, las propiedades de Emergencia ($E$), Auto-organización ($S$), Complejidad ($C$) y Homeostasis ($H$) fueron calculadas para los datos generados en 1000 iteraciones adicionales. Los paquetes BoolNet (Müssel et al., 2010) y entropy (Hausser and Strimmer, 2012), del proyecto R (www.r-project.org), fueron usados como apoyo.

Las series de tiempo de nodos simples fueron evaluadas, y con ello se obtuvo el promedio de los resultados para la red. Para la medida de homeostasis $H$, $d$ fue medida entre los estados $t$ y $t-1$, para la serie de tiempo total.

### 4.1.3 Resultados para Redes Booleanas

La figuras 4-1 y 4-2 muestran los resultados de las *RBA* con 100 nodos y conectividad $K$ variable entre $0-5$. Se observa que para bajos $K$, existe alta $S$ y $H$ ($S, H \cong 1$), y baja $E$ y $C$ ($E, C \cong 0$). Esto refleja el régimen ordenado de *RBA*, que tienen alta robustez y pocos cambios. Se puede decir que en esta fase existe poca o ninguna información emergente y que hay alto grado de auto-organización y homeostasis, debido a que los estados de los nodos tienen alta regularidad, de manera que el estado siguiente no trae información nueva, es decir, hay una alta probabilidad que el estado más común se repita.

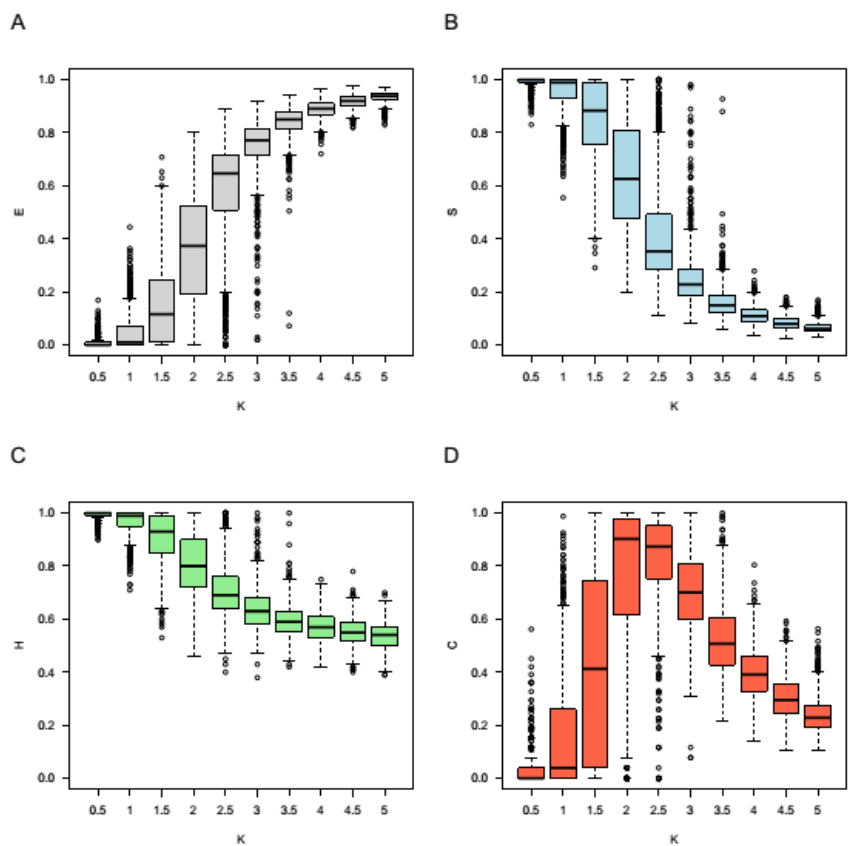

**Figura 4-1 Diagramas de Cajas para los resultados de 1000 RBA, N=100 con variación de la conectividad K. A. Emergencia B. Auto-organización (S). C. Homeostasis y D. Complejidad** (Gershenson and Fernández, 2012a).

Para altos $K$, existe alta $E$, baja $S$ y $C$, y una homeostasis de 0.5. Esto refleja los rasgos del régimen caótico (alta variabilidad y aleatoreidad) de las *RBA*, donde existe alta fragilidad en la red y muchos cambios. Casi cada bit (un nuevo estado para muchos nodos) trae información emergente nueva. En este sentido, los constantes cambios de estado en la red implican baja organización ($S = 0; E = 1$) o regularidad, y baja complejidad o capacidad de adaptación.

Para conectividades medias ($2 \leq K \leq 3$) existe un balance entre $E$ y $S$ (cercanas a 0.5), lo que conduce a alta $C$. Esto corresponde con el régimen crítico de *RBN*, el cual ha sido asociado con complejidad y la posibilidad de vida (Kauffman, 2000).

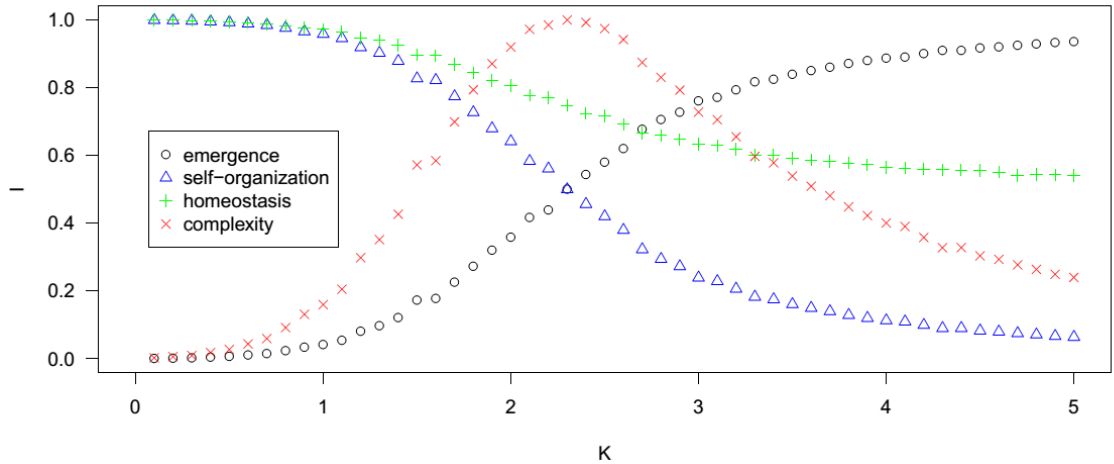

**Figura 4-2 $E, S, C$ y $H$ promedio para 1000 *RBA*, $N = 100$ nodos variando el promedio de conectividad $K$ (Gershenson y Fernández, 2012).**

En síntesis se puede decir que la emergencia aumenta con la conectividad, mientras que la auto-organización disminuye con ella, y la complejidad se hace máxima a niveles intermedios de $K$. El aumento de la conectividad hace que surja mayor probabilidad que el estado de un nodo, por la afección de otros, cambie en una mayor proporción.

Los experimentos de *RBA* también fueron llevados a cabo para múltiples escalas de observación. Si bien existen las escalas espaciales y temporales, también existen las escalas numéricas. El cambio de escala se logra a través de un cambio de la base, cómo se explica en el capítulo de formulaciones matemáticas, en el ítem de "Múltiples Escalas y Normalización". Las escalas en las que se evaluaron las medidas fueron 2, 4, 6, 8, 16, 32 y 64.

El análisis a múltiples escalas se dio debido a que la información puede cambiar drásticamente con la escala observada. Por ejemplo, una cadena 10101010 tiene $I_1 = 1$, pero a una escala mayor, por ejemplo base 4, la cadena anterior se vuelve 2222 y su información es $I_2 = 0$.

Sobre la anterior base, se calcularon $E, S, C$ y $H$ a múltiples escalas. Los resultados obtenidos no mostraron cambios importantes entre escalas, excepto por una pequeña elevación del valor máximo de $C$, el cual que se incrementó moderadamente con la escala. Para $E$, se halló un pequeño decremento a altas escalas, que puede ser atribuido al pequeño tamaño de las cadenas evaluadas (1000 por nodo). Los valores de $H$ decrecieron con la escala.

Para el cálculo de autopoiesis ($A$), que relaciona la complejidad del sistema con el medio ambiente, de manera que se pueda observar su grado de autonomía del primero sobre el segundo, se modeló a través de dos redes acopladas que representaron el sistema y su ambiente. La primera *RBA*, fue una red interna que simuló el sistema, con $N_i$ nodos y $K_i$ conexiones promedio. La segunda, fue una red "externa", con $N_e$ nodos y $K_e$ conexiones promedio, que simuló el ambiente. Se considera una *RBA* acoplada como $N_c = N_i + N_e$ nodos y $K_i$ conexiones. En cada iteración, la *RBA* externa evolucionó independientemente. Sin embargo, su estado fue copiado a los $N_e$ representados en la *RBA* acoplada, los cuales ahora evolucionan parcialmente dependiendo de la *RBA* externa. Así, los $N_i$ nodos en la *RBA* acoplada representaron la red interna que puede ser afectada por la dinámica de la *RBA* externa, pero no viceversa. Se simularon 50 redes acopladas con 96 nodos en la externa y 32 en la interna.

Como base para el cálculo de la autopoiesis se obtuvo la complejidad $C$ de cada nodo, que luego fue promediada para cada red. De allí se obtuvo una complejidad interna $C_i$ y una externa $C_e$. La autopoiesis $A$ fue obtenida desde la relación $A = C_i/C_e$.

Los valores resultantes de $A$ se reportan en la tabla 4.1 y se ilustran en la figura 4.3. En ellas, la autopoiesis $A$ cambia con la conectividad. Se espera, entonces, tener $A \approx 1$ cuando $Ki \approx Ke$. Cuando la complejidad externa $C_e$ es alta ($Ke = 2$ o $Ke = 3$), entonces el ambiente ($RBA_e$) domina los patrones de la red acoplada, conduciendo a $A < 1$. Esto es debido a que el sistema al tener menor complejidad que su ambiente no puede adaptarse a los cambios del mismo por su menor autonomía. Es por ello que los patrones que muestra el ambiente se imponen a los patrones que muestra el sistema.

En caso contrario, cuando $C_i > C_e$, o sea $C_e$ es más baja ($Ke < 2$ o $Ke < 3$), los patrones producidos por el sistema no son afectados, de mayor manera, por su ambiente; entonces $A > 1$, siempre que $K_i < K_e$ (de otra forma, el sistema es más caótico que su ambiente, y la complejidad de los patrones vienen de fuera).

$A$ no trata de medir cuanta información emerge en el sistema o en el ambiente, sino como los patrones que son interna o externamente producidos se afectan entre sí. Alta $E$ significa que no hay patrones, pues existe un cambio constante. Alta $S$ implica que el patrón es estático. Alta $C$ refleja patrones complejos (balance entre la inexistencia de patrones y patrones estáticos). En este trabajo, hemos estado interesados en una medición de $A$ a partir de la razón de la complejidad de los patrones que se producen por un sistema, *comparada* con la complejidad de los patrones producidos por su ambiente. $A$ en este aspecto, es una expresión de la autonomía del sistema respecto de su ambiente, sobre la base de su complejidad que le profiere adaptabilidad. Un sistema más autónomo no será influido por los patrones del ambiente.

**Tabla 4-1 Autopoiesis (A) promedio para 50 conjuntos de redes Ne=96, Ni=32. En rojo: redes acopladas con $A < 1$, en azul redes acopladas con $A > 1$ Los resultados son los mismos que se ven en la fig. 4-3** (Fernández et al., 2014b).

|  |  | $K_e$ ($N_e = 96$) | | | | |
|---|---|---|---|---|---|---|
|  | $A = {C_i}/{C_e}$ | **1** $S=1; C=0$; R.ordenado | **2** | **3** $S=E; C=1$; R. crítico | **4** | **5** $E=1; C=0$; R. caótico |
| $K_i$ ($N_i = 32$) | **1** $S=1; C=0$; R.ordenado | 0.4464025 | 0.5151070 | 0.7526248 | 1.6460345 | 3.4081967 |
|  | **2** | 1.6043330 | 0.9586809 | 1.1379227 | 2.0669794 | 3.2473729 |
|  | **3** $S=E; C=1$; R. crítico | 2.4965328 | 0.9999926 | 0.9355231 | 1.3604272 | 2.6283798 |
|  | **4** | 2.1476247 | 0.7249803 | 0.6151742 | 0.8055051 | 1.38900630 |
|  | **5** $E=1; C=0$; R. caótico | 1.8969094 | 0.4760027 | 0.3871875 | 0.4755580 | 0.8648389 |

En la figura 4-3 el tamaño de los círculos es relativo a su valor negativo (en rojo) o positivo (en azul).

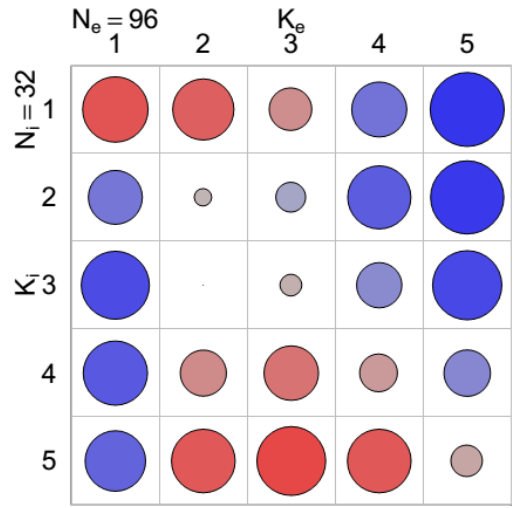

**Figura 4-3 Autopoiesis (A) promedio para 50 conjuntos de redes Ne=96, Ni=32. Valores de A<1 en rojo. Valores en azul para A>1. El tamaño de los círculos indican que tan lejos está A de A=1. Los valores numéricos se muestran en la tabla 4-1** (Fernández et al., 2014b).

## 4.2 Autómatas Celulares Elementales (ACE)

Los autómatas celulares elementales son modelos matemáticos para representar sistemas dinámicos que evolucionan en pasos discretos. Los *ACE* han sido estudiados de manera igualmente extensa (Wolfram and Gad-el-Hak, 2003; Wolfram, 1984; Wuensche and Lesser, 1992). Ellos pueden ser vistos como casos particulares de *RBA* (Carlos Gershenson, 2002; Wuensche, 1998), donde los nodos tienen la misma función (regla) y la estructura es regular. Esto es, cada nodo tiene $K = 3$ entradas: ellos mismos y sus dos vecinos más cercanos. Dado que hay 8 posibles estados binarios para las 3 celdas de vecinos en una célula dada, existen 256 "reglas" posibles. Es decir, las posibles combinaciones de funciones booleanas para tres entradas $22^{\wedge}3 = 2^8 = 256$. Cada regla puede ser indexada por un número binario de 8 dígitos (Wolfram and Gad-el-Hak, 2003; Wolfram, 1984; Zenil, 2009).

Por ejemplo, la regla 30 en notación binaria es como sigue: $30_2 = 00011110$. Su regla de evolución se ilustra en la figura 4-4. En ella se puede observar los valores de los conjuntos de tres vecinos en la parte superior que resultan en la condición inferior de la celda central en la siguiente generación.

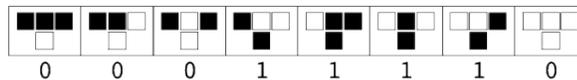

**Figura 4-4 Evolución para la Regla 30 desde condiciones iniciales para grupos de tres vecinos.**
http://reference.wolfram.com/language/ref/CellularAutomaton.html

La evolución de la regla 30 para dos pasos, a partir del acostumbrado estado inicial que parte de una única celda central en estado 1, será numéricamente así:

- Inicial:  0,0,0,1,0,0,0
- Paso 1:   0,0,1,1,1,0,0
- Paso 2:   0,1,1,0,0,1,0

Pasados 50 pasos o iteraciones, su forma gráfica es como se ve en la siguiente figura.

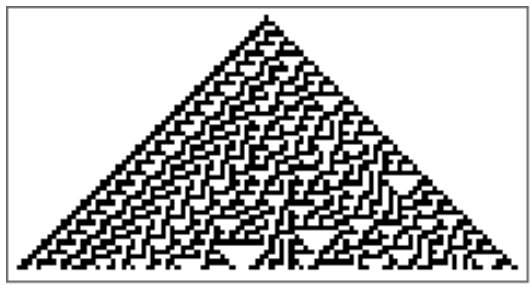

**Figura 4-5 Evolución de la regla 30 en 50 iteraciones**
http://reference.wolfram.com/language/ref/CellularAutomaton.html

Entre las reglas existentes hay 88 clases que son inequivalentes (Li and Packard, 1990; Wuensche and Lesser, 1992).

Han existido varias clasificaciones para la dinámica de los *ACE*, la más popular ha sido realizada por Stephen Wolfram, quien distinguió el comportamiento evolutivo cualitativo a través del establecimiento de 4 clases (Wolfram, 1984). Para Wólfram, las clases I y II evolucionan hasta aproximarse a dinámicas ordenadas de configuraciones estables, homogéneas, simples y/o periódicas, que terminan con celdas de un mismo valor. Las reglas de la clase I tienden a atractores puntuales desde todos los estados iniciales (p.e. reglas 0, 248 y 32), mientras que las reglas de la clase II tienden a atractores cíclicos (p.e. reglas 34, 1 y 4 definidas cómo periódicas). Las reglas clase III evolucionan a estados caóticos (p.e. reglas 90, 30 y 165). La evolución de las reglas clase IV generan dinámicas críticas (p.e. 54, 110 y 193); algunas de ellas han sido definidas como capaces de realizar computación universal (Cook, 2004).

### 4.2.1 Experimentos con *ACE*.

Los experimentos de *ACE* evaluaron 50 instancias de algunas reglas de las cuatro clases de Wolfram. Se consideraron 256 celdas, $2^{12}$ pasos iniciales y $2^{12}$ pasos adicionales fueron usados para generar los resultados. Finalmente, 18 reglas de *ACE*, escogidas como representativas de cada una de las clases, fueron evaluadas. Una relación de ellas se da en las tablas 4-2 a 4-5.

**Tabla4-2 Reglas Incluidas de la Clase I. Condiciones de Evolución y Representación de 20 Iteraciones.**

| ACE  Clase I | Evolución |
|---|---|
| Regla Cero 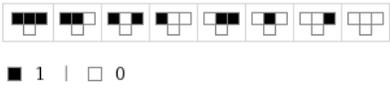 ■ 1  \| □ 0 | 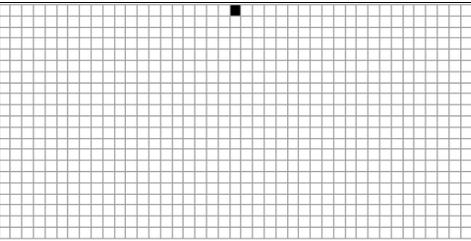 |
| Regla 8 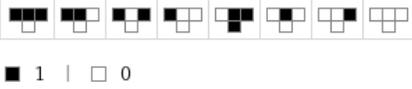 ■ 1  \| □ 0 | 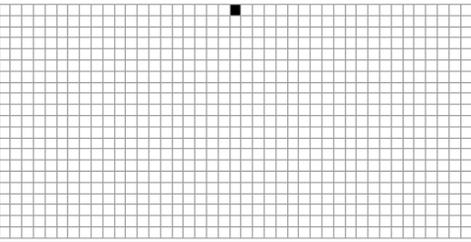 |
| Regla 32 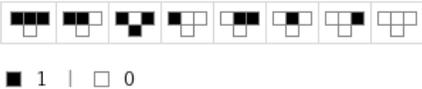 ■ 1  \| □ 0 | 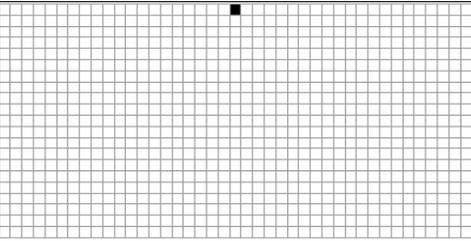 |
| Regla 40 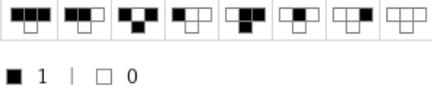 ■ 1  \| □ 0 | 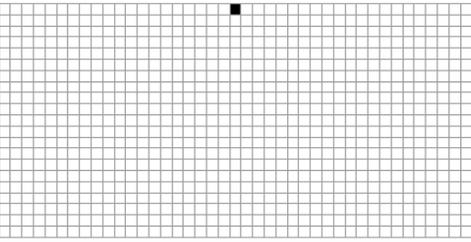 |
| Regla 128 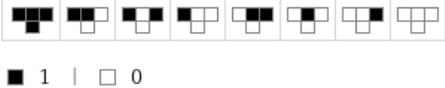 ■ 1  \| □ 0 | 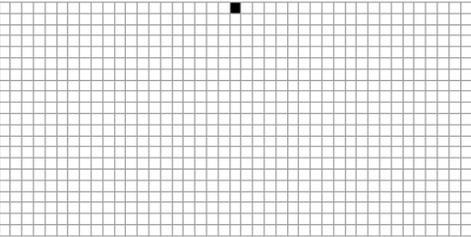 |

http://reference.wolfram.com/language/ref/CellularAutomaton.html

**Tabla 4-3 Reglas Incluidas de la Clase II. Condiciones de Evolución y Representación de 20 Iteraciones**

| *ACE Clase II* | *Evolución* |
|---|---|
| Regla 1 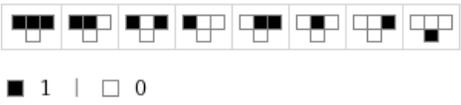 ■ 1 \| □ 0 | 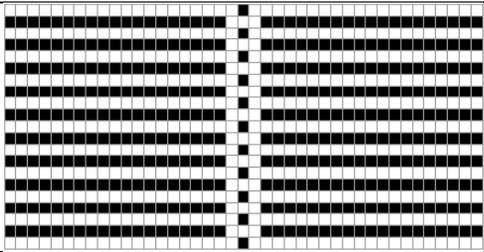 |
| Regla 2 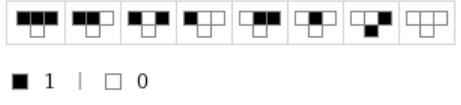 ■ 1 \| □ 0 | 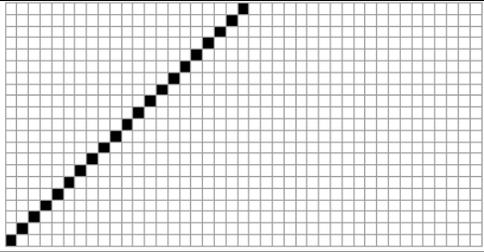 |
| Regla 3 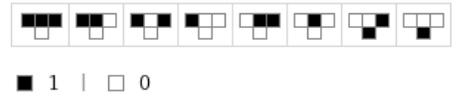 ■ 1 \| □ 0 | 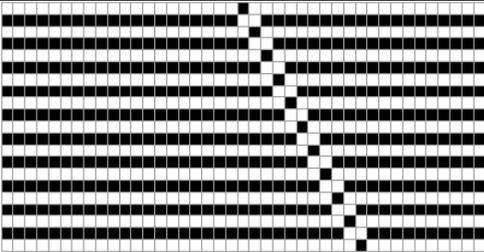 |
| Regla 4 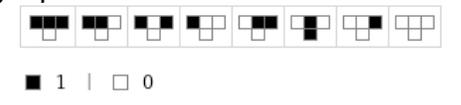 ■ 1 \| □ 0 | 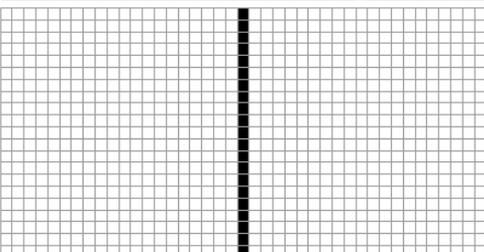 |
| Regla 5 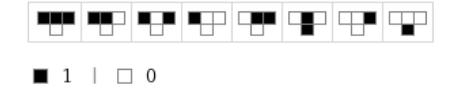 ■ 1 \| □ 0 | 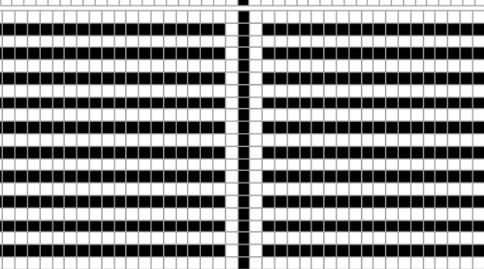 |

http://reference.wolfram.com/language/ref/CellularAutomaton.html

**Tabla 4-4 Reglas Incluidas de la Clase III. Condiciones de Evolución y Representación de 20 Iteraciones**

| *ACE Clase III* | *Evolución* |
|---|---|
| Regla 41 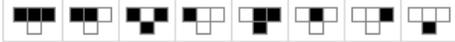 ■ 1 \| □ 0 | 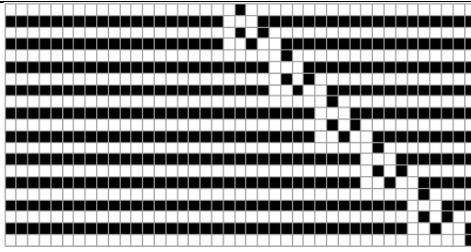 |
| Regla 54 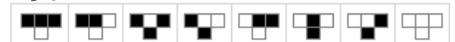 ■ 1 \| □ 0 | 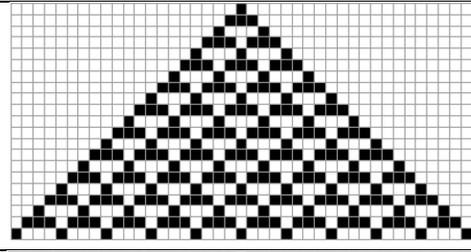 |
| Regla 106 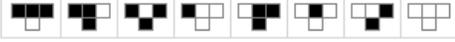 ■ 1 \| □ 0 | 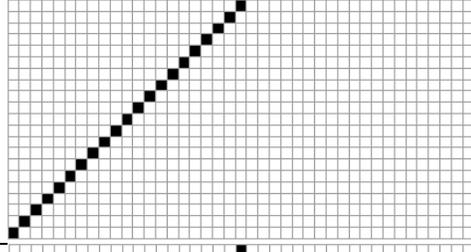 |
| Regla 110 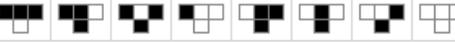 ■ 1 \| □ 0 | 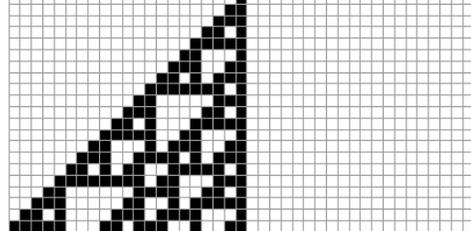 |

http://reference.wolfram.com/language/ref/CellularAutomaton.html

**Tabla 4-5 Reglas Incluidas de la Clase IV. Condiciones de Evolución y Representación de 20 Iteraciones**

| *ACE Clase IV* | *Evolución* |
|---|---|
| Regla 18 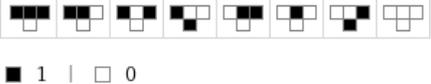 ■ 1 \| □ 0 | 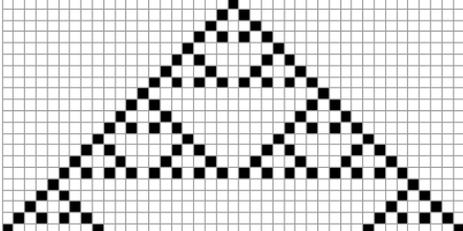 |
| Regla 22 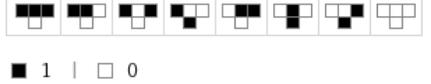 ■ 1 \| □ 0 | 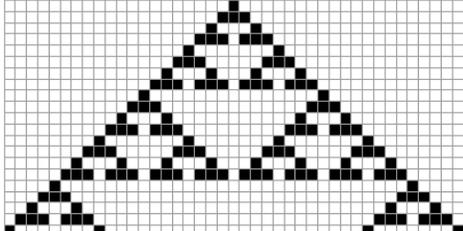 |
| Regla 45 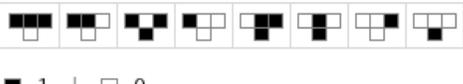 ■ 1 \| □ 0 | 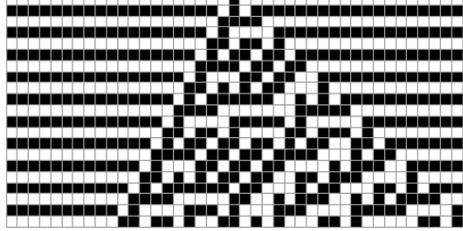 |
| Regla 161 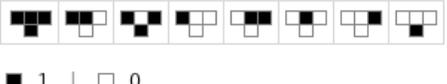 ■ 1 \| □ 0 | 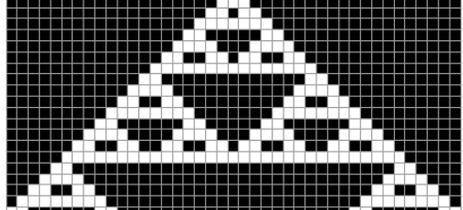 |

http://reference.wolfram.com/language/ref/CellularAutomaton.html

Las simulaciones fueron programadas en R con el uso de los paquetes BoolNet (Müssel et al., 2010), CellularAutomaton (Hugues, 2012) y entropy (Hausser and Strimmer, 2012).

### 4.2.2 Resultados *ACE*

Los resultados para diferentes clases ($I_1$, $I_2$, $I_4$, e $I_8$) se muestran en la figuras 4-6 a 4-9. En ellas se puede observar que las reglas de la clase I tienen $E = C = 0$ y $S = H = 1$ para todas las escalas. Esto es debido a que en la evolución de los nodos no hay cambio de estado, así que el *ACE* alcanza un estado estable. El tiempo que toma ello, es el de básicamente una iteración.

Algunas reglas de la clase II se comportaron de manera similar a la clase I, por ejemplo la regla 4. Otras, como la 1 y 5, tuvieron emergencias medianamente altas ($E > 0,5$ en $b = 1$). Esto se debío a que la mayoría de las células oscilaron entre cero y uno en cada iteración. Sin embargo, al incrementar la escala ($b \geq 2$), los patrones se tornaron más regulares y se comportaron como las reglas de la clase I ($E = C = 0$ ; $S = H = 1$). Dado que medimos el cambio de la información por nodo, las reglas cómo la 2 y 3 parecen tener alta $C$, a pesar de que sus patrones tienden a la regularidad (Gershenson and Fernández, 2012a).

Las reglas de la Clase III, tienden a tener alta $E$. La regla 18 es particular, dado que tiene una proporción mucho mayor de ceros que unos, que se acumulan en triángulos. Su $E$ es menor y su $S, C,$ y $H$ es mayor, comparativamente con las otras de la clase III. Esto es notable aún más a escalas mayores, en razón a que las otras reglas de la clase III tienen un porcentaje más balanceado de ceros y unos. La regla 161 está en el medio, y tiene un porcentaje ligeramente mayor de unos que ceros.

Las reglas de la Clase IV se comportan de manera similar a las de la clase III para $b = 1$, con alta $E$, baja $S$ y $C$. La regla 54 es una excepción para $H$, donde la alternancia de patrones del entorno (éter) genera una $H$ anti-correlacionada. Es decir que los patrones espaciales en un tiempo determinado comparado con el siguiente, son completamente opuestos. El éter regular de la regla 54, y la regularidad de sus deslizadores (gliders), genera un decremento en $E$ a altas escalas, por razones similares a las de las reglas 2 y 3. Para la regla 41 la $E$ decrece y se incrementa su $S$ y $C$ con la escala, y se mantiene una $H$ no correlacionada. Esto revela patrones más regulares en altas escalas, similar a la regla 18. La regla 110 se comporta similar a la regla 41: $E$ es baja, $S$ y $C$ incrementan para escalas altas. La diferencia está en $H$. Para la regla 110, $H$ es anti-correlacionada para $b \geq 4$. Esto puede ser nuevamente explicado por lo que pasa en el éter, el cual tiene un periodo de 7 para la regla 110.

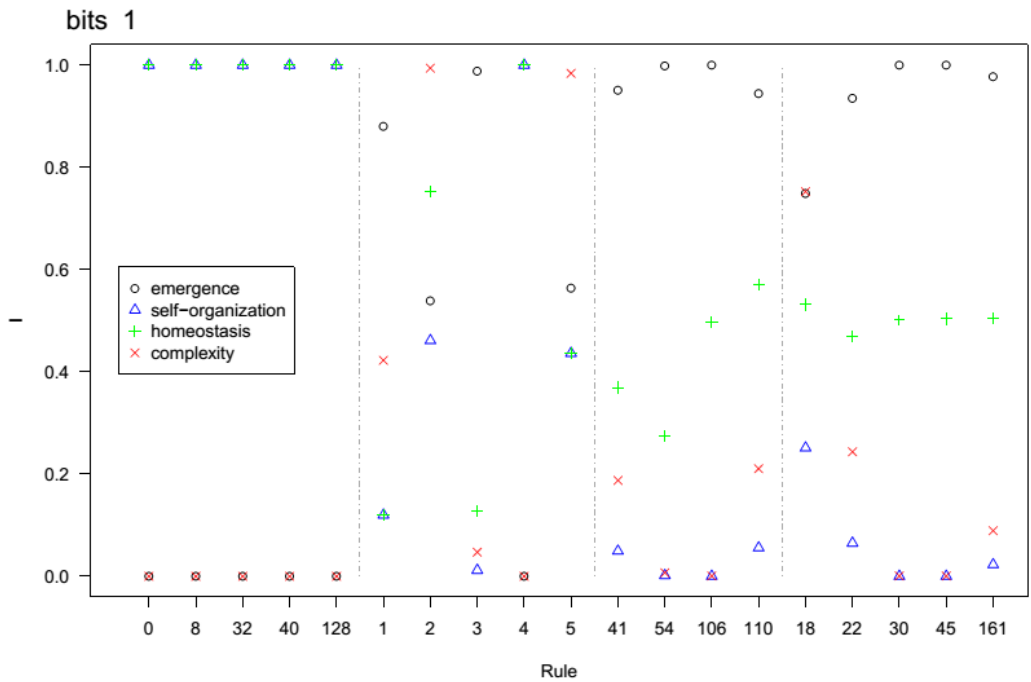

**Figura 4-6 Promedios de 50 ECA para la 19 reglas, N=256, b=1.** (Gershenson and Fernández, 2012a)

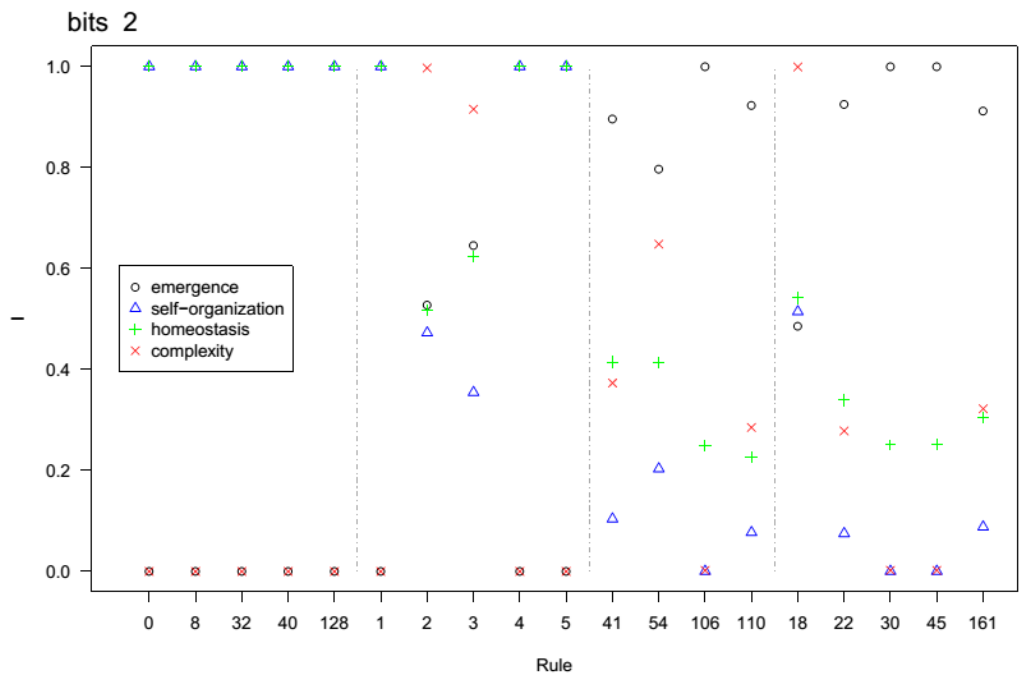

**Figura 4-7 Promedio de 50 ACE para 19 reglas, N=256, b=2** (Gershenson and Fernández, 2012a)**.**

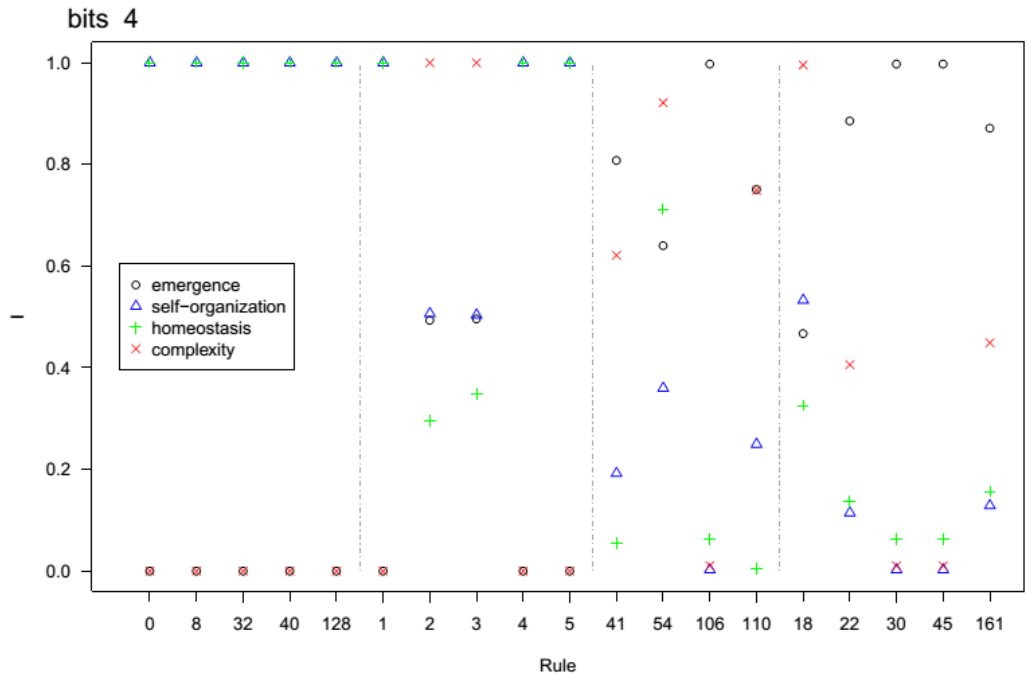

**Figura 4-8 Promedio de 50 ACE para 19 reglas, N=256, b=4** (Gershenson and Fernández, 2012a).

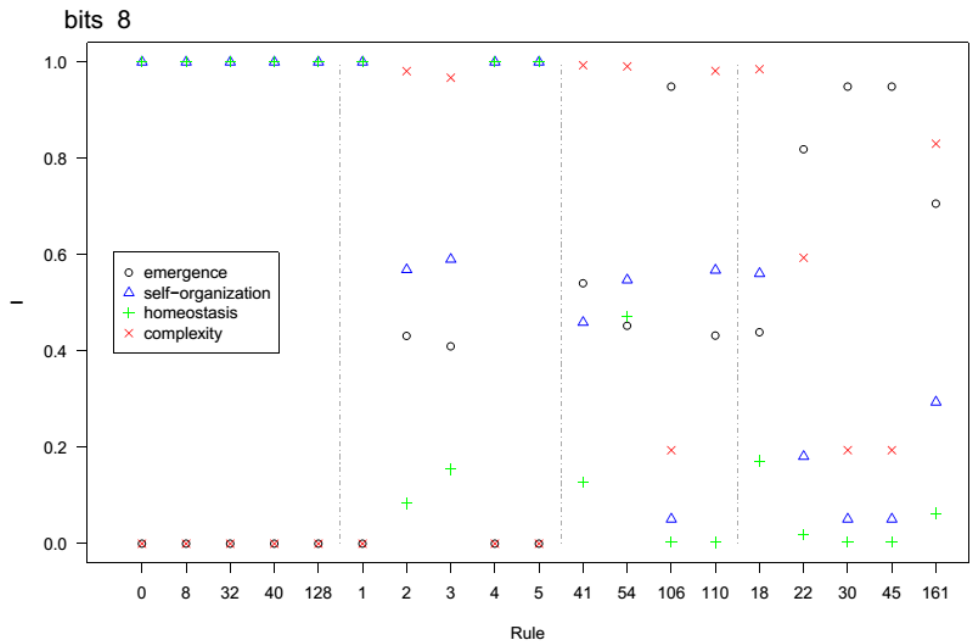

**Figura 4-9 Promedio de 50 ACE para 19 reglas, N=256, b=8** (Gershenson and Fernández, 2012a).

El comportamiento de $E, S, C$ y $H$ a múltiples escalas (bits 1,2,4,8) se puede ver en la figura 4-10. La regla 0 no cambia con la escala. La regla 1 tiene alta $E$, baja $S$ y $H$, y media $C$ para $b = 1$, pero es similar a la regla 0 para escalas grandes. En la regla 11, la $E$ y $H$ decrecen y se incrementan su $S$ y $C$ con la escala. La regla 30 tiene alta $E$, baja $S, H$ y $C$, para todas las escalas.

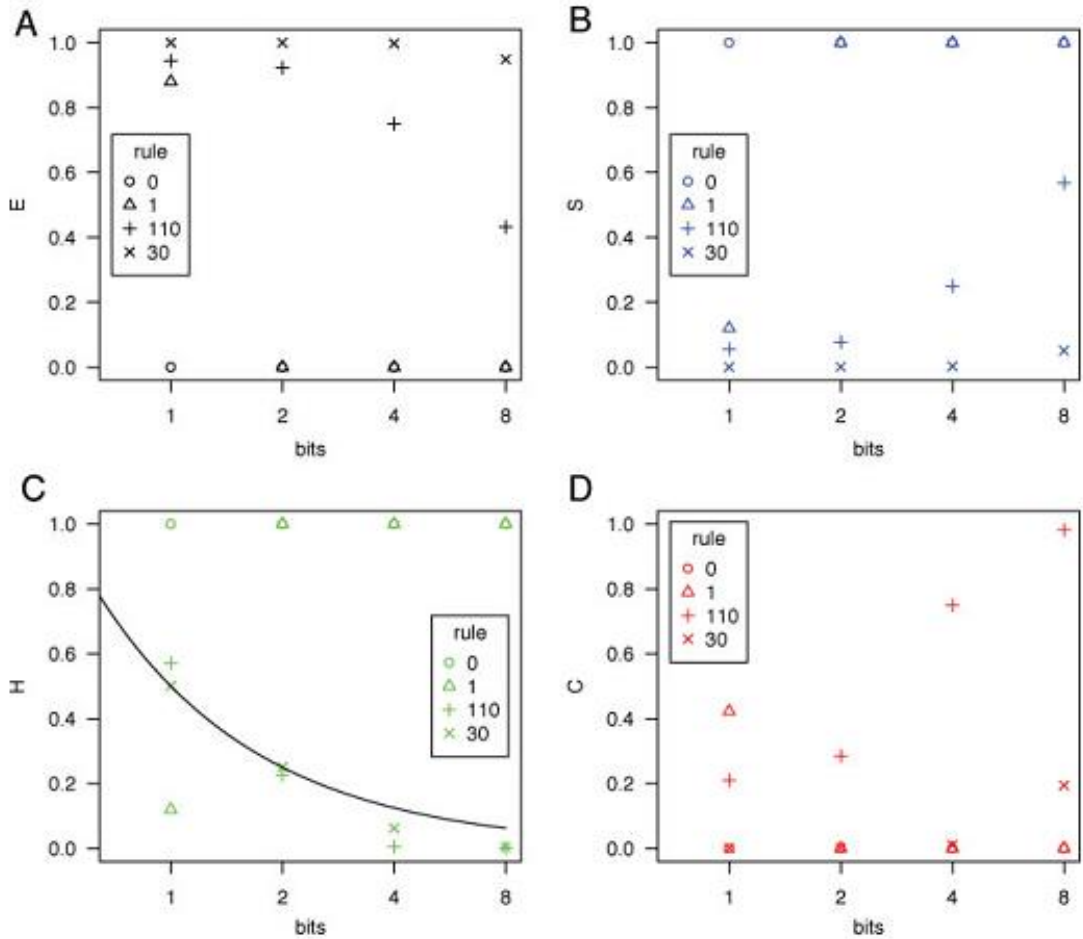

**Figura 4-10 Perfiles de E,S,C y H a Escalas 1,2,4,8. .** (Gershenson and Fernández, 2012a)

## 4.3 Medición de la Complejidad en Sistemas Urbanos- Semáforos Auto-organizantes

### 4.3.1 La Movilidad Vehicular en las Ciudades Modernas y los Métodos Auto-Organizantes para su Solución

Vivir en un mundo urbanizado ha traído grandes ventajas a sus habitantes, como la solución a muchas necesidades básicas que se deben satisfacer. Se estima que esta satisfacción impulsa a las personas a buscar cada vez más las ciudades, tanto que para el 2050 el 70% de las personas vivirán en ciudades. En la actualidad, la mitad de las ciudades en el mundo tienen entre 100 y 500 mil personas, y el 10% de la población mundial viven en megalópolis, ciudades con más de 10 millones de habitantes (http://goo.gl/FY0IfP).

Ante el gran número de personas que necesitan desplazarse en las ciudades, su planeación ha tenido cómo uno de sus ejes centrales la movilidad vehicular. La movilidad vehicular es un problema complejo en el que el flujo que cambia constantemente, interactúa con transeúntes y la sincronización de los semáforos. Al ser un problema no estacionario, se requiere de un enfoque adaptativo para su solución, como por ejemplo un enfoque auto-organizante propuesto por Gershenson (Gershenson, 2005). Las preguntas que surgen son: ¿Cómo la auto-organización puede ser orientada para lograr flujos eficientes de tráfico? ¿Cuál puede ser su régimen deseado? ¿Cómo puede ser medida en términos de complejidad?

En el contexto anterior, se presenta un modelo de auto-organización guiada. Además, se hace una comparación de la complejidad entre los métodos tradicionales de sincronización, comparado con los semáforos auto-organizantes (Gershenson, 2005). Los métodos y resultados presentados son acordes con lo expresado en Zubillaga et al. (2014).

### 4.3.2 Modelo de Tráfico

El modelo de tráfico usado fue desarrollado previamente por (Gershenson and Rosenblueth, 2012), basado en autómatas celulares elementales explicados en la sección 4.2 (Wolfram and Gad-el-Hak, 2003; Wuensche and Lesser, 1992). Este modelo es determinístico, en tiempo y espacio discreto, velocidad uno o cero y aceleración infinita. Su condición sintética permite medir claramente las diferentes fases dinámicas. Se aclara que el modelo es descriptivo, no predictivo.

Cada calle es representada por un autómata celular elemental (*ACE*), acoplado en las intersecciones. Cada *ACE* contiene un número de celdas que toman valor de cero (vacío) o uno (vehículo). El estado de la celda se actualiza en correspondencia con los estados previos de la celda y de los vecinos más cercanos. Muchas celdas vacías adelante del vehículo permiten su avance.

El comportamiento fue modelado con la regla 184, la cual es uno de los modelos más simples de tráficos en autopistas. Sus reglas de actualización según sus dos vecinos se dan en la tabla 4-7.

**Tabla 4-6 Reglas de Actualización de ACE**

| $t-1$ | $t_{184}$ | $t_{252}$ | $t_{136}$ |
|:-----:|:---------:|:---------:|:---------:|
| 000   | 0         | 0         | 0         |
| 001   | 0         | 0         | 0         |
| 010   | 0         | 1         | 0         |
| 011   | 1         | 1         | 1         |
| 100   | 1         | 1         | 0         |
| 101 | 1 | 1 | 0 |
| 110 | 0 | 1 | 0 |
| 111 | 1 | 1 | 1 |

Los modelos ACE-184 se mueven a la derecha. Otras direcciones son obtenidas cambiando las posiciones de los vecinos con el uso de la misma regla. En las intersecciones se usaron dos reglas más. El no avance de los vehículos antes de una luz roja, se modeló con la regla ACE-252. La regla ACE-136 se utilizó para modelar los vehículos que cruzaban la intersección después de una luz roja. Las calles con luz verde utilizaron la regla ECA-184. La figura 4-11 muestra detalles de las reglas.

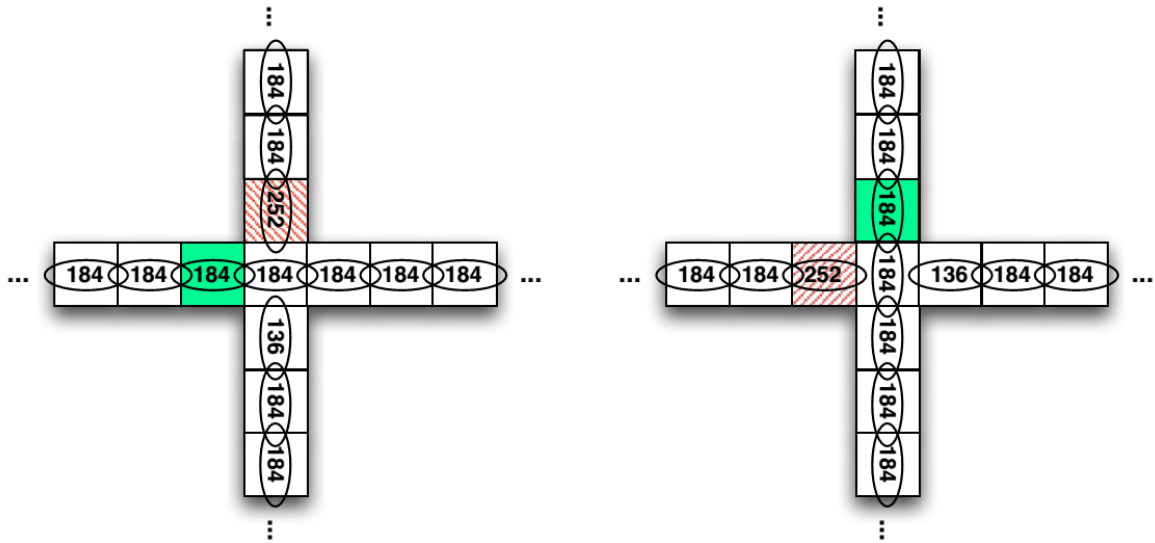

**Figura 4-11 Reglas de autómatas celulares elementales usadas en el modelo de tráfico y su función cómo interruptores de los semáforos** (Zubillaga et al., 2014).

El modelo de tráfico es conservativo. Esto es, la densidad de vehículos $\rho$ (proporción de celdas con un valor de 1) es constante en el tiempo. El promedio de velocidad $v$ se calcula por el número de celdas que cambian desde cero a uno (movimiento ocurrido entre el tiempo $t$ y $t+1$, dividido por el número total de vehículos (celdas con 1). $v = 0$ sí el vehículo no se mueve, $v = 1$ cuando se mueven.

El flujo $J$ se define como la velocidad $v$ multiplicada por la densidad $\rho$; $J = v * p$. $J = 0$ cuando no hay flujo. Es decir, no hay vehículos en la simulación ($\rho = 0$), o los vehículos están parados ($v = 0$). Una $J$ máxima ocurre cuando todas las intersecciones son cruzadas por vehículos. En el estudio de escenarios $J_{max} = 0.25$. Esto es debido a que para una calle única (ACE-184), los vehículos necesitan espacio para avanzar, limitando la $J_{max} = 0.5$. Sin embargo, cuando dos calles se intersectan, una de ellas tiene el pare mientras la otra fluye, reduciendo $J_{max}$ a 0.25.

Teóricamente, la velocidad y flujo óptimos para una proporción dada en un problema de coordinación, deben ser los mismos $v$ y $J$ para las intersecciones aisladas. Esto es un límite superior. Lo que implica que cada intersección interactúa con sus vecinos si es tan eficiente como cuando está vacía. Esto puede ser visto con curvas de optimización (definidas como el mejor rendimiento para una intersección aislada, para diferentes densidades, menos el rendimiento actual del sistema de tráfico (Gershenson and Rosenblueth, 2012), que visual y analíticamente compara los diferentes resultados con el óptimo teórico.

Las fronteras fueron no orientables (como en una banda de Möbius o una botella Klein) por sugerencia de Bedau and Humphreys (2008), lo que mantiene la entrada y la densidad de vehículos. Si bien estos son deterministas, en la práctica se destruyen las correlaciones que se dan en fronteras cíclicas. El resultado es equivalente a una sola calle con 100 intersecciones. Una muestra de la aplicación de las fronteras no orientables se observa en la figura 4-12.

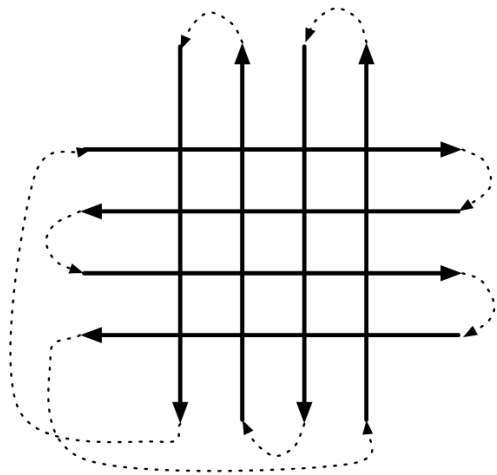

**Figura 4-12 Fronteras no orientables para una grilla de cuatro por cuatro** (Zubillaga et al., 2014)

También es posible extender el modelo a una grilla hexagonal para permitir intersecciones más complejas (Gershenson and Rosenblueth, 2012).

El problema de coordinación de los semáforos es EXPTIME-Completo (Gershenson, 2005). Esto implica que la optimización de grandes redes de tráfico es inviable. Además es no-estacionario. Aún si se encontrara una solución óptima, sería obsoleta en segundos. Como alternativa se tiene la adaptación, siendo la auto-organización un método útil para construir sistemas adaptativos.

La solución común es la "ola verde", donde los semáforos son sincronizados con el mismo periodo y fase se ajusta acorde con la velocidad esperada de vehículos. Esto es bueno para dos direcciones, mientras que las otras quedan detenidas por estar sus semáforos en rojo, es decir en fases anti-correlacionadas. En densidades medias, la ola verde causa largas esperas y colas. Igualmente, cuando las densidades cambian, los vehículos no van a la velocidad esperada. Las fases halladas en la ola verde son dos: intermitentes (algunos vehículos detenidos, algunos se mueven) y embotellamiento (todos los vehículos parados).

Por su parte el método auto-organizante se basa en 6 reglas, simples y por intersección, con el que se alcanza un óptimo global de coordinación (22,33,38,40). A continuación se esquematiza una intersección y sus parámetros (Fig. 4.13).

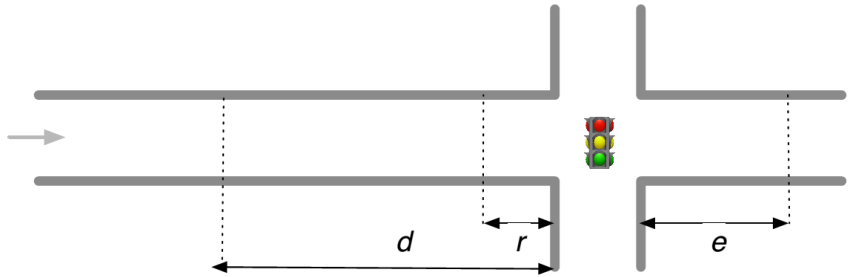

**Figura 4-13 Esquema de una intersección y parámetros de medición para la definición de reglas en semáforos auto-organizantes** (Zubillaga et al., 2014)

Las reglas son:

i. En cada tic se adiciona a un contador el número de vehículos que se aproximan o esperan en una luz roja a una distancia $d$. Cuando el contador excede el umbral $n$, se enciende la luz verde. Cuando la luz verde se enciende el contador se reinicia a cero.
ii. La luz se mantiene en verde un tiempo mínimo $u$
iii. Si los vehículos son pocos ($m$ o pocos, pero más que cero) y están a una distancia corta $r$, la luz no pasará a verde.
iv. Sí no hay ningún vehículo que se aproxime a la luz verde dentro de una distancia $d$, y hay por lo menos un vehículo aproximándose, en la otra dirección, a la luz roja dentro de la distancia $d$, la luz cambiará a verde para este último.
v. Sí hay un vehículo parado en una calle a una distancia corte $e$, ante una luz verde, entonces la luz pasará a roja.
vi. Si hay vehículos parados en ambas direcciones, a una distancia corta $e$, entonces ambas luces pasan a rojo. Una vez, una de las direcciones esté libre, se restablecerá la luz verde en esa dirección.

### 4.3.3 Resultados

Los resultados de $v$ y $J$ para fronteras no orientables se muestran en la figura 14, en la que se indica con la línea punteada las transiciones de fase del método auto-organizante, cuando se incrementa la proporción. Estas fases fueron halladas en estudios previos para fronteras cíclicas y se colocan para propósitos ilustrativos. Ellas son: (i) flujo libre (ningún vehículo detenido $v = 1$), (ii) cuasi-libre (casi ningún vehículo detenido), (iii) subutilizado-intermitente (algunos vehículos detenidos, las intersecciones no alcanzan su flujo máximo), (iv) capacidad completa-intermitente (algunos vehículos detenidos, $J_{max}$, todas las intersecciones tienen vehículos siempre), (v) sobre-utilizado-intermitente (algunos vehículos parados, las intersecciones tienen que restringir el flujo en ambos sentidos, usando la regla 6), (vi) cuasi-embotellamiento (casi todos los vehículos parados, pero con algunos espacios de "pelotones" o convoys), (vii) embotellamiento (todos los vehículos parados y las intersecciones bloqueadas desde condiciones iniciales; $v = 0$). En fronteras cíclicas, las transiciones de fase se dieron a valores de $\rho$: 0.15, 0.22, 0.38, 0.63, 0.77 y 0.95.

En la figura 4-14 se observa claramente que el método auto-organizante (en azul) se acerca estrechamente con el óptimo teórico y mantiene mayores velocidades y flujos que la ola verde (en verde). Con ello se evidencia la gran ventaja y bondades del método auto-organizante. Entre tanto, para la ola verde la fase de embotellamiento llega rápidamente a bajas densidades.

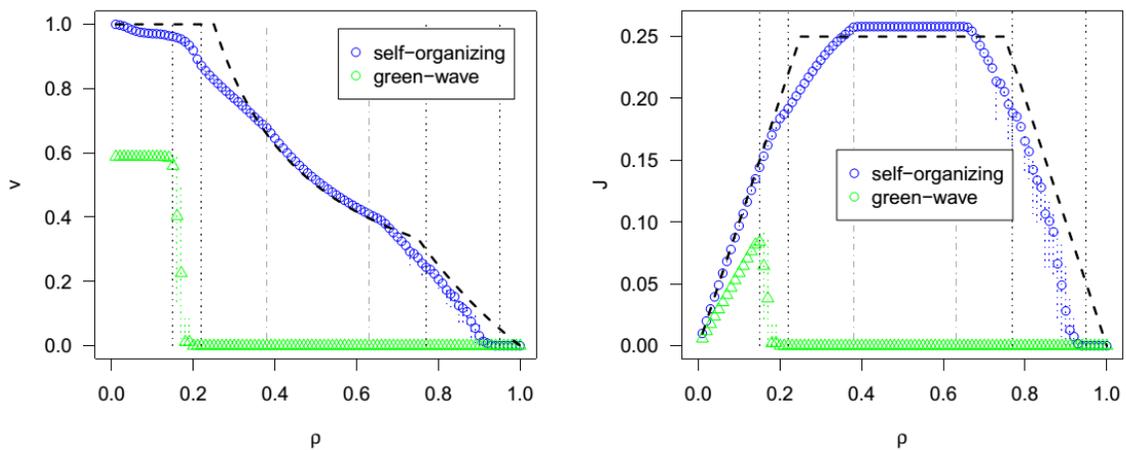

**Figura 4-14 Resultado de la simulación para fronteras no-orientables para la ola verde y método auto-organizante: promedio de velocidad $v$ y promedio de flujo $J$ para diferentes densidades de $\rho$. Las curvas óptimas se muestran con línea punteada discontinua** (Zubillaga et al., 2014).

Se puede ver que el método auto-organizante está por encima de la curva optima en la fase de capacidad completa-intermitente ($J > J_{max}$). Esto es debido a que un vehículo en una calle, puede alcanzar la luz verde para avanzar sin dejar espacio entre él y el último vehículo en la calle que alcanzó la luz roja.

Para una clara diferenciación de los métodos en términos de complejidad, las métricas $E, S,$ y $C$ fueron aplicadas a tres componentes del modelo: (i) los intervalos de cambio de luz en los semáforos-ICLS, (ii) los intervalos de autos en la intersección-ICI y (iii) los intervalos de carros en la calle-ICC. Los resultados se presentan en las figuras 4-15 a 4-17

En la figura 4-15 se puede ver que para ICLS el método de la ola verde tiene periodos cíclicos para los semáforos ($E = C = 0$ y $S = 1$, puntos verdes). Se aclara que nuestra medida de $S$ no distingue entre organización interna (auto-organización) o externa, cómo en este caso que el método depende de un control central. Teniendo un extremo de regularidad en los intervalos en los semáforos representado por $S = 1$, se entiende que no puede haber adaptación a los cambios en el flujo de tráfico. Entre tanto, el método auto-organizante se adapta constantemente a los cambios en la demanda (puntos azules), como puede ser visto desde la variación de las medidas para diferentes densidades ($\rho$). Los intervalos en el semáforo son más irregulares (muy alta $E$, muy baja $S$) en las fases de flujo cuasi-libre ($\rho[0 - 0.15]$) y cuasi-embotellamiento ($\rho[0.77 - 0.95]$), mientras que alcanza regularidad (muy baja $E$ y $C$, y máxima $S$) para la fase de capacidad-completa intermitente, dado que la demanda viene en todas direcciones. $C$ es alta para todas las fases, pero en las fases de capacidad-completa intermitente ($\rho[0.22 - 0.38]$) y embotellamiento ($\rho[0.95 - 1]$) exhibe un comportamiento similar a la ola verde en estas dos fases.

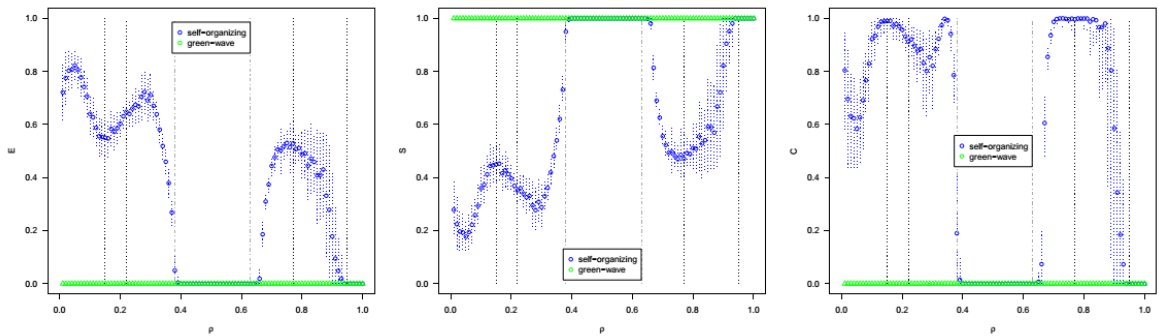

**Figura 4-15 Emergencia (E), Auto-organización (S) y complejidad para los intervalos de cambio de luz en semáforos para fronteras no-orientables** (Zubillaga et al., 2014). **Las líneas verticales señalan la transición de cada una de las 7 fases del flujo (libre, cuasi-libre, subutilizado, capacidad completa, sobre-utilizado, cuasi-embotellamiento, embotellamiento)**

Para ICI (Fig. 4-16), las medidas de $E, S,$ y $C$ para diferentes densidades $\rho$ se observó que son similares en las fases de capacidad-completa-intermitente y embotellamiento, dado que hay vehículos que constantemente cruzan la intersección o hay vehículos detenidos. $E$ y $C$ son altas para bajas densidades en ambos métodos, y se hacen más regulares con $\rho$ (incremento de $S$), hasta alcanzar un máximo en la fase de capacidad-completa-intermitente para el método auto-organizante y para la fase de embotellamiento en la ola verde. El método auto-organizante nuevamente incrementa $E$ y $C$ hasta la fase de cuasi-embotellamiento ($\rho > 0.6; p < 0.9$), dado que los vehículos cruzando son menos regulares y se forman pelotones espaciados, solo para decrecer de nuevo en la fase de embotellamiento.

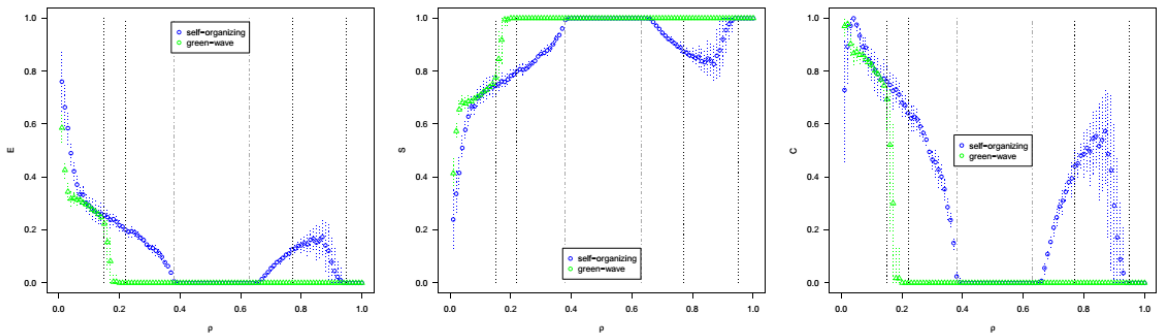

**Figura 4-16 Emergencia (E), Auto-organización (S) y complejidad para los intervalos en intersecciones en semáforos para fronteras no-orientables** (Zubillaga et al., 2014)**. Las líneas verticales señalan la transición de cada una de las 7 fases del flujo (libre, cuasi-libre, subutilizado, capacidad completa, sobre-utilizado, cuasi-embotellamiento, embotellamiento)**

En las últimas pruebas (Fig. 4-17), los intervalos de los vehículos en diferentes calles (no intersecciones) se escogieron aleatoriamente en cada simulación. El comportamiento de las medidas fue similar al descrito para las intersecciones. Sin embargo, $S = 1$ se alcanza sólo en la fase de embotellamiento ($\rho > 0.95$). Parece haber un mínimo de $E \approx 0.2$ para la fase de capacidad-completa-intermitente. A pesar que se forman pelotones grandes, existen espacios grandes entre ellos. $C$ es alta para todas las fases, excepto para el embotellamiento para ambos métodos.

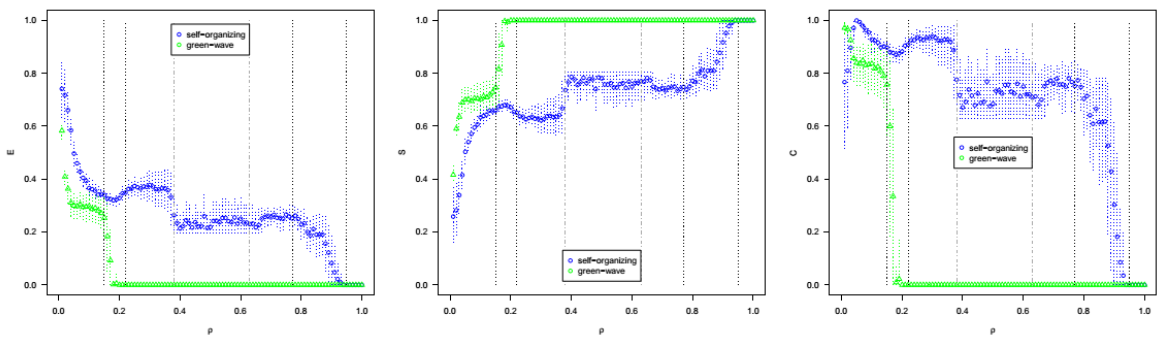

**Figura 4-17 Emergencia (E), Auto-organización (S) y complejidad para los intervalos en las calles en semáforos para fronteras no-orientables** (Zubillaga et al., 2014)**. Las líneas verticales señalan la transición de cada una de las 7 fases del flujo (libre, cuasi-libre, subutilizado, capacidad completa, sobre-utilizado, cuasi-embotellamiento, embotellamiento)**

Se pudo constatar que el flujo en tráfico urbano es muy complejo. No obstante, se pueden identificar diferencias entre algunas fases del método auto-organizante. En otras palabras, la fase dinámica puede tener un impacto directo sobre $E, S$ y $C$, lo cual puede ser potencialmente utilizado para identificar las fases dinámicas en sistemas más reales de tráfico donde las transiciones no son tan fáciles de identificar.

## 4.4 Discusión

La siguiente discusión abordará los tópicos de la utilidad general de las medidas a partir de lo observado en sistemas discretos. La relación entre complejidad e información. Los resultados de las múltiples escalas y la relación con otras medidas. Algunos tópicos serán abordados de nuevo en las siguientes aplicaciones, con el fin de ahondar en su significado a partir de los resultados obtenidos en cada una de ellas.

### 4.4.1 La Aplicación de Las Medidas

Aunque pueda ser deducible, nuestras medidas no se enfocan en qué elemento interactúa con cual, cómo y cuándo. En contraste, estamos más interesados en las propiedades globales y tendencias de los estados de los elementos. En este sentido, queremos capturar dinámicas globales a través de medidas que la sintetizan en forma adecuada. Igualmente, estamos interesados en su cambio con la escala. Si bien en al emplear *E,S,C, A* y *H* se debe considerar la perdida de información subyacente en el fenómeno que se describe, también se debe considerar el tipo de información en el que estemos interesados, ya que la relevancia es también parcialmente dependiente del observador (Gershenson, 2002).

Para el caso de semáforos auto-organizantes, se recalca que dada la generalidad y simplicidad de *S*, ella no puede distinguir entre un sistema auto-*organizado* (p.e. por un controlador externo) y un sistema auto-*organizante*, el cual se adapta constantemente a las demandas de cambio. Esto puede ser comparado con el caso de la temperatura, la cual no indica sí el calor es generado externa o internamente en el sistema, pero sería muy útil saberlo. *S*, cómo medida, tiene que ser interpretada adecuadamente en el contexto específico para explotar su utilidad. Una vez definido el contexto, el complemento de las medidas de *E* y *C*, son de gran utilidad para la caracterización de fases dinámicas en sistemas con y sin control central.

Podemos afirmar que actualmente no hay un acuerdo en la comunidad científica sobre como los términos de complejidad, emergencia y auto-organización deben ser definidos y usados, dada la evidente contradicción entre autores. En este aspecto, nuestras medidas muestran desde los resultados obtenidos su potencial y esclarecen su utilidad, por lo que vale la pena emplearlos. No obstante, acatamos las restricciones que puedan tener y que el revisor manifiesta. En este sentido, no nos apartamos que las sugerencias mostradas por el revisor sirvan para motivar el desarrollo de una mejor terminología; situación que también es una contribución.

### 4.4.2 La Complejidad cómo Balance

Cómo ha sido propuesto por varios autores y confirmado en nuestros experimentos en sistemas discretos, la complejidad puede ser vista como balance entre el orden y el desorden (Gershenson and Fernández, 2012; Kaufmann, 1993; Langton, 1990; Lopez-Ruiz et al., 1995).

Desde esta perspectiva, $C$ como un balance entre $E$ y $S$, cumple con esta condición, que ha sido definida también para la llamada criticalidad.

Podría sonar cómo contradictorio o contra-intuitivo definir la emergencia cómo el opuesto de auto-organización, dado que ambas están presentes en muchos fenómenos complejos. Sin embargo, cuando se toma un extremo de los dos ($E$ o $S$), la expresión del otro es muy baja. Es decir, si se tiene emergencia muy alta, se tendrá auto-organización muy baja. Es precisamente cuando $E$ y $S$ están en balance, que la complejidad se maximiza, pero esto no significa que ambos hayan sido máximos.

Un punto importante para resaltar es que existen visiones opuestas de información. Shannon definió la información como entropía (Equivalente a nuestra $E$). Wiener (1948) y von Bertalanfy (1968), definieron información como lo contrario, *negentropía* (Equivalente a nuestra $S$). En este sentido, nuestra medida de complejidad $C$ reconcilia estas dos visiones opuestas, como balance entre la regularidad u orden ($S$) y el cambio o caos ($E$), donde la complejidad es máxima. Los sistemas dinámicos cómo autómatas celulares y redes booleanas, tienen máxima $C$ en la región donde su dinámica es considerada cómo más compleja.

### 4.4.3 Múltiples Escalas y Perfiles

En general, para redes booleanas aleatorias-*RBA* los resultados para $E, S, C$ y $H$ cambian poco con la escala, precisamente por ser generadas de manera aleatoria. Los autómatas celulares elementales-*ACE* son mucho más regulares en cuanto a estructura y función (expresada en la dinámica). Por ejemplo, *ACE* con dinámicas ordenadas y caóticas (Clase I y III, respectivamente), no cambian mucho con la escala. En el medio (Clase II), existen reglas de interesantes patrones que demuestran que la escala tiene importancia.

Para *RBA* y *ACE* con dinámicas caóticas existe una ligera reducción de la emergencia con la escala. Este es un efecto de tamaño finito, debido a que cadenas cortas tienen una emergencia más reducida que las largas. Es interesante, dado que muchos estudios en sistemas dinámicos, especialmente analíticos, están hechos asumiendo sistemas infinitos. Sin embargo, muchos sistemas reales son mejor modelados como finitos, y aquí hemos visto que la longitud de la cadena -no sólo la escala- puede jugar un papel relevante en determinar la emergencia y la complejidad de los sistemas. Una de las implicaciones del efecto finito, es que a grandes escalas se requiere menor información para describirlo. Extrapolando las "escalas mayores" ( $b \to \infty$) implica no información ($I \to 0$): sí todo esta contenido, cuando hay información no necesaria para describirlo (Gershenson, 2012b).

Los perfiles de $E, S, C$ y $H$ a múltiples escalas (figura 4-10), dan una mejor percepción que las escalas particulares o por separado (figura 4-6 a 4-9). Esto se pudo ver claramente en las reglas de la clase II, donde la $E$ es alta a una escala, pero mínima a escalas altas. El cambio de las medidas a través de las escalas, es una herramienta de gran utilidad en el estudio de sistema dinámicos (Weeks, 2010; Wolpert and Macready, 1999), cómo se ha formalizado por la teoría de Krohn-Rhodes (Egri-Nagy and Nehaniv, 2008; Rhodes, 2009). Por ejemplo, un sistema con alta (pero no máxima) $E$, a través de las escalas sería más emergente que un sistema con $E$ muy alta en una escala, pero con $E$ muy baja en otra escala.

Con soporte en los resultados visuales de *ACE,* se puede observar como que existe una condición inicial de una única celda "viva" con emergencia muy baja ($E$). Este es el caso de la regla cero de la clase I. No importaría con cuanta información se alimente al autómata, esta se perdería en una sola iteración. La regla 1 de la clase II produce gran cantidad de información, y emerge un patrón estriado (Tabla 4-3). No obstante, para base 2, este patrón desaparece. La emergencia, entonces, es mínima para escalas base 2 (Figura 4-7). La regla 110 de la clase IV produce estructuras y patrones complejos, con deslizadores que interactúan de manera regular en su éter. Se puede decir, que estos patrones son más emergentes, dado el incremento de información que se produce. La Regla 30 de la clase III produce aún más patrones, dado que su comportamiento seudo-aleatorio produce máxima información. La emergencia muy alta será la que produce mayor información partiendo de una menor a la inicial.

La auto-organización ($S$) definida como reducción de la información de Shannon, es mínima cuando un patrón ordenado es convertido en un patrón desordenado. Los sistemas caóticos, cómo la regla 30, tienen mínima $S$. La máxima auto-organización ocurre cuando un patrón altamente desordenado se convierte en altamente ordenado. Sistemas que tienden a estados estacionarios, tales como los de la clase I de *ACE*, exhiben máxima auto-organización(Gershenson and Fernández, 2012a; Wolfram, 1984).

Casos como el perfil de complejidad de Bary-Yam mide la complejidad cómo la cantidad de información requerida para describir un sistema en diferentes escala (Bar-Yam, 2004b). Desde nuestra perspectiva, esta puede ser mejor llamado "perfil de información" o "perfil de emergencia". Inspirados en el perfil de complejidad (Información) de Bar-Yam, el perfil sigma fue propuesto para medir la organización a múltiples escalas (Carlos. Gershenson, 2011). Siguiendo el enfoque de Lopez-Ruiz et al. (1995) que fue usado en nuestros desarrollos, estos dos perfiles pueden ser combinados para desarrollar un nuevo perfil de complejidad. Esto podría ilustrarse por las propiedades de complejidad a diferentes escalas.

### 4.4.4 La Complejidad su Diferencia con el Caos

Algunos enfoques relacionan la complejidad con la alta entropía (para nosotros, alta emergencia), p.e. el contenido de información Bar-Yam 2004; Delahaye and Zenil 2007. Sin embargo, es importante destacar que la complejidad no debe ser confundida con Caos (Gershenson 2013). El Caos es alta entropía (Información de Shanon) (Prokopenko et al., 2009). Una alta entropía (alta emergencia $E$) implica mucho cambio, lo que genera que patrones complejos sean destruidos. De otra parte, baja entropía (alta auto-organización $S$) previene que patrones complejos emerjan.

En los experimentos de *ACE* puede ser visto que las reglas caen generalmente en dos categorías: las que computan algo ($E > 0$) o las que no ($E = 0$). Esto está relacionado con el principio de equivalencia computacional de Wolfram (Wolfram and Gad-el-Hak, 2003; Zenil, 2009), el cual conjetura que los sistemas dinámicos son capaces (sistemas complejos) o no (sistemas simples) de hacer computación universal[8]. Desde los *ACE*, sólo la regla 110 ha sido capaz de producir computación universal (Cook, 2004), pero otras reglas también posiblemente sean capaces. No solo las de la clase IV, también reglas de la clase III; sin embargo, la variabilidad les hace difícil guardar información. Aun así, hay técnicas para lograr dinámicas complejas de sistemas caóticos, por ejemplo, con memoria (Martínez et al., 2012) o con señales externas regulares (Luque and Solé, 1998, 1997) (Control del Caos). La clase II podría ser considerada muy regular para computar universalmente, pero adicionando ruido o una señal compleja podría ser posible que ella llevara a cabo computación universal. La clase I es muy estática, ella podría requerir más de la computación que viniese de fuera del sistema. En general, por medio de diferentes técnicas se puede direcciona un sistema dinámico a la alta complejidad.

### 4.4.5 Homeostasis

En muchos casos, la homeostasis ($H$) ha sido relacionada con auto-organización. Alta $S$ indica baja variabilidad, lo cual es característico de una alta $H$, esto fue visible en *RBA* con *K* menores a 2. Por otra parte, baja *S*, en muchos casos, está acompañada por una $H$ no correlacionada $H \approx 1/2_b$. Esto también es un rasgo de sistemas con dinámicas caóticas (Alta *E*). Sin embargo, algunos *ACE* con baja *S* tienen una $H$ correlacionada $H > 1/2_b$ o anti-correlacionada $H < 1/2_b$. Esto es característico de estructuras complejas que interactúan en un éter regular. Al respecto, diferentes reglas de *ACE* pueden tener alta *C* en una escala particular debido a $S \approx 0.5$. No obstante, reglas con una $H$ que se desvía de $1/2_b$ muestran un rasgo de estructura compleja sobre un éter. El éter facilita la computación, de manera que la universalidad puede ser explorada más fácilmente. La desviación de $H \approx 1/2_b$ podría ser usada para explorar el inmenso espacio posible de los sistemas dinámicos.

Por otra parte, al implementar un análisis de componentes principales, encontramos que para las *RBA* la desviación estándar de $H$ está inversamente correlacionada con *C* ($E \times Ao$ en

---

[8] Capacidad de realizar cualquier proceso computable que se le pida, sin necesidad de cambiar su estructura, ni su función de transición local.

fig. 4-18). *RBAs* en un régimen ordenado tienen consistentemente alta $H$. Entre tanto, *RBAs* en régimen crítico tienen una $H$ variable.

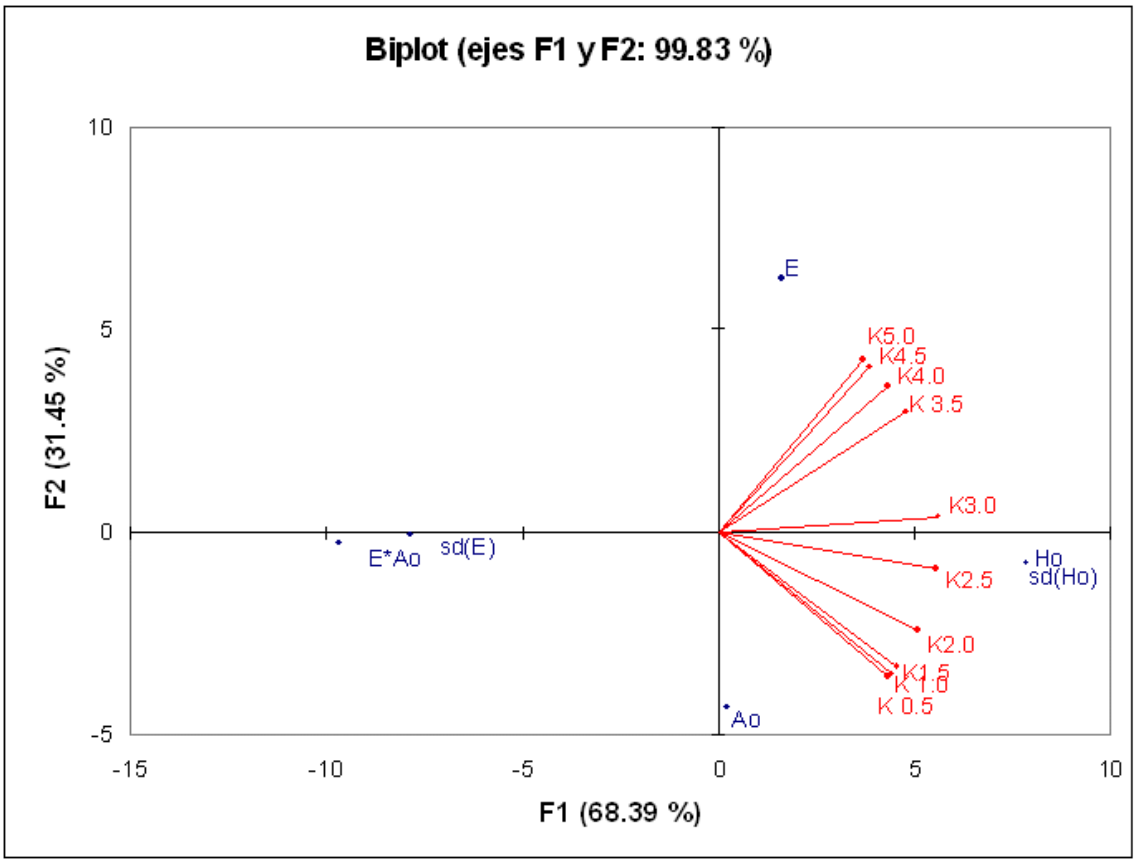

**Figura 4-18 Análisis de Componentes Principales. Las propiedades en Azul y las conectividades en Rojo.**

## 4.4.6 Autopoiesis y el Requisito de Variedad de Ashby.

Nuestro enfoque de autopoiesis cómo la razón entre las complejidades del sistema y el ambiente, nos lleva a considerar la ley de la Variedad Requerida de Ashby (Ross Ashby, 1960a). En ella se establece que un controlador activo requiere mucha variedad (número de estados), para que el sistema controlado sea estable. Por ejemplo, si un sistema puede estar entre cuatro estados, su controlador debe discriminar entre los cuatro estados para regular la dinámica del sistema.

Desde esta perspectiva, la medida propuesta de autopoiesis está relacionada con el requisito de variedad requerida. Cuando un sistema tiene $A > 1$, tiene una complejidad (variedad) mayor que su ambiente, lo que refleja una mayor autonomía. Es decir, la complejidad del sistema es mayor que la del ambiente, en consecuencia, un controlador exitoso para este sistema debería tener $A > 1$ (a múltiples escalas; (Gershenson, 2011)). Para que un controlador fuese más eficiente debería tener $A \to 1$, lo que le daría más variedad que el sistema. Lo ideal es que el sistema tenga la capacidad de acoplar su complejidad a la del entorno, o sistema que se está tratando de controlar. Si el entorno cambia mucho, el sistema también.

En cuanto a los sistemas vivos desde la perspectiva de la variedad requerida, la complejidad y la autopoiesis existen elementos de importancia. Primero los seres vivos requieren de cumplir con la ley de variedad requerida. Segundo, se ha observado que la variedad puede ser vista como un sinónimo de complejidad (Bar-Yam 2004). Tercero, los organismos pueden ser descritos como sistemas de control. Estos elementos nos llevan a afirmar que la ley de variedad requerida también se encuentra en la vida, dado que los organismos deben hacer frente a la complejidad o variedad de su ambiente a diferentes escalas. Sobre las anteriores bases, la relación de la ley de variedad requerida se relaciona con nuestra medida de autopoiesis. Aún más, todo ello ha dado pie para que Gershenson (2014), haya armonizado y generalizado estos elementos en lo que puede ser denominado la ley de complejidad requerida: un activo y eficiente controlador requerirá por lo menos la misma complejidad que la complejidad de lo controlado. Es decir, un controlador para un sistema complejo requiere ser por lo menos tan complejo como el sistema que intenta controlar. Según Gershenson (2014), en la práctica se requiere de un balance entre predictibilidad y adaptabilidad del controlador para enfrentar la emergencia y auto-organización de lo controlado.

## 4.5 Síntesis

Hemos ilustrado a partir de experimentos computacionales, en sistemas dinámicos discretos (*RBA* y *ACE*), y en sistemas de tráfico vehicular, la utilidad de la valoración de la regularidad, cambio, adaptabilidad y equilibrio dinámico, a partir de las medidas propuestas de auto-organización, emergencia, complejidad y homeostasis.

La complejidad representa un balance entre auto-organización y emergencia, desde la relación establecida entre orden y caos (Kaufmann, 1993; Langton, 1990). Esta relación, es otra expresión de diversos balances hallados en la naturaleza, como por ejemplo el balance de especies obtenido por selección natural. Desde este conocimiento, el balance entre cambio y orden puede ser alcanzado por diseñadores que buscan explotar los beneficios de la complejidad.

La homeostasis brindó información sobre la estabilidad de los estados en el tiempo y su transición. La autopoiesis brindó información acerca de la autonomía del sistema respecto de su ambiente, en términos de complejidad-adaptabilidad.

Dado que todo puede ser descrito en términos de información, se hace posible la evaluación de la complejidad, emergencia, auto-organización, homeostasis y autopoiesis, en todo lo que nos rodea. En este sentido, existen diferentes tópicos de investigación en los cuales las presentes ideas pueden ser extendidas. Se propone la continuidad de su evaluación a través de la aplicación en diferentes sistemas, como por ejemplo: a las máquinas de Turing, a las máquinas $\epsilon$-psilon, a los sistemas tecnológicos, ecológicos y sociales, entre otros.

Desde una mirada compuesta se puede decir que combinaciones de medidas pueden ser usadas para un mejor entendimiento y combinación de sistemas dinámicos. Observando solo $C$ o $H$, no es posible distinguir la complejidad desde reglas caóticas, pero su combinación revela que el éter es requerido para la interacción de estructuras complejas. Existe también la pregunta de cómo considerar y medir la información usada por las medidas. Por ejemplo, en *ECA* diferentes reglas son obtenidas si una cadena de bits es considerada verticalmente (cómo lo hicimos), horizontalmente o diagonalmente.

# Capítulo 5: APLICACIONES EN SISTEMAS ECOLÓGICOS

## Resumen

*El estudio de la complejidad en Sistemas Ecológicos es un tópico de gran interés actual. Sin embargo, las interpretaciones que se hacen de ella han demostrado que se requiere de clarificaciones debido a la novedad del tema. En este capítulo, estudiaremos la complejidad ecológica de lagos y mamíferos terrestres. En estos dos casos de estudio, se analizaron las métricas de emergencia, auto-organización, complejidad, homeostasis y autopoiesis, desarrolladas en este trabajo. Adicionalmente, se desarrolla una solución analítica para la emergencia ecológica por medio de la programación genética, y se prueba su confiabilidad en lagos. Nuestros resultados muestran que la dinámica y complejidad ecológica puede ser efectivamente caracterizada en términos de información. Este enfoque, constituye una manera complementaria de entender la dinámica ecológica.*

## 5.1 Visiones Tradicionales y Complejidad Ecosistémica

Existen diferentes vías para describir el estado de un ecosistema y su dinámica. En ecología, lo natural es describir el sistema a través de sus atributos de abundancia, diversidad, frecuencia y/o riqueza de especies. Adicionalmente, los estudios ecológicos buscan establecer la estructura jerárquica a través de análisis de clasificación, que parten de medidas de similitud, y dan como resultados arboles de afinidad de especies y de sitios o dendrogramas (Legendre and Legendre, 1998). Igualmente, se tienen opciones de realizar análisis de ordenación de correspondencia canónica en los que las especies, los sitios y los parámetros fisicoquímicos del entorno, son correlacionados y asociados. El resultado de la ordenación son grupos de variables asociadas de manera significativa (Borcard et al., 2011).

En el anterior contexto y desde los resultados obtenidos, se sugiere que la medición de la complejidad ecológica (o biológica en general), puede complementar la descripción de ecosistemas y la explicación de la dinámica de especies. Sobre esta base, es posible ahondar en la descripción sintética la dinámica de los procesos ecológicos en términos de regularidad, cambio y adaptabilidad.

Las anteriores consideraciones, toman mayor importancia si se considera que el campo de la complejidad ecológica está en construcción y el interés por su estudio es creciente. Se destaca que ya se han dado esfuerzos que tratan de relacionar la complejidad con los descriptores ecológicos comunes, o al menos hacer una distinción entre unos y otros (Boschetti, 2010, 2008; Parrott, 2010, 2005; Proulx and Parrott, 2008).

Desde un principio, la complejidad biológica se trató de describir desde la caracterización de los componentes estructurales y sus interacciones funcionales (Pond, 2006). No obstante, se observó la dificultad de describir el comportamiento global de organismos y sistemas. Desde inicio de siglo, se estimó que en la asociación y modularidad de componentes podría darse algunas explicaciones al fenómeno de la complejidad. Desde allí, se estimó que la auto-organización juega un papel importante en la evolución de la complejidad biológica (Camazine et al., 2001; Kauffman, 2000). Enfoques desde la teoría de la información, han relacionado la complejidad de la vida como la cantidad de información que tiene o guarda un organismo o sistema en su genómica acerca de su ambiente (Adami et al., 2000). Se podría pensar que un sistema entre más información contenga, más complejo será. En consecuencia, se pensó que un organismo con alta información será muy auto-organizado, cómo lo propone el enfoque de la *negentropía* de (von Bertalanfy, 1968). Esta visión ha sido más coincidente con el pensamiento biológico, y se ha contrapuesto a la visión física de la entropía como producción de información nueva, como Shannon lo propuso.

Por otra parte, desde la teoría de la información se han desarrollado formalismos alternativos que han relacionado la complejidad con el grado de aleatoriedad de los estados de los sistemas ecológicos (Anand et al., 2010; Ulanowicz, 2011, 2004). Sin embargo, esta visión resultaría ser no la más adecuada, pues sería difícil demostrar que la complejidad es igual a la emergencia. Más aún, la complejidad no se podría igualar al caos, cómo ya se discutió en 4.4.4.

En el anterior contexto, y basados en que nuestra medida de complejidad puede reconciliar las dos visiones opuestas (entropía equivalente a nuestra $E$, y negentropía, equivalente a nuestra $S$), este capítulo mide la complejidad de variables y sistemas ecológicos.

### 5.1.1 Fundamentos de Limnología

Los ecosistemas acuáticos continentales superficiales son estudiados por la Limnología (*λίμνη*, limne, "lago" y *λόγος*, logos, "conocer"). Entre ellos se hallan dos tipos: los *lénticos*, como los lagos, lagunas, estuarios, esteros, pantanos embalses; y los *lóticos*, como quebradas, riachuelos y ríos.

Al ser sistemas abiertos, los lagos interactúan con su cuenca de drenaje y la atmosfera. A raíz de ello, un lago integran las relaciones funcionales de crecimiento, adaptación, ciclado de nutrientes, productividad biológica y composición de especies. En su estudio, se describen y evalúan cómo los ambientes químicos, físicos y biológicos regulan estas interacciones (Wetzel, 2001).

Los lagos presentan una zonación particular, que se presenta en la figura 5-1. Relacionada con la porción terrestre, en el litoral, se halla la zona de macrófitas (macrophytes zone). Esta zona está compuesta por plantas acuáticas que pueden estar enraizadas, sumergidas o flotantes, y constituye una frontera entre los sistemas terrestres y la zona planctónica (planktonic zone). Por su función fotosintética, las macrófitas son de gran importancia en la producción primaria.

La zona planctónica corresponde con las aguas más allá de la zona de macrófitas, donde los organismos que fotosintetizan (Fitoplancton) flotan pasivamente, y se dejan llevar por corrientes o gradientes horizontales y/o verticales. Entre tanto, los organismos que los pastorean (zooplancton) los siguen en sus migraciones.

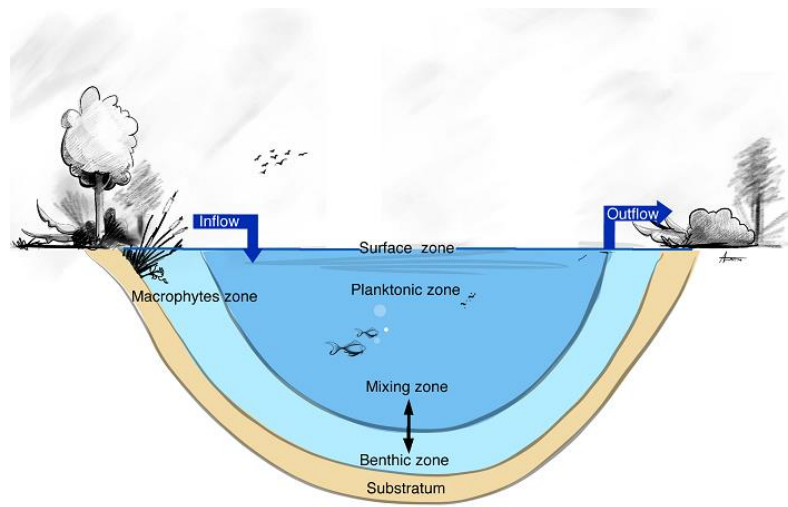

**Figura 5-1 Zonación de Un Lago. La figura muestra las zonas Superficial, planctónica, de macrófitas acuáticas, bentónica y el substrato.**(Fernández et al., 2014b)**.**

La zona bentónica (Benthic zone) tiene que ver con el fondo litoral y el fondo profundo del lago. Está relacionada con el sedimento (substratum) y la capa subsuperficial (ubicada a 20 centímetros arriba del substrato). Es bentónico todo organismo que esté en las inmediaciones del fondo, bien sea para escarbarlo, penetrarlo, o posarse en él. La zona de mezcla (Mixing zone), es una zona de intercambio entre las aguas de las zonas planctónica y bentónica.

En las zonas descritas arriba, uno o más componentes o subsistemas pueden ser valorados para determinar la dinámica ecológica. Para nuestro primer caso de estudio ecológico, se consideraron tres componentes.

(i)  El físico-químico (*PC*) referido a la composición física y química del agua. El *PC* es afectado por varias condiciones y procesos de naturaleza geológica, por la disolución y dispersión, por el ciclo hidrológico, por la generación de solutos y sólidos (de la fotosíntesis, por ejemplo) y por la sedimentación. En este componente, se destacan procesos cómo el equilibrio del *pH,* que afecta, entre otros, el intercambio de elementos entre un organismo y su ambiente. También, la regulación de la temperatura, que se relaciona con el calor específico del agua.

(ii) El componente de nutrientes limitantes (*LN*), que si bien está relacionado con el componente físico-químico (*PC*), es básico para la fotosíntesis. *LN* está asociado con los ciclos bio-geoquímicos del nitrógeno, carbono y fósforo, que permiten la adsorción de gases en el agua o la disolución de los nutrientes.

(iii) El componente de biomasa (*Bio*), considera el peso por unidad de volumen de organismos como productores primarios (fotosintéticos), consumidores primarios (que pastorean los primarios), consumidores secundarios (que predan los consumidores primarios), y los descomponedores.

Las variables estudiadas de cada componente se relacionan con la zonificación de un lago en las siguientes tablas.

**Tabla 5-1 Variables del Componente Físico-químico-PC**

| Variable (Español) | Acrónimo | Unidades |
|---|---|---|
| Luz Superficial | *SL* | MJ/m$^2$/día |
| Luz Planctónica | *PL* | MJ/m$^2$/día |
| Luz Bentónica | *BL* | MJ/m$^2$/día |
| Temperatura Superficial | *ST* | °C |
| Temperatura Planctónica | *PT* | °C |
| Temperatura Bentónica | *BT* | °C |
| Flujo de Entrada (Afluente) y Salida (Efluente) | *IO* | m$^3$/sec |
| Tiempo de Retención | *RT* | days |
| Evaporación | *Ev* | m$^3$/día |
| Mezclado entre zonas | *ZM* | %/día |
| Conductividad del afluente | *ICd* | μS/cm |
| Conductividad planctónica | *PCd* | μS/cm |
| Conductividad bentónica | *BCd* | μS/cm |
| Oxígeno superficial | $SO_2$ | mg/litro |
| Oxígeno planctónico | $PO_2$ | mg/litro |
| Oxígeno bentónico | $BO_2$ | mg/litro |
| Oxígeno de sedimento | $SdO_2$ | mg/litro |
| *pH* del afluente | *IpH* | Unidades de *pH* |
| *pH* planctónico | *PpH* | Unidades de *pH* |
| *pH* bentónico | *BpH* | Unidades de *pH* |

**Tabla 5-2 Variables del Componente de Nutrientes Limitantes-LN (Todas en mg/litro)**

| Variable (Español) | Acrónimo |
|---|---|
| Silicatos Afluente | *IS* |
| Silicatos Planctónico | *PS* |
| Silicatos Bentónico | *BS* |
| Nitratos Afluente | *IN* |
| Nitratos Planctónico | *PN* |
| Nitratos Bentónico | *BN* |
| Fosfatos Afluente | *IP* |
| Fosfatos Planctónico | *PP* |
| Fosfatos Bentónico | *BP* |
| Dióxido de Carbono Afluente | *ICD* |
| Dióxido de Carbono Planctónico | *PCD* |
| Dióxido de Carbono Bentónico | *BCD* |
| Detrito Planctónico | *Pde* |
| Detrito Bentónico | *Bde* |

**Tabla 5-3 Variables del Componente de Biomasa-Bio (Todas en mg/m³)**

| Variable (Español) | Acrónimo |
|---|---|
| Diatomeas Planctónicas | *PD* |
| Cianobacterias Planctónicas | *PCy* |
| Algas verdes Planctónicas | *PGA* |
| Planctónicas | *PCh* |
| Diatomeas Bentónicas | *BD* |
| Cianobacterias Bentónicas | *BCy* |
| Algas verdes Bentónicas | *BGA* |
| Macrófitas de superficie | *SurM* |
| Macrófitas sumergidas | *SubM* |
| Zooplancton Herbívoro | *HZ* |
| Zooplancton Carnívoro | *CZ* |
| Bentónicos Herbívoro | *BH* |
| Bentónicos Carnívoro | *BD* |
| Peces Planctonicos | *PlF* |
| Peces Bentónicos | *BF* |
| Peces Piscivoros | *PiF* |

### 5.1.2   Tipos de Lagos Estudiados

La escogencia de los cuatro lagos estudiados se dio en razón a su posición en el gradiente latitudinal ártico-trópico. Ellos correspondieron a un lago ártico (*Ar*), un lago de alta montaña en zona templada (*NH*), un lago de tierras bajas en zona templada (*NL*), y un lago tropical (*T*). El anexo A muestra la descripción ecológica detallada de cada uno de los lagos. El criterio de escogencia de los lagos se dio por su disposición latitudinal, que genera un gradiente de *Ar* a *T*, donde las condiciones de luz, temperatura e hidrología tienen amplia variación.

### 5.1.3   Simulaciones y Cálculo de las Medidas

Los datos de los lagos *Ar, NH, NL* y *T* fueron obtenidos usando *The Aquatic Ecosystem Simulator-AES* (Randerson and Bowker, 2008). Las variables consideradas para cada componente fueron definidas en las tablas 5-1 a 5-3.

El modelo para la obtención de valores diarios de las variables de los tres componentes, fue determinista, de manera que no hubo variación entre sus diferentes ejecuciones o corridas. Para todas las variables y lagos fueron obtenidos valores diarios.

Los datos resultantes de las simulaciones fueron normalizados a base 10, para ello se utilizó la siguiente ecuación para todos los puntos x de todas las variables $X$.

$$f(x) = \left\lfloor b \cdot \frac{x - \min X}{\max X - \min X} \right\rfloor \quad (5.1)$$

Donde $|x|$ es la función piso de $x$.

El software de apoyo para este fin fue R: *Comin*, desarrollado por Villate et al. (2013). Se destaca que en la comparación entre lagos, el rango máximo y mínimo para todas las variables fue tomado como la mayor y menor cantidad de todos los valores de todos los lagos. En casos individuales de comparación, se tomó solamente el máximo y mínimo de cada variable en el respectivo lago.

Una vez las variables transformadas en un alfabeto finito (para $b = 10$), se calculó la emergencia, auto-organización, complejidad, homeostasis y autopoiesis (Fernández et al., 2014b; Gershenson and Fernández, 2012a). Para esto se definió una clasificación para las medidas de acuerdo con las categorías, rangos y colores de la tabla 4. Estos rasgos fueron inspirados en los existentes para índices de calidad y contaminación del agua (Fernández et al., 2005; Ramírez et al., 2003).

**Tabla 5-4 Categorías, Rangos y Colores de Clasificación de las Propiedades E,S y C**
(Fernández et al., 2014b)

| Categoría | Muy Alta | Alta | Media | Baja | Muy Baja |
|---|---|---|---|---|---|
| Rango | [0.8 − 1] | [0.6 − 0.8] | [0.4 − 0.6] | [0.2 − 0.4] | [0.0 − 0.2] |
| Color | Azul | Verde | Amarillo | Naranja | Rojo |

## 5.1.4 Resultados

- Complejidad en un Ecosistema Ártico (*Ar*)

*Ar* tiene una marcada dinámica de la temperatura e hidrología que ejerce gran influencia en los componentes estudiados (Fig. 6-1 Anexo A). Este comportamiento se resume en la figura 5-2, que muestra la distribución de cada una de las variables del componente físico-químico en su escala original. Allí, es notoria la amplitud de las variables como temperatura y luz, los altos valores y dispersión del tiempo de retención y la evaporación, los cortos rangos de la conductividad y *pH*. Por razones como estas, que dificultan la visualización, es que se indican procedimientos de estandarización y normalización.

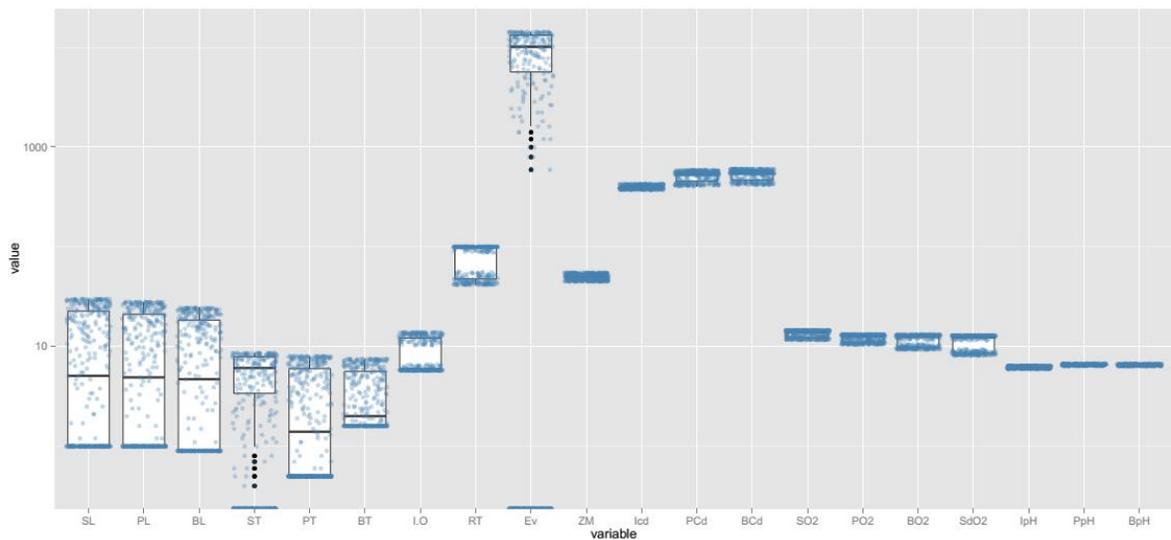

**Figura 5-2 Cajas de bigotes para las variables físico-químicas** (Fernández and Gershenson, 2014; Fernández et al., 2014b). En ella se representan cada uno de los valores de cada variable representados por puntos. Con ello se observa la distribución real de los valores. Las abreviaciones se dan en las tablas 5-1

En anterior figura (Fig. 5-2) se puede observar las variables transformadas a base 10 y la distribución de cada uno de sus 365 valores en las clases 0 a 9. La aglomeración o distribución de los puntos en una u otra categoría, da indicios del comportamiento regular o cambiante de la variable. Basados en esta distribución, algunos rasgos del comportamiento emergente o auto-organizado de cada variable, pueden ser descritos y comparados. Variables con distribuciones más homogéneas producirán más información, debido a que puede ubicarse en cualquier clase de la escala. Esto generará altos valores de emergencia. En cambio, variables con distribuciones más heterogéneas serán más regulares en sus estados, y por lo tanto, producirán valores de auto-organización más altos. La complejidad, sin embargo, requiere de una mirada más detallada, y no puede ser fácilmente deducida. Para ello, se requerirá de su cálculo y clasificación en las categorías propuestas en la tabla 5-3.

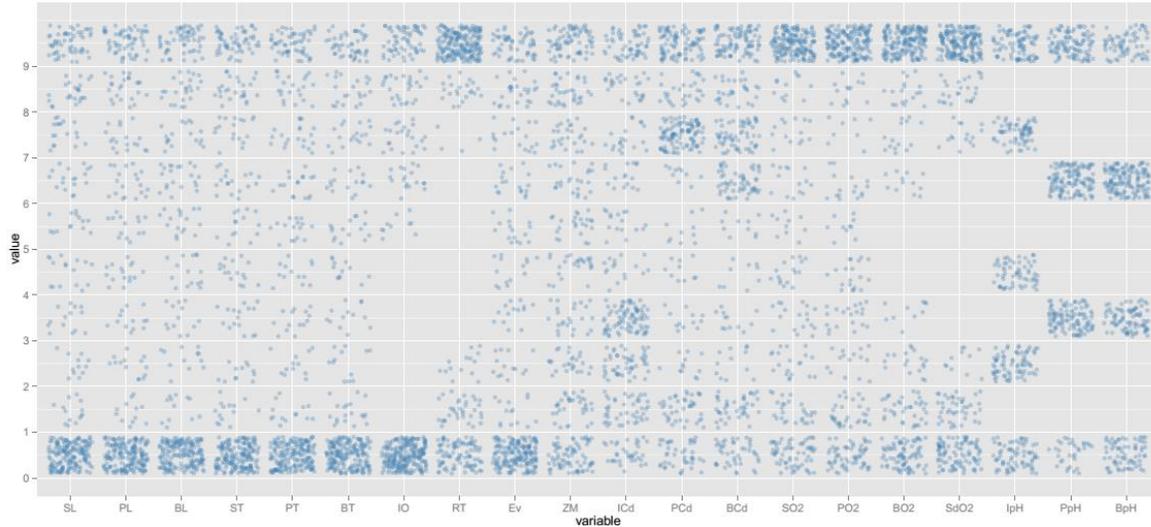

**Figura 5-3 Distribución de los valores normalizados a base 10 de las variables físico-químicas en las clases $0 - 9$.** (Fernández and Gershenson, 2014; Fernández et al., 2014b)**.** Variables con valores distribuidos de manera similar entre las clases **0** a **9**, tendrán mayor información. Es decir, alta emergencia pues la variable puede tomar cualquier valor. Variables con puntos distribuidos de manera más concentrada en una u otra clase, producirán menos información. Es decir, alta auto-organización pues tales clases tendrán una probabilidad alta..

Las probabilidades obtenidas para cada clase en cada variable, será la base para calcular la cantidad de información, y por ende, la emergencia, que será el primer cálculo en la aplicación de las medidas desarrolladas, explicadas en el capítulo 3.

El resultado de la aplicación de la métrica propuesta se observa en las figuras 5-4 a 5-6, donde se muestran los valores de emergencia $E$ (emergence), auto-organización $S$ (self-organization) y complejidad $C$ (complexity) para el sistema físico-químico (PC). Variables con muy alta complejidad $C \in [0.8 - 1]$ (según categorías de la tabla 5-4) reflejan un balance entre cambio/caos (emergencia) y regularidad/orden (auto-organización). Este es el caso del $pH$ planctónico y bentónico ($BpH;PpH$), flujo de entrada y salida ($IO$) y tiempo de retención hidráulica ($RT$). Variables con altas emergencias ($E > 0.92$), como conductividad ($ICd$) y zona de mezcla ($ZM$), cambian constantemente en el tiempo, una condición necesaria que exhibe el caos. Para las demás variables, los valores de auto-organización se clasifican como bajos y muy bajos ($S < 0.32$), lo que refleja baja regularidad. Se hace interesante observar que el lago ártico no tiene variables con alta auto-organización o baja emergencia. Esto nos conduce a sugerir que un sistema con periodos climáticos muy marcados, el sistema tenderá a su cambio periódico, debido a las fuertes transiciones de temperatura e hidrología, y mantendrá cierta estabilidad en los periodos intermedios. Por ende, la complejidad fisicoquímica, y la biológica se verá limitada.

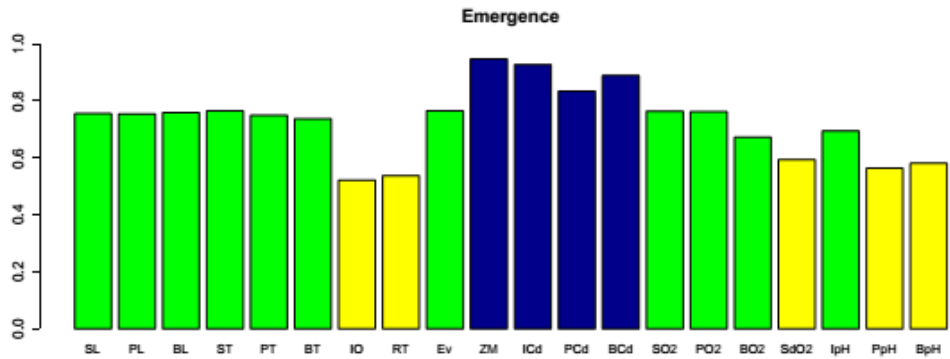

**Figura 5-4 Emergencia para las variables físico-químicas de un lago ártico** (Fernández and Gershenson, 2014; Fernández et al., 2014b)

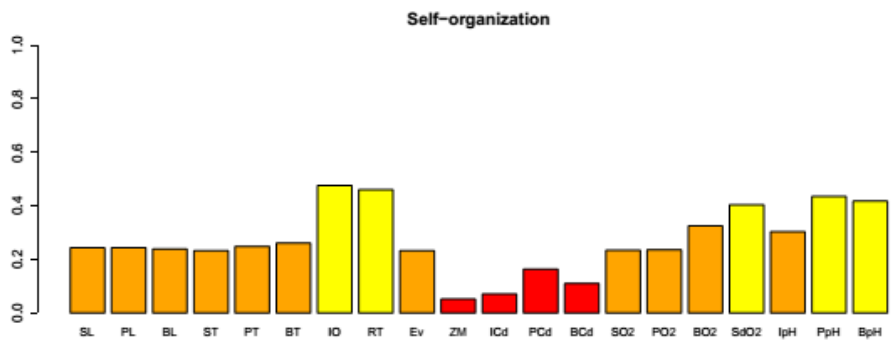

**Figura 5-5 Auto-organización de las variables físico-químicas de un lago ártico** (Fernández and Gershenson, 2014; Fernández et al., 2014b)

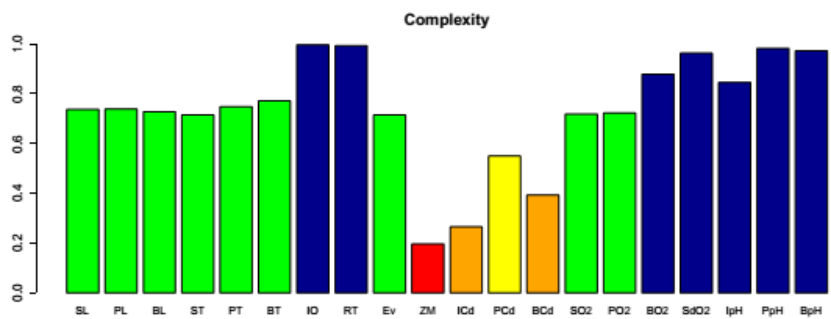

**Figura 5-6 Complejidad de las variables físico-químicas de un lago ártico** (Fernández and Gershenson, 2014; Fernández et al., 2014b)

Dado que $E, S, C \in [0,1]$, se puede hacer una clasificación de acuerdo a las categorías presentadas en la tabla 4 y definidas por (Fernández et al., 2014b), como sigue.

- **Variables con muy alta complejidad**: $C \in [0.8 - 1]$ (En Color Azul). Existen dos grupos de variables con un balance entre S y E: (i) *pH* del flujo de entrada, planctónico y bentónico (B*pH*, P*pH*), flujo de entrada y salida (*IO*), y tiempo de retención (*RT*). (ii) El oxígeno en el sedimento y el oxígeno Bentónico. Es interesante ver cómo estos dos grupos se correlacionan de manera inversa. Es decir, el incremento del régimen hidrológico durante el verano genera, por influencia de la temperatura, estratificación, y la consecuente depleción del oxígeno disuelto ($SO_2$, $BO_2$). En este sentido, se puede observar cómo el comportamiento complejo de variables que inciden en otras, generan respuestas igualmente complejas por asociación.

- **Alta Complejidad**: $C \in [0.6 - 0.8]$ (En Color Verde). Este grupo incluye 11 de las 21 variables que tienen alta $E$ y baja $S$. Estas 11 variables que muestran un comportamiento preferentemente caótico y baja regularidad, están altamente influenciadas por la radicación solar. Ellas son Oxígeno superficial y planctónico ($PO_2$, $SO_2$), temperatura y luz planctónica y bentónica (*ST, PT, BT, PL, BL*), y Evaporación (*Ev*). Por su parte, las conductividades en la zona planctónica y bentónica (*PCd, BCd*) están influenciadas por el régimen hidrológico que se desprende de la temperatura.

- **Muy Baja Complejidad**: $C \in [0 - 0.2]$ (En Color Rojo). Variables claramente *caóticas* o cambiantes caen en esta categoría, como se desprende de su muy alta $E$ y muy baja $S$. Incluye la conductividad en el flujo de entrada (*IO*) y el porcentaje de mezcla entre las zonas bentónicas y planctónicas (*ZM*). Las conductividades se aumentan debido al incremento del flujo de entrada, que genera un arrastre de iones y cationes en temporadas de deshielo. Posteriormente, en temporadas de congelación de la capa superior se ve reducida. El flujo también influencia de manera directa la zona de mezcla. Lo anterior hace que se asocien flujo, conductividades y mezclas, como las variables que mayor cambio tienen producto de la estacionalidad.

Desde la clasificación realizada, es notorio que en el ártico, dadas las fuertes condiciones climáticas que marcan los periodos de invierno y verano, el cambio físico-químico prevalece sobre la regularidad. Es decir, existe alta emergencia y baja auto-organización en su ambiente físico y químico. Sin embargo, lo pequeños valores de auto-organización permiten estos espacios altamente emergentes, que son la base de su alta complejidad. Alta complejidad físico-química, marcará la pauta para la adaptabilidad y la evolución biológica.

- Homeostasis ($H$)

La homeostasis fue calculada comparando los valores diarios de todas las variables según la ecuación 3.8. Como resultado se obtuvo la variación diaria, como se ve en la figura 5-7. La escala temporal es importante porque a través de la homeostasis podemos considerar si comparar estados cada minuto, días, semanas, mes o año.

El lago ártico tuvo un valor promedio de 0.9574 con una desviación de 0.0649. La homeostasis más baja fue de 0.6 y la más alta de 1. Si bien, se podría apreciar que la homeostasis general es alta, el ciclo anual mostró 4 periodos diferentes, cómo se observa en la figura 5-7, que se soportan en la descripción hecha de los lagos en el Anexo A. Estos periodos muestran valores dispersos de homeostasis como resultados de las transiciones entre invierno y verano, y nuevamente invierno.

Para el periodo de invierno (primeros y últimos meses del año), se da una homeostasis alta debido a los valores regulares de todas las variables en el lago. El periodo abarca del día 212 hasta el 365, y del 1 al 87, y se caracteriza por la baja luz, temperatura, máximos periodos de retención debido al cubrimiento de la superficie con hielo, bajo flujo y mezcla, bajas conductividades, bajos $pH$ y altos oxigenos presentes (ver anexo A).

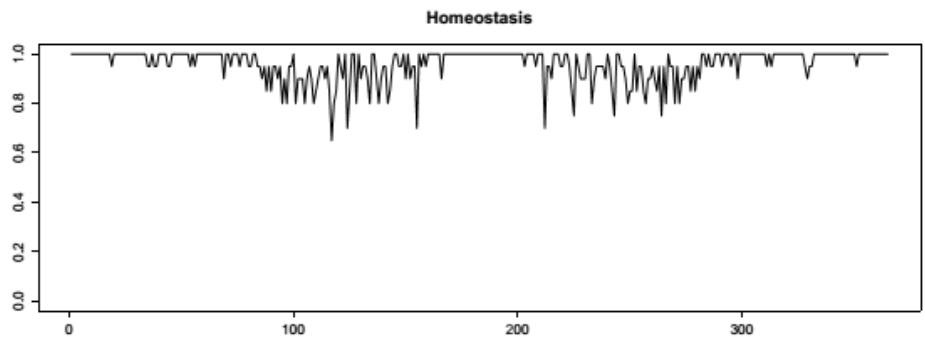

**Figura 5-7 Homeostasis del componente físico-químico de un lago ártico en un ciclo anual.** (Fernández and Gershenson, 2014; Fernández et al., 2014b)

El segundo periodo comienza en el día 83, y va hasta el 154. Se caracteriza por el incremento del $pH$ en la zona béntica, el incremento de la mezcla, el afluente y el efluente. Las fluctuaciones son fuertes debido al incremento de luz y temperatura por el advenimiento del verano. La homeostasis, por tanto, fluctúa y alcanza su mínimo de 0,6 en el día 116. Al final de este segundo periodo se incrementa la evaporación y el oxígeno decae en el fondo del lago.

El tercer periodo va entre los días 155 a 162, y refleja la estabilización de las condiciones de verano. Significa que la evaporación, temperatura, luz, mezcla, conductividad y $pH$ son máximos; mientras el oxígeno alcanza su más bajo nivel.

El cuarto periodo (días 163 − 211), en el que la homeostasis fluctúa en altos niveles cerca de 0,9, corresponde al periodo de transición entre el verano y el invierno.

Cómo se puede ver, la utilidad de la homeostasis se da en el seguimiento y estudio de la dinámica estacionaria.

- Autopoiesis (*A*)

El análisis de *A* estuvo enfocado a observar la autonomía de la biomasa, con respecto a la fisicoquímica (*PC*) y nutrientes limitantes, en las zonas planctónica cómo bentónica. Para ello, se establecieron las relaciones más estrechas entre organismos fitoplanctónicos y su entorno funcional directo, representadas por las variables listadas en la tabla 5-5, donde también se muestran los organismos escogidos.

Como primera condición para calcular *A*, se calculó la complejidad de cada grupo de variables, y se clasificaron según las categorías propuestas de muy alta a baja complejidad (Tabla 5-4).

**Tabla 5-5 Componentes tenidos en cuenta para el cálculo de la complejidad y posterior autopoiesis, en las zonas planctónica y bentónica del lago ártico**

| Componente/Zona | Zona Planctónica | Zona Bentónica |
|---|---|---|
| **Físico-químico** | Luz, temperatura, conductividad, *pH* | Luz, temperatura, conductividad, Oxígeno, Oxígeno en Sedimento, *pH* |
| **Nutrientes Limitantes** | Silicatos, Nitratos, Fosfatos, Dióxido de Carbono | Silicatos, Nitratos, Fosfatos, Dióxido de Carbono |
| **Biomasa** | Diatomeas, Cianobacterias, Algas verdes, Clorófitas | Diatomeas, Cianobacterias, Algas verdes |

La complejidad calculada fue promediada para cada componente y zona, como se observa en la figura 5-8. Así, los nutrientes limitantes en ambas zonas tuvieron baja complejidad ($C \in [0.2 - 0.4]$; color naranja). Las variables fisicoquímicas de la zona planctónica son de alta complejidad ($C \in [0.6 - 0.8]$; color verde). La biomasa y las variables fisicoquímicas de la zona bentónica están en la categoría de muy alta complejidad ($C \in [0.8 - 1]$; color azul).

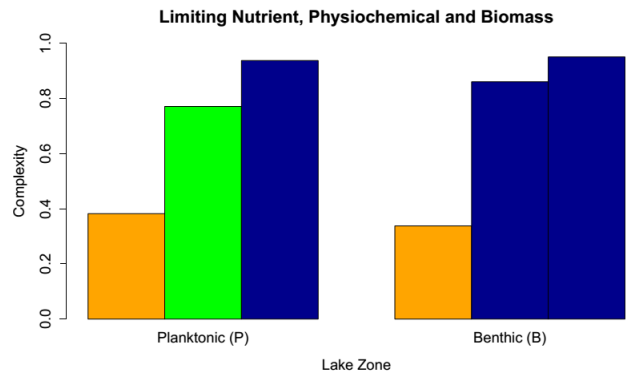

**Figura 5-8 Complejidad para los 3 componentes desde las variables elegidas en la tabla 5, para dos zonas de un lago ártico** (Fernández and Gershenson, 2014; Fernández et al., 2014b)

A partir de los valores de $C$ para los organismos y su ambiente (ecuación 3.5), se calculó $A$ según la ecuación 3.13, para los organismos en relación a su ambiente físico-químico y los nutrientes limitantes. Todos los valores resultantes fueron mayores a 1 (Figura 5-9). Esto significa que las diatomeas, cianobacterias, algas verdes y clorófitas que se hallan en el fitoplancton (biomasa planctónica y bentónica), tienen mayor complejidad que las variables de su ambiente, representado en *PC* y *LN*. A raíz de ello, se puede sugerir que algunas variables *PC*, incluidas *LN*, tienen un efecto pequeño sobre la dinámica biomasa bentónica y planctónica, pues su grado de complejidad, y por tanto adaptabilidad, es mayor que la de su ambiente. De allí se estima que la biomasa ártica es más autónoma (adaptable) comparada con su ambiente físico y químico. Los altos valores de complejidad de la biomasa, implican que estos organismos pueden adaptarse a cambios de su ambiente debido al mayor balance entre su emergencia y auto-organización.

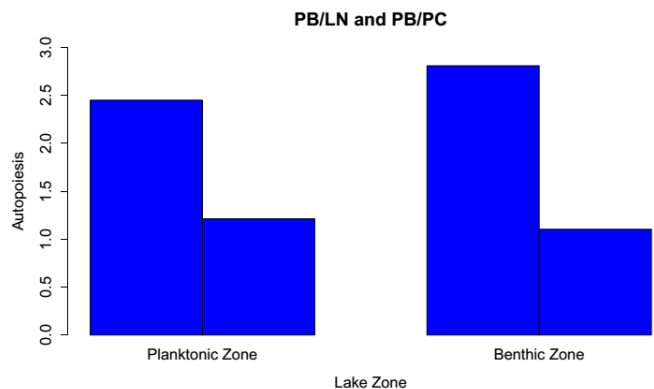

**Figura 5-9 Autopoiesis de la Biomasa Planctónica y Bentónica respecto de los nutrientes limitantes (PB/LN) y las variables fisicoquímicas (PB/PC), escogidas en tabla 5** (Fernández and Gershenson, 2014; Fernández et al., 2014b)**.** Para ambos casos fue superior a uno por lo que se les da el color azul

## 5.2 Complejidad comparativa en el Gradiente Altitudinal Ártico-Trópico (Ar-T).

Para el análisis de complejidad comparativa se consideraron, para la transformación a base 10 y la definición de los máximos y mínimos, los valores completos de cada variable para los cuatro lagos. En este sentido, el cálculo queda normalizado para todos los lagos. Una vez obtenidos los valores de complejidad de cada variable, se obtuvo el promedio del componente para un ciclo anual, y su dispersión, cómo se observa en la figura 5-10.

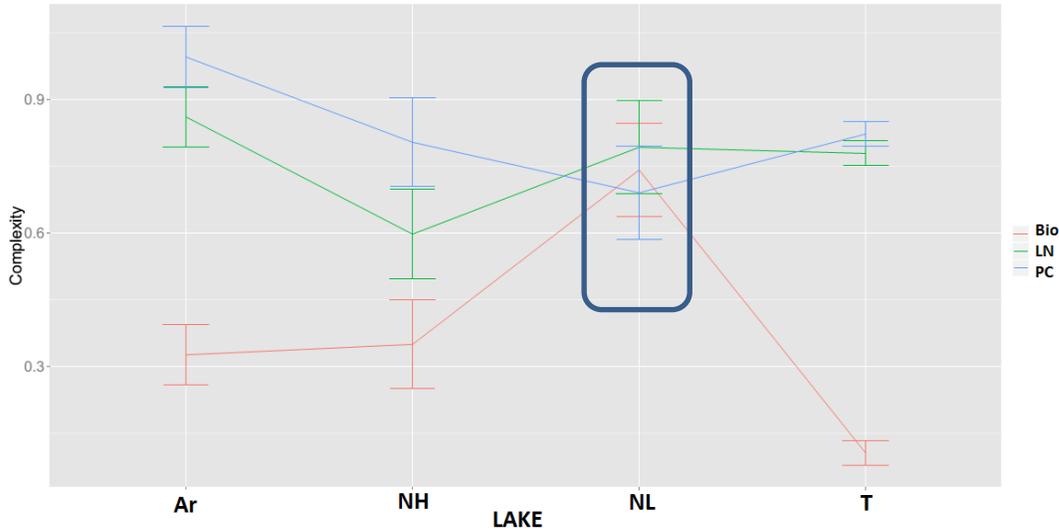

**Figura 5-10 Complejidad Comparativa en cuatro ecosistemas acuáticos- Ártico-Ar, Templado tierras altas-NH, Templado tierras bajas-NL y Tropical-T** (Fernández et al., 2014d)

En general, la complejidad fisicoquímica y de nutrientes limitantes se reduce desde el ártico al trópico, al igual que su variación. En tanto la complejidad biológica aumenta desde *Ar* a *NL*. Para ambos casos, podemos considerar al *Ar* y *T* cómo los extremos de la complejidad ecosistémica en el gradiente latitudinal. En el trópico se nota que la adaptabilidad de la biomasa, deducida de su complejidad comparativa que se requiere para responder al ambiente, se reduce debido a que en el trópico las condiciones son más estables.

En general en el ártico, la dinámica es más caótica y emergente, y en el trópico más regular y auto-organizada. La dinámica del ártico, da como resultado especies de más amplia tolerancia ambiental (Eurítípicas) que en el trópico (especies más estenotípicas, o de estrecha tolerancia ambiental). Al haber más auto-organización en el trópico, pero con la suficiente emergencia, se hace posible los procesos de especiación. Es por ello que en el trópico hay más diversidad y riqueza de especies. Esto es, las especies contarán con la suficiente posibilidad de cambio (emergencia, aunque sea poca) para que emerja una nueva especie como solución, si el ambiente cambia de forma considerable. La especie original que no pudo adaptarse, tendrá entonces el aislamiento a un lugar determinado donde pueda tolerar las condiciones ambientales, o el fin como destino dado su alta regularidad u auto-organización. En los dos casos, la especie originaria da paso a una nueva con nuevos rangos para poderse adaptar al nuevo ambiente (Rundell and Price, 2009; Schluter, 1996). Desde esta perspectiva, ser muy auto-organizado disminuye la posibilidad de adaptación biológica (Fernández et al., 2014d).

En el recuadro de la figura 5-10, se puede observar que los lagos templados de tierras bajas (*NL*) parecen representar un punto de cambio para la tendencia de la complejidad. A partir de él, la complejidad físico-química se pasa de alta a muy alta en el trópico. La complejidad de la biomasa va de alta en *NL* a muy baja en el trópico, y la complejidad de los nutrientes limitantes disminuye ligeramente, pero se mantiene en la categoría de muy alta. Se podría estimar que *NL* constituyen un punto crítico, o de mayor complejidad, entre dinámicas emergentes del *Ar* y auto-organizadas del *T*. Las variables responsables del cambio de la complejidad en el gradiente fueron, las macrófitas, clorofíceas y fósforo planctónico.

En promedio, para los cuatro lagos se puede ver la mayor complejidad de los componentes *PC* y *LN*, que *Bio* muestra que la respuesta de los componentes biológicos tiene menor autonomía dada su menor complejidad. Se puede decir que la complejidad fisicoquímica es mayor que la complejidad de los nutrientes, y ellas dos mayores que la complejidad de la biota ($C_{PC} > C_{LN} > C_{Bio}$). No obstante, es apropiado ver que la complejidad promedio de la biomasa, siendo más baja, tiene una mayor dispersión. Con ello, la biomasa tiene posibilidad de responder a su ambiente, acorde con la ley de variedad requerida de Ashby, dentro de su rango de complejidad ($0.382 \pm 0.22$).

Adicionalmente, análisis estadísticos de comparación de promedios múltiples (Prueba de Tukey) demostraron que la complejidad promedio de *PC* y *LN* entre lagos no tuvo diferencias significativas ($p = 0.85; p > 0.05$), lo que si sucedió para la biomasa con *PC* y *LN*. Esto demuestra que en promedio la complejidad de *PC* y *LN* es muy diferente de la *Bio* que es mucho menor. Este hecho ratificaría que la influencia del ambiente fisicoquímico y de nutrientes sobre la biota tiene una incidencia alta, por lo que los organismos deben hacer, en promedio, un esfuerzo mayor para adaptarse a ello.

Lo anterior contrasta con lo sucedido en el lago ártico, donde al observar la complejidad de un grupo de organismos no de toda la biota, como fue el caso de la complejidad comparativa, que fue mayor que la del ambiente *PC* y *LN*. Este hecho evidenció que grupos de organismos particulares pueden tener más complejidad local que global, y responder mejor dentro de sus condiciones particulares. Estos resultados sugieren dos formas de ver la complejidad, (i) por grupos de organismos (caso del ártico) y (ii) como un componente completo (caso de la complejidad comparativa), situaciones que pueden verse como complementarias.

## 5.3 Complejidad del Número Incremental de Especies de Mamíferos

Un análisis de gran interés en ecología con fines de conservación de especies, es el relativo a los análisis de ocupancia[9] de especies en un territorio (Bailey et al., 2014; Mackenzie and Royle, 2005; Royle, 2006), acorde con su detectabilidad[10] (Garrard et al., 2013; Tang et al.,

---

[9] Uso y ocupación del territorio por una especie.
[10] Probabilidad de detección de la especie, que tiene efecto en los registros de ella en los muestreos por diferentes medios como por ejemplo cámaras trampa o registro de rastros (huellas, heces, restos de consumo de alimentos, etc)

2006). Así, con el fin de ahondar en los aspectos de la dinámica de la presencia-ausencia de especies animales, se realizaron simulaciones computacionales de ocupancia en una comunidad de mamíferos, con número incremental de especies (Fernández et al., 2013). Para ello se diseñó una formulación jerárquica de dos procesos o ecuaciones estocásticas ligadas, como sigue:

$oc \sim Binomial(n, psi)$ (5.2)
$y \sim Binomial(i, oc * p)$ (5.3)

Donde $oc$ es la ocurrencia de especies en un sitio dado, $n$ es el número de individuos de especies, $psi$ es la ocupancia. $y$ es la detectabilidad de las especies donde: $i$ es el número de sitios con cámaras trampa, y $p$ es la probabilidad de detección. Para los dos casos, se partió de una distribución de probabilidad binomial. La comunidad simulada tuvo 10 sitios de muestreo con cámara-trampa y un ciclo de 1000 iteraciones variando en $psi$ y $p$ en cada iteración, como números aleatorios entre 0.1 y 0.9. Una vez generadas las poblaciones multi-específicas vía simulación, se aplicaron las medidas de $E, S$ y $C$ para el número incremental de especies en cada uno de los sitios.

Los resultados se muestran en la figura 5-11. Desde allí se puede establecer que existen para la emergencia niveles altos para la complejidad, niveles medios para la emergencia y bajos para la auto-organización. Lo anterior muestra mucha más variabilidad que regularidad en las apariciones o ausencias de las especies. Esta variabilidad se mantiene con el incremento de especies. Es decir, no es común encontrar a todos los especies de manera constante (alta probabilidad de 1) o no encontrarlos (alta probabilidad de cero). Más bien, existe una combinación entre mediana y altamente variable de su presencia y ausencia, lo que genera su alta complejidad.

Se observa, además, que no por incluir más especies en los 10 sitios se eleva la auto-organización, por suponer que se podría incrementar el número de detecciones. Se hace interesante observar cómo al incrementar el número de especies, la complejidad se mantiene prácticamente estable y tiende a mantener y consolidar valores, en la categoría de alta. La línea que representa la dispersión confirma la poca variación entre sitios, al hacerse cada vez más pequeña con el incremento de 1 a 1000 especies. Resultados como los anteriores son muy estimulantes, dado que muestran una tendencia estable en la dinámica de poblaciones de mamíferos.

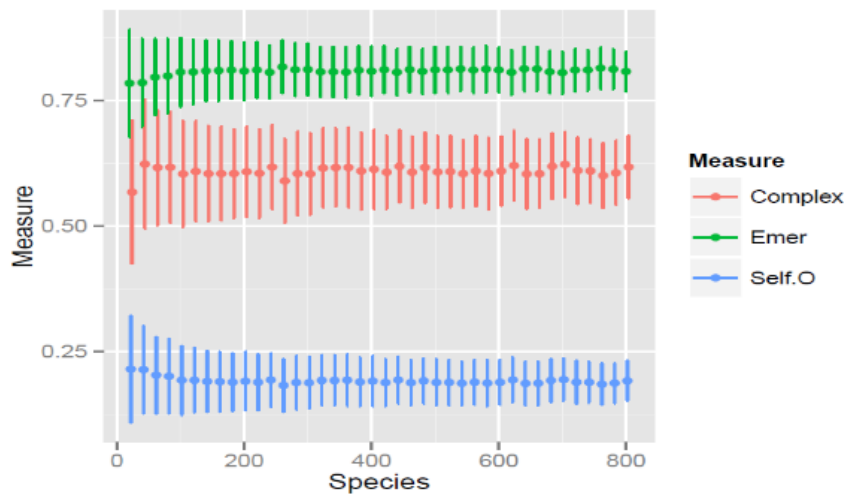

**Figura 5-11 Tendencia de la complejidad en una comunidad simulada de mamíferos para 10 sitios. Se muestran las primeras 800 especies en 1000 iteraciones** (Fernández et al., 2013).

## 5.4 Modelado y Especificación de la Emergencia Acuática a través del uso de programación genética

Uno de los mayores esfuerzos en ecología está en entender los aspectos de la emergencia asociada a la complejidad. Al respecto, soluciones analíticas son requeridas. Para llegar a ellas, se hace conveniente la integración del conocimiento y teorías desde las ciencias físicas hasta las sociales. En este apartado, utilizamos la programación genética para develar las relaciones entre la emergencia ecológica y los procesos complementarios de auto-organización, homeostasis y autopoiesis. Los métodos y resultados expuestos se presentan acorde con lo desarrollado en (Fernández et al., 2014c).

Respecto de la programación genética (*PG*) y su aplicación en ecología, se puede estimar que la generalidad, realismo y certeza de los modelos desarrollados para ecología hacen de ella una herramienta apropiada para entender y expresar el comportamiento complejo de los ecosistemas. Basada en la forma en que se da el procesamiento de la información biológica, la PG define un algoritmo evolutivo para encontrar programas de computador que busquen soluciones analíticas más apropiadas a través de procesos de entrecruzamiento y evolución de funciones. Esencialmente, la PG tiene un conjunto de instrucciones y una función de ajuste para medir la bondad del resultado de la tarea llevada a cabo (John R Koza, 1994). La PG tiene ventajas comparativas con otras técnicas de aprendizaje de máquina, debido a que necesita un conocimiento pequeño a priori, que se obtiene de un conjunto de datos del sistema en estudio (Koza and Poli, 2003). La literatura revela los beneficios de la PG dentro de un marco de identificación del sistema (Koza 1994), razón por la cual fue escogida para llegar a las soluciones analíticas que requirió la determinación de la emergencia ecológica a partir de los fenómenos de auto-organización, homeostasis y autopoiesis.

El modelo desarrollado tuvo los siguientes componentes básicos:

- **Los individuos**. Descritos por los árboles que representaron las ecuaciones matemáticas. Los nodos de las hojas son variables o constantes. Los demás, son funciones matemáticas u operadores.
- **Funciones y terminales**. Las funciones fueron operadores matemáticos básicos como: adición, sustracción, multiplicación y división. Para los terminales (u hojas), se usaron las variables que representan la emergencia, auto-organización, homeostasis y autopoiesis.
- **Los operadores genéticos**. Los principales fueron el cruzamiento y la mutación. El primero fue un operador de punto único para dos individuos. El segundo, tomó parte del árbol (rama) de un individuo y generó uno nuevo.
- **El ajuste del modelo**. Se definió una función de evaluación de ajuste del modelo, entre los valores observados de la emergencia del lago ártico comparado con los valores calculados por la programación genética para el mismo lago. La métrica para el ajuste del modelo, fuero el coeficiente de correlación y análisis de residuos; para ello se usó el paquete *CAR* (Companion to Applied Regression) de R.

Las condiciones experimentales tuvieron un número de generaciones de 3000, con excepción de la biomasa en que 2000 generaciones fueron usadas. El tamaño de la población fue de 50 y la probabilidad de uso del operador de cruzamiento fue de 1.0, la probabilidad de mutación 0.08.

Los mejores organismos obtenidos se muestran a continuación en las figuras 5-12 a 5-14.

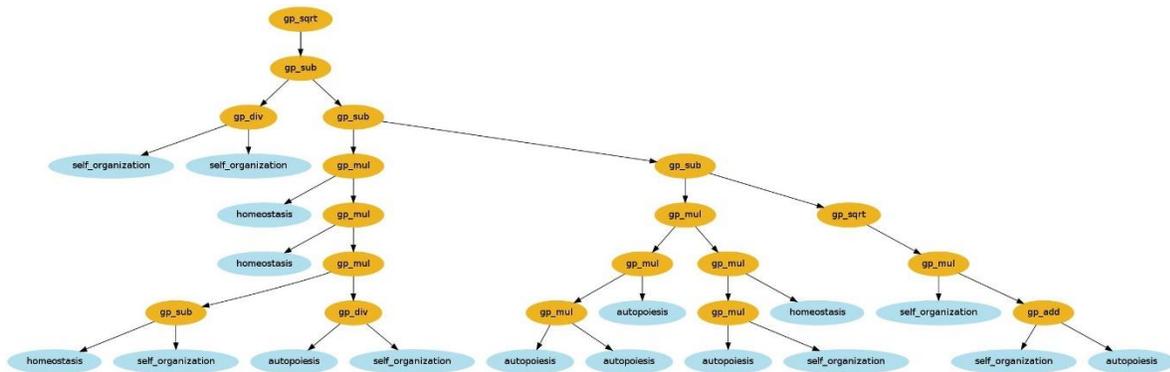

**Figura 5-12 Mejor Solución Obtenido para Nutrientes Limitantes**

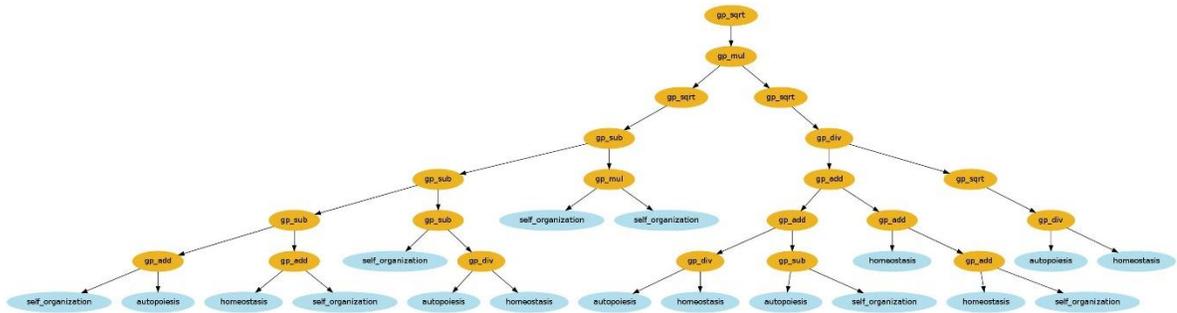

**Figura 5-13 Mejor Solución Obtenido para Biomasa**

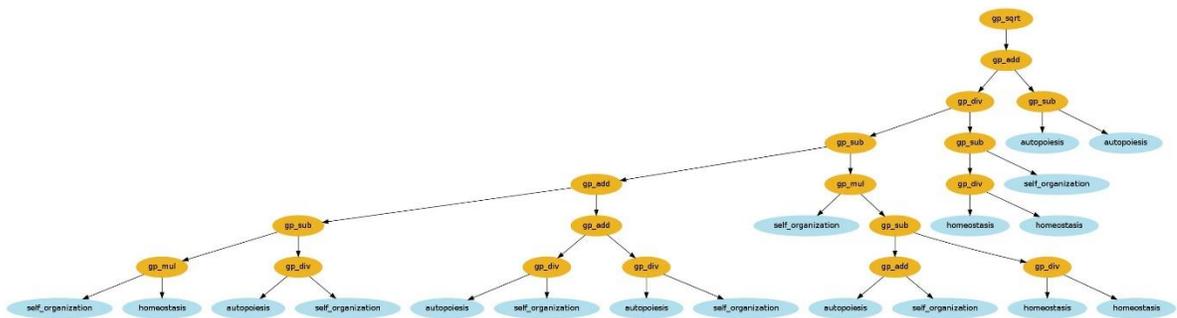

**Figura 5-14 Mejor Solución Obtenido para el Componente Físico-químico**

Estos individuos obtuvieron el mínimo error a partir de la función objetivo entre las $E$ observadas y calculadas. Las funciones halladas fueron las siguientes:

Para biomasa:

$$f(S,A,H) = \sqrt{1 - \left(\frac{H^2 A(H-S)}{S}\right) - (A^4 SH) - \sqrt{S(S+H)}} \qquad (5.4)$$

Para nutrientes limitantes

$$f(S,A,H) = \sqrt{\frac{(SH + A/S) - (S(A+S) - 1}{1-S}} \qquad (5.5)$$

Para el componente físico-químico:

$$f(S,A,H) = \sqrt{\sqrt{\left(\left(A - H - S - A/H\right)\right) - S^2 \sqrt{\frac{A/H + A + 2H}{\sqrt{A/H}}}}} \quad (5.6)$$

Donde $f(S,A,H), A, S, H$ son la emergencia, autopoiesis, auto-organización y homeostasis, respectivamente.

La robustez del modelo se evalúa al observar las gráficas 5-15 a 5-17, el ajuste de los datos calculados y predichos en la regresión y la desviación de los residuales a la misma. En ellas, la línea roja indica el ajuste perfecto de 1. Cómo se observa, los puntos que representan las parejas entre calculados y predichos se ubican de manera cercana a la línea de regresión. El análisis se complementa con la graficación de los residuales, representados en la línea verde. Para todos los casos los residuales fueron muy pequeños, lo que se puede notar en la escala del eje $Y$. Esto denota una pequeña diferencia en la predicción del modelo, que no es considerada como significativa, como se confirma en otros indicadores como el error absoluto promedio del modelo que se muestra más adelante.

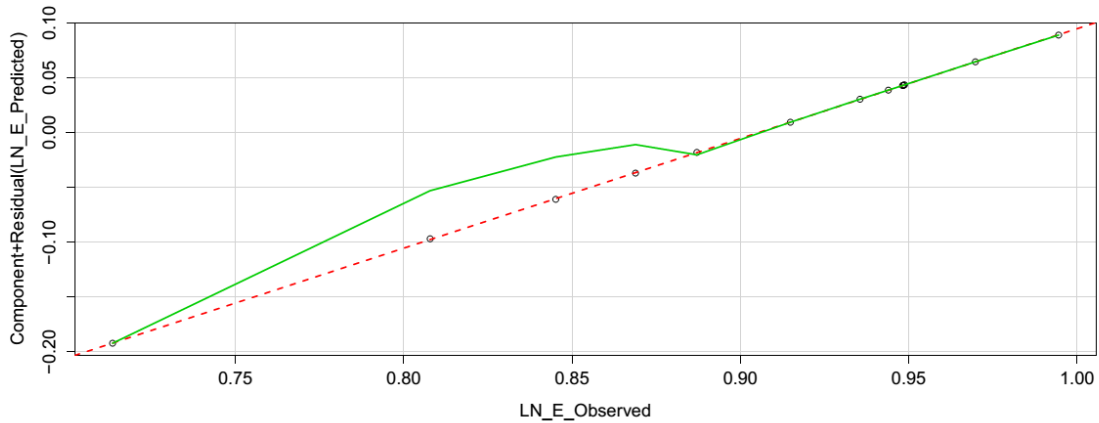

**Figura 5-15 Prueba de Ajuste del Modelo PG al Componente de Nutrientes limitantes**

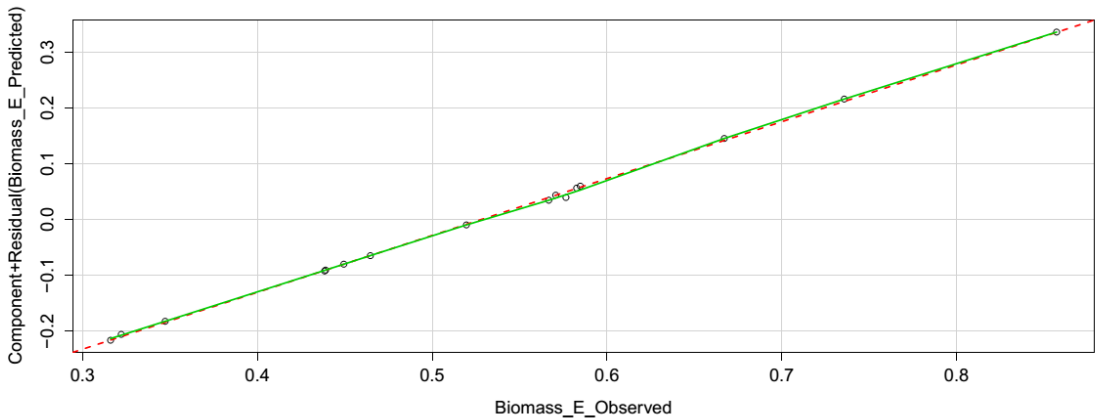

**Figura 5-16 Prueba de Ajuste del Modelo PG a la Biomasa**

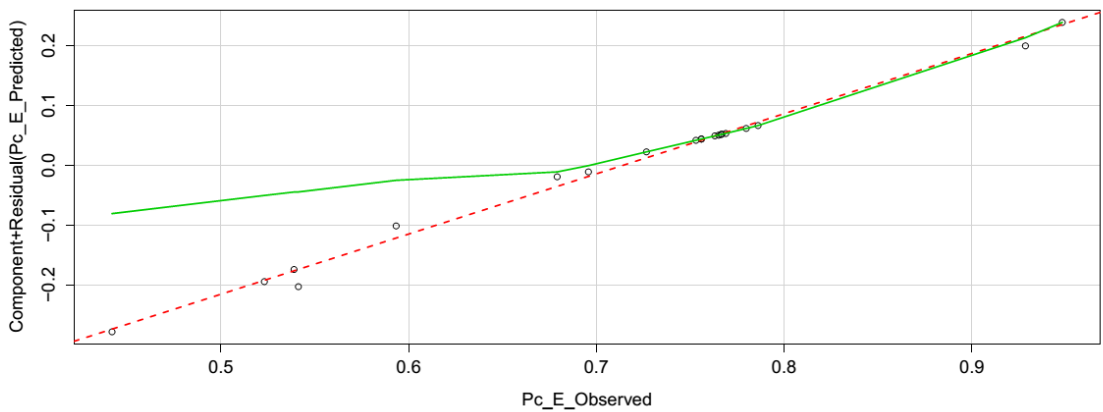

**Figura 5-17 Prueba de Ajuste del Modelo PG al componente Físico-químico**

Se puede observar que en las representaciones gráficas, el adecuado ajuste de las soluciones obtenidas por *PG* a pesar de la gran diferencia de los lagos. Lo anterior puede ser confirmado por los coeficientes de correlación obtenidos que fueron de 2; 0.9995 y 0.9937 para nutrientes, biomasa y físico-químicos. Los errores medios absolutos para los mismos componentes fueron 0.0004169; 0.005368 *y* 0.007129. Lo anterior demuestra la gran exactitud del modelo

En términos ecológicos, los resultados alcanzados son de gran importancia. En particular, debido a las grandes diferencias que tienen los lagos ártico y tropical. Se podría decir que están en verdaderos extremos ecológicos, el ártico con una alta variabilidad lo que conduce a mayores emergencias de sus variables, y el tropical con alta regularidad, lo que incrementa su auto-organización. A manera ilustrativa se da el ejemplo entre el comportamiento del componente físico-químico de los dos lagos, en las figuras 5-18 y 5-19, lo que confirma su diferencia, en especial en cuanto a su homeostasis, la cual muestra mayor equilibrio del sistema tropical. Además, para complementar esta ilustración, en el anexo A se pueden ver la descripción de cada uno de los lagos y sus regímenes hidro-climáticos.

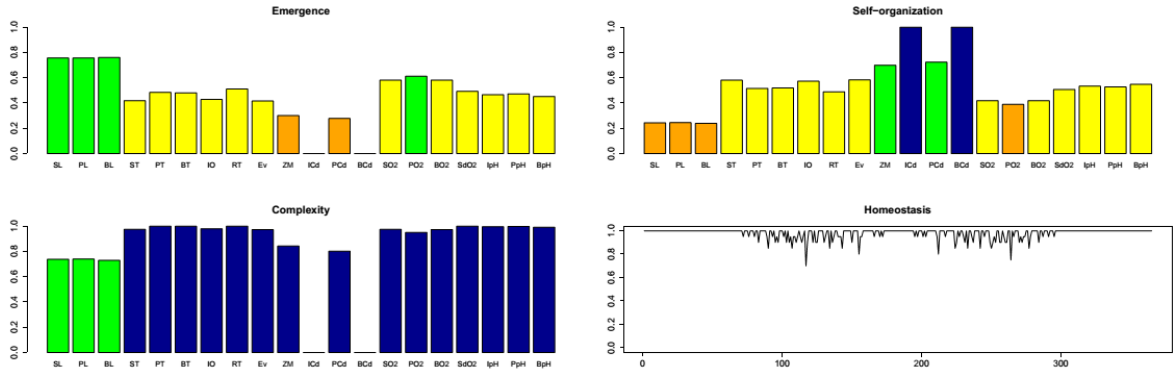

**Figura 5-18 Emergencia, Auto-organización, Complejidad y Homeostasis para un el Componente Físico-químico de un Lago Ártico**

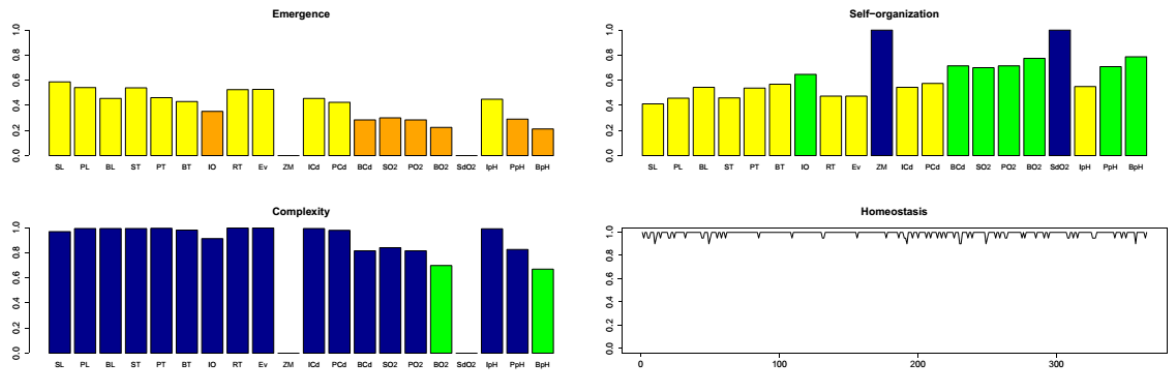

**Figura 5-19 Emergencia, Auto-organización, Complejidad y Homeostasis para un el Componente Físico-químico de un Lago Tropical**

Las anteriores diferencias entre los lagos ártico y tropical, demuestran por un lado la importancia de la emergencia como una expresión que mide la nueva información, y por el otro la importancia de las expresiones halladas que permiten medirla desde los procesos auto-organizantes, que se dan sobre la base de procesos homeostáticos y autopoiéticos.

## 5.5 Discusión

### 5.5.1 Descripción de los Ecosistemas y la Inclusión de la Complejidad como Indicador Ecológico

El uso de la complejidad como indicador ecológico nos permite ver cuánto una entidad ecológica es más compleja que otra, como el caso comparativo entre variables fisicoquímicas, o entre lagos en diferentes latitudes. También podemos generar perfiles de complejidad espacial (latitudinal, latitudinal, entre otros). Podemos identificar entidades que representan cambios de complejidad en los gradientes. Por ejemplo, basados en los valores promedios de $C$ para los tres componentes, se pudo observar que en el gradiente altitudinal, $NL$ representa un punto intermedio donde la complejidad latitudinal cambia. Sí bien los resultados muestran dinámicas diferenciales por componentes, se requiere de más estudios para determinarlo con precisión. En este mismo sentido, será apropiado definir si existen variables que exhiben gradientes relacionados con este cambio de complejidad. Por ejemplo para la biomasa, podemos estimar que podría haber un cambio en el transecto *Ar-T*, en *NL*, para las cianobacterias en la zona planctónica.

El uso de la complejidad como indicador ambiental, también permite ver como en los trópicos la alta $S$ y baja $C$ pueden conducir a baja $A$ en el componente de biomasa. La alta regularidad muestra que los organismos son estables a escalas anuales. En estos ambientes, la biomasa reduce su autonomía y adaptabilidad, motivo por lo que se favorece la especiación, por los cortos rangos que soportan las especies. Esto se asocia con el grado de robustez que tienen los organismos, que al ser alta da baja adaptabilidad, por lo que la generación de nuevas especies se promueve. Todas estas apreciaciones, complementan de manera apropiada lo que se sabe de los trópicos en cuanto a su regularidad, con la ventaja que le dan un sustento numérico para su explicación.

En adición a lo anterior, podemos usar la complejidad para ver en qué condiciones los organismos pueden tener rasgos de mayor autonomía al ser más complejos que su ambiente. Esto fue explícito en los resultados para el componente fitoplanctónico de los lagos.

La medición de la complejidad del número incremental de especies, vía simulación, es un punto interesante de partida para la deducción de relaciones con aspectos como la riqueza y número de especies, con fines de conservación.

### 5.5.2 Sobre los Aspectos Computacionales de la Medición de la Complejidad Ecológica

Entre los aspectos computacionales de la complejidad ecológica se debe considerar que los sistemas complejos como sujeto de estudio, cuentan con el aporte de biólogos, ecólogos, computistas, matemáticos, físicos, economistas y muchos otros científicos. Como resultado, existe una mezcla de enfoques que puede generar confusión en los conceptos. En cada campo, un enfoque distinto es tomado. Los biólogos, por ejemplo, están interesados en la auto-organización y la emergencia, debido a que vienen trabajando con estos conceptos desde Darwin y los trabajos de Mendel. Por otra parte, los científicos de la computación quieren relacionar la medida de complejidad para sistemas complejos con la complejidad que ellos ya han definido. Por ejemplo, con la complejidad de Chaitin, donde la complejidad de un objeto/problema es la longitud del programa mínimo que necesita una máquina universal de Turing para reproducirlo/resolverlo. En este contexto, la teoría de la información es muy útil para obtener una medida de $C$, comenzando con la noción de $E$ cómo un tipo de entropía, descrita por la medida de información de Shannon, que es equivalente a la entropía de Boltzman. Cómo la temperatura, que caracteriza las propiedades de la energía cinética de las moléculas o fotones, una medida global de la complejidad ecológica puede caracterizar la dinámica de las interacciones espacio-temporales (Fernández and Gershenson, 2014). Por ejemplo, a pesar que cada componente de un sistema ecológico tiene diferentes (micro) estados que son afectados por el estado de los demás elementos (interacciones), su dinámica promedio y los diferentes regímenes dinámicos pueden ser representados en forma sintética en nuestra medida de complejidad. Esto se pudo observar en la aplicación de la complejidad comparativa de lagos.

Considerando que $E$ puede ser entendida cómo la información producida por un proceso o ecosistema, una $E = 0$ nos permite interferir sobre la predictibilidad de los estados, dado que no se produce nueva información. Es decir, dado que los estados presentan alta regularidad ($S = 1$), lo más probable es que el estado que viene sea el mismo que los anteriores. Entre tanto, cuando $E$ es máxima ($E = 1$) los estados son irregulares o seudo-aleatorios, y por lo tanto, el siguiente estado traerá nueva información. Dado que la regularidad es prácticamente nula ($S = 0$), será muy difícil predecir el próximo estado.

La integración de los aspectos de $S$ y $E$ en nuestra medida de $C$, tiene ventajas como el observar la dinámica compleja de un ecosistema como un balance entre su regularidad y cambio. En este contexto, podemos definir qué variable, componente o proceso, es más complejo-adaptable que otro. Esta pregunta de gran importancia en ecología, nos puede conducir a generar categorías de complejidad de procesos, poblaciones, comunidades y ecosistemas. También nos puede brindar una medida de la adaptabilidad de nuestra biosfera, y en general, de la complejidad biológica.

Igualmente, la medida de $C$ puede ser útil cuando se desea relacionar el cambio evolutivo y la sostenibilidad de ecosistemas. Es únicamente cuando $C$ es máxima y mayor que el ambiente que la evolución potencial en los sistemas puede ser mejorada, dado que los niveles intermedios de $E$ y $S$ le dan al sistema el suficiente potencial de cambio, pero con la suficiente robustez para perdurar. Para ello se parte de una condición más o menos estable, que para el momento será el estado más probable del sistema.

Cómo los sistemas vivos requieren de un balance entre adaptabilidad-emergencia y estabilidad-auto-organización (Kaufmann, 1993), $C$ es útil para caracterizar sistemas vivos. A partir de ella, se puede relacionar la complejidad del sistema con la del ambiente, para obtener su autopoiesis. En este sentido, la caracterización del comportamiento del componente biótico, ante los cambios o disturbios ambientales, en términos de complejidad, puede dar información acerca de la autonomía de los organismos. Desde esta perspectiva, la complejidad, la homeostasis y la autopoiesis, pueden ser de utilidad en estudios de cambio climático global. En ellos podría confirmarse nuestra hipótesis que sistemas con alta complejidad tienen el suficiente grado de robustez y adaptabilidad para enfrentar los cambios ambientales. Esto lo soportamos en los datos de autopoiesis que se dan en ambientes extremos, cómo el lago ártico.

### 5.5.3 Comparación con otras Medidas de Complejidad.

Desde las aplicaciones realizadas hemos hallado diferencias con algunos autores. Por ejemplo, Shalizi et al.(2004) define auto-organización cómo un incremento interno de la complejidad estadística. Esto puede suceder en sistemas con alta $C$ y máxima $S$. Entre tanto, para nosotros consideramos una máxima $S$ para sistemas con mayor regularidad y de patrones más estáticos. Alineados con la termodinámica y la teoría de la información, creemos que un cristal y un estado congelado, son más organizados que complejos. Sin embargo, un sistema con una $S$ mínima, lo más probable es que sea no organizado. En este aspecto creemos que la medida de Shalizi es más apropiada para estudiar la dinámica de sistemas auto-organizados, más que emergentes (Camazine et al., 2001; Gershenson, 2007). Se destaca que un sistema de baja auto-organización podría elevarla bajo la influencia de un factor externo, en lo que podemos considerar como auto-organización guiada (Fernández et al., 2014b).

Desde la simplicidad de nuestras definiciones, la auto-organización puede ser vista como el opuesto de la emergencia. La auto-organización es alta para sistemas de dinámicas ordenadas, en tanto la emergencia es alta para sistemas altamente caóticos (en todas las escalas). Qué la auto-organización y emergencia sean opuestos puede ser visto como contra-intuitivo, dado que hay muchos sistemas que presentan alta complejidad al *parecer* por tener alta emergencia y alta auto-organización. Precisamente, estos sistemas, son los más interesantes para la ciencia. Para nosotros estos sistemas son los que tienen alta $C$ , a partir de una combinación de alta auto-organización y baja emergencia, o viceversa. Cabe destacar, que en nuestra propuesta en valores opuestos y altos de $S$ y $E$ encontramos el extremo del orden y del desorden, de allí su carácter opuesto.

La información (y emergencia) puede ser vista, en una variable binaria, como un balance de ceros y unos ($P(0) = 1 - P(1) \Leftrightarrow P(0) = P(1) = 0.5$). Así, la complejidad puede ser vista como un balance de emergencia y auto-organización ($I = 0, S = 1 - E; Max(C) \Leftrightarrow S = E = 0,5$). Ya ha sido discutido en la comunidad científica que la complejidad es máxima cuando existe una compensación entre persistencia de información (memoria) y su variabilidad (exploración, computación) (Bar-Yam, 2004a; Langton, 1990). También ha sido argumentado, que existe una tendencia de auto-organización o evolución a estados que incrementan su balance y alta complejidad (a múltiples escalas) (Bak et al., 1988; Gershenson, 2012b; Kauffman, 2000). Lo anterior está en acuerdo con el trabajo de Escalona-Moran et al.(2012).

Aun cuando nuestra formulación es diferente, $C$ está estrechamente relacionada con la complejidad estadística de Crutchfield and Young (1989), que mide cuanta computación es requerida para producir información. Cadenas ordenadas o aleatorias si bien requieren de computación, cadenas complejas requieren más. Esta complejidad estadística, puede también ser definida como información mínima requerida para predecir el futuro de un proceso (Shalizi et al., 2004). Para nuestra $C$, baja $I$ implica cadenas que son fácilmente predecibles, mientras altas $I$ implican cadenas que no tienen mucho que predecir. Alta $C$ ocurre cuando $I$ es media, lo cual corresponde con más computación requerida para producir o predecir un patrón determinado. Aun cuando $C$ y la complejidad estadística, exhiben comportamientos similares, $C$ es mucho más simple de calcular y explicar.

### 5.5.4 La Búsqueda de Soluciones analíticas para la Emergencia

El principal aspecto a resaltar de este tipo de modelado y especificación de la emergencia en sistemas acuáticos, producto del uso de la programación genética, es que se obtuvieron tres expresiones generales para la emergencia basada en sus propiedades asociadas de auto-organización, homeostasis y autopoiesis. Si se considera que la emergencia es información nueva, y que por las interacciones se hace difícil de predecir, debido a los cambios ambientales, esta expresión cobra especial interés.

El primer punto de partida para la comprobación de la generalización de las expresiones obtenidas por *PG* en el lago ártico, fue su comprobación de funcionamiento correcto en la estimación la emergencia en el lago tropical, en los componentes de *PC*, *LN* y *Bio*. Este inicio, sienta las bases para la extensión de la generalización de las expresiones a otros lagos. También nos promueve a seguir en la evaluación de otras propiedades sintéticas, cómo la complejidad ecostémica, además de otros componentes ecosistémicos, de manera que se vaya confirmando su generalización.

## 5.6  Síntesis y Comentario Final.

A partir de las medidas desarrolladas de $E, S, C, H$ y $A$, basadas en la teoría de la información y compuestas por expresiones matemáticas simples, se pudieron capturar aspectos complementarios a los tradicionales estudiados para la dinámica ecológica. Esto es, indicadores cómo la abundancia, diversidad, riqueza de especies, etc.

En contraste con los anteriores indicadores, y otros análisis ecológicos cómo el de estructura jerárquica, nuestro enfoque permite analizar propiedades sintéticas de los ecosistemas, así como sus tendencias, en términos de la información. Esto puede ser aplicado a cualquier variable en diferentes componentes, al componente como tal, o a uno o más ecosistemas. Para ello se requiere que previamente las variables hayan sido transformadas la escala (base) deseada. Con esto, se logra una forma complementaria a la tradicional de observar los sistemas ecológicos sobre la base de su regularidad, cambio y balance.

Las aplicaciones en sistemas acuáticos descritas, han considerado: (i) las dinámicas de ecosistemas por separado en cuanto a sus componentes *PC, LN* y *Bio*. (ii) Los análisis comparativos entre ecosistemas (*Ar, NH, NL* y *T*). (iii) Estudios de la ocupancia y detectabilidad de especies, un tópico de interés en estudios de conservación. Desde estas últimas, se pueden dar otras extensiones, como por ejemplo, estudios de patrones de movimiento.

Los resultados obtenidos muestran como las medidas $E, S, C, H$ y $A$, pueden ser vistas como indicadores ecológicos útiles a diferentes escalas, que permiten describir cómo los ecosistemas se mantienen cerca de ciertos estados (atractores), pero con la suficiente variabilidad para poderse adaptar de manera autónoma al medio. En este aspecto, la regularidad de los estados puede ser observada en la auto-organización como sucedió para la biomasa en el lago tropical, la variabilidad en la emergencia como en el componente fisicoquímico del ártico, la adaptabilidad en la complejidad como todos los componentes del lago ártico, y la autonomía en la autopoiesis como en el fitoplancton ártico.

Como se ha demostrado, es posible describir la dinámica ecológica en términos de información, como también ha sucedido en estudios previos (Fernández and Gershenson, 2014; Fernández et al., 2014b, 2014d, 2013), lo que muestra un campo creciente y una visión complementaria en ecología. Ello sienta las bases para inspeccionar el significado ecológico y aplicación de las medidas propuestas, así como buscar su validación progresiva en otros procesos y sistemas ecológicos.

Basados en nuestra expresión de complejidad, puede ser posible dar soporte y explicación al porqué los organismos tendiendo al incremento del orden por el almacenamiento de información, tienen la posibilidad de tener un componente aleatorio que les permite explorar e innovar, para persistir. Lo anterior constituye la base de la adaptabilidad de las especies y ecosistemas (Walker, 2005).

Por su parte, los resultados obtenidos desde la programación genética, han sentado las bases para la continuación en la búsqueda de soluciones analíticas, que desde los resultados obtenidos para la emergencia, nos invita a continuar en la inspección de expresiones con las que se puedan caracterizar y especificar las propiedades estudiadas.

# CONCLUSIONES Y TRABAJO FUTURO

## 6.1  Logros Generales

Esta tesis logró el desarrollo de un enfoque metodológico para la especificación y estudios de sistemas dinámicos con un gran número de agentes, desde la perspectiva de su auto-organización, emergencia, complejidad, homeostasis y autopoiesis. Para tal fin, se partió de una definición propia de sistema dinámico que expresa la inclusión de múltiples agentes, interacciones, organización propia, emergencias y complejidad. Desde allí se ahondó en la descripción de un sistema multi-agente como un sistema dinámico, que incorporó la teoría de los sistemas dinámicos con el abordaje de los aspectos de su complejidad. De esta manera, fue posible incluir mayores aspectos para la descripción de un sistema dinámico sobre la base de un número mayor de propiedades, más allá de la auto-organización (cómo orden) y la emergencia (como cambio). Es decir, considerando la homeostasis (como equilibrio dinámico) y autopoiesis (como autonomía). Estas propiedades, por su carácter sintético, permitieron capturar diversos aspectos de la dinámica global desde las interacciones locales. Centrados en las interacciones, se logró la formalización de un sistema dinámico como una red de agentes. Con ello, se pudo integrar la teoría de sistemas dinámicos, la teoría multi-agentes y la teoría de grafos. Estos últimos resultados teóricos sientan las bases para posteriores aplicaciones.

Como segundo propósito, la tesis propuso nuevas nociones y formalismos de auto-organización, emergencia, complejidad, homeostasis y autopoiesis. El objeto estuvo en la clarificación de estas medidas y sentar las bases para la posterior formalización. Se quiso desde un principio que las nociones y medidas representasen lo que manifestaban. El alcance de estas medidas contribuye a la desambiguación de los conceptos. No obstante, son un enfoque desde la teoría de la información, en el que se debe tener en cuenta su trasfondo probabilístico. De la emergencia resaltamos su carácter de información nueva, pues corresponde con propiedades que surgen desde las interacciones locales en una escala dada. De la auto-organización como una expresión del grado de orden, sin la participación de un control central, resaltamos su carácter opuesto a la emergencia. De la complejidad, resaltamos ser el resultado del balance entre auto-organización y emergencia, o balance entre orden y caos. De la homeostasis resaltamos su capacidad para determinar si el sistema mantuvo a través del tiempo los mismos o diferentes estados, como producto de su equilibrio dinámico. Su valor agregado está en que es útil para identificar los ritmos del sistema a través de la dinámica temporal, como muestra de su funcionamiento en su zona óptima. Finalmente, de la autopoiesis, su capacidad para enfrentar los cambios ambientales.

En cuanto a las diversas aplicaciones realizadas, es destacable que los formalismos desarrollados permiten describir muchos fenómenos en términos de información a diversas escalas. Las aplicación en redes booleanas y autómatas celulares elementales, definieron y categorizaron dinámicas ordenadas, críticas y caóticas, de forma apropiada. La aplicación en tráfico, permitió determinar de forma comparativa con el sistema tradicional de la ola verde, que el sistema auto-organizante generó para unos casos un comportamiento regular y una mayor complejidad. Lo anterior se expresó en un número mayor de fases, lo que demostró su adaptabilidad a diferentes flujos y demandas de tráfico.

Respecto a las aplicaciones en sistemas ecológicos, se puede estimar que la medición de la complejidad en ella dio interesantes resultados. En términos prácticos, las medidas propuestas contribuyen a darle significado e interpretación a la complejidad ecológica. Como hemos argumentado, en la literatura no se ha definido una única medida de complejidad. La discusión se mantiene en las diferentes interpretaciones dadas a la complejidad ecológica (complejidad funcional y estructural). Como clarificación se puede decir que acorde con Heylighen (2011, 2000), la complejidad estructural es el resultado de la diferenciación espacial y de la selección del ajuste de las interacciones entre los componentes. Por otra parte, la complejidad funcional, está relacionada con la variedad de acciones y procesos para enfrentar las perturbaciones ambientales. Ambos procesos producen una jerarquía de sistemas anidados, subsistemas o meta-sistemas, que tienden al auto-refuerzo. Desde nuestro enfoque de la complejidad como balance entre $E$ y $S$, está asociada con la caracterización de la complejidad funcional, por cuanto es producto de la dinámica de los estados de los elementos del sistema. Estas dinámicas pueden ser ordenadas (auto-organizadas), caóticas (emergentes) o críticas (complejas). Desde esta condición, los lagos en el gradiente latitudinal pueden se clasificados en uno u otro régimen. Igualmente, la medida de autopoiesis está asociada con la complejidad funcional, dado que refleja la adaptabilidad y autonomía ante los cambios ambientales en el tiempo. Un aspecto interesante que se desprende de esta conclusión, es que surgen nuevas oportunidades de contrastar la complejidad funcional con la complejidad estructural. Esta última puede ser definida desde las diversas topologías de redes de interacción propuestas por la teoría de redes. La base para esto lo constituyen los experimentos realizados con redes booleanas aleatorias, donde las dinámicas ordenadas, caóticas y críticas, estuvieron relacionadas con la conectividad.

Sobre las anteriores bases, además de los formalismos propuestos, el uso de la programación genética (ítem 5.5.5) permitió obtener expresiones analíticas que complementan el estudio de los sistemas ecológicos. Esto quedó demostrado por la exactitud del modelo desarrollado, que obtuvo valores de emergencia apropiados para un lago tropical a partir de los datos de un lago ártico, un lago con una dinámica muy diferente. Así, el enfoque de programación genética nos permitió hallar expresiones generales para caracterizar ecosistemas ecológicos, y contribuye al entendimiento de las interacciones entre procesos ecológicos de auto-organización, homeostasis y autopoiesis. La utilidad de este enfoque puede ser evaluado gradualmente en otros ecosistemas acuáticos.

Se podría pensar que la metodología desarrollada, en especial las métricas propuestas, tienen limitaciones por su generalidad. Sin embargo, vemos que esto también es una virtud, pues puede sintetizar diversos fenómenos, los cuales pueden ser descritos en términos de información. Desde las aplicaciones y los resultados obtenidos, surge la motivación y la posibilidad de hacer extensiones de la metodología a otros casos, e incluso lograr perfeccionamientos y formalizaciones adicionales.

## 6.2 Trabajo Futuro

Existen varias situaciones que sería interesante explorar, por ejemplo:

- Nuestras medidas pueden ser generalizadas para redes computacionales (Gershenson, 2010), con potenciales aplicaciones al estudio de redes complejas de agentes (Fernández et al., 2012b; Newman, 2003).
- En consecuencia con lo anterior, se puede llegar a la implementación de una plataforma de modelado basada en agentes que explore el surgimiento de las propiedades de auto-organización, emergencia, complejidad, homeostasis y autopoiesis, desde la representación de una red computacional de agentes.
- Es necesario continuar la aplicación de los formalismos propuestos, a grandes series de tiempo de sistemas reales, para ahondar más en su significado.
- La información de Shannon ha sido recientemente usada como una medida de la complejidad espacial (Batty et al., 2012), con aplicaciones en áreas urbanas. Será interesante comparar este tipo de trabajos con nuestras medidas de complejidad.
- Es necesario ahondar en el ajuste de indicadores homeostáticos y su aplicación, debido a que a que hemos notado su asociación directa con el proceso de auto-organización.
- Se debe finalizar un paquete de software libre para el cálculo y graficación ágil de las métricas propuestas. La primera versión llegó hasta la programación de una función para R, que hemos llamado *Comin* (Complexity and Information) (Villate et al., 2013).
- Nuevos casos de estudio pueden ser abordados. Por ejemplo, la complejidad biológica desde la genómica, la proteómica, la metabolómica y la enfermedómica. Desde la base de su medición, se pueden explorar los aspectos de modularidad y complejidad a múltiples escalas de organización. Por otra parte, surgen tópicos especiales como la complejidad del ritmo cardiaco en diferentes cohortes de edad, pues hemos notado, en inspecciones preliminares, que el ritmo cardiaco en los niños es más emergente que en los adultos. Nace también el interés por explorar la complejidad de ciertas patologías cardiacas.

- Y finalmente, desde la perspectiva de la complejidad como balance, creemos que podemos abordar tópicos relativos a la cognición, la cual puede tener elementos como el balance entre el aprendizaje y la memoria. El flujo, entendido como aquel estado de plenitud y complacencia en que existe felicidad, también puede ser considerado como un balance entre los retos y las habilidades. Este último hecho favorecería el aprendizaje complejo, que puede ser estimado como un nuevo modelo educativo (Fernández et al., 2014a).

# Anexo: Caracterización de los Lagos Estudiados

## 7.1 Lago Ártico (Ar)

Los lagos árticos están localizados en el círculo polar ártico. Su temperatura superficial (ST) promedio es de 3°, su máxima de 9°C y su mínima de 0°C.

En general, los sistemas *Ar* se clasifican cómo oligotróficos debido a su baja producción primaria, representada en valores de clorofila de 0.8-2.1 mg/m³. Su zona limnética, o columna de agua, está bien mezclada, lo que significa que no tiene estratificaciones (capas con diferentes temperaturas). Durante el invierno (Octubre a Marzo), la superficie del lago está cubierta de hielo. Durante el verano (Abril a Septiembre), el hielo se derrite y el flujo de agua y evaporación se incrementa. Estos dos regímenes climáticos tienen una fuerte influencia en el comportamiento hidrológico de los lagos, que a su vez influencia la físico-química y la biota.

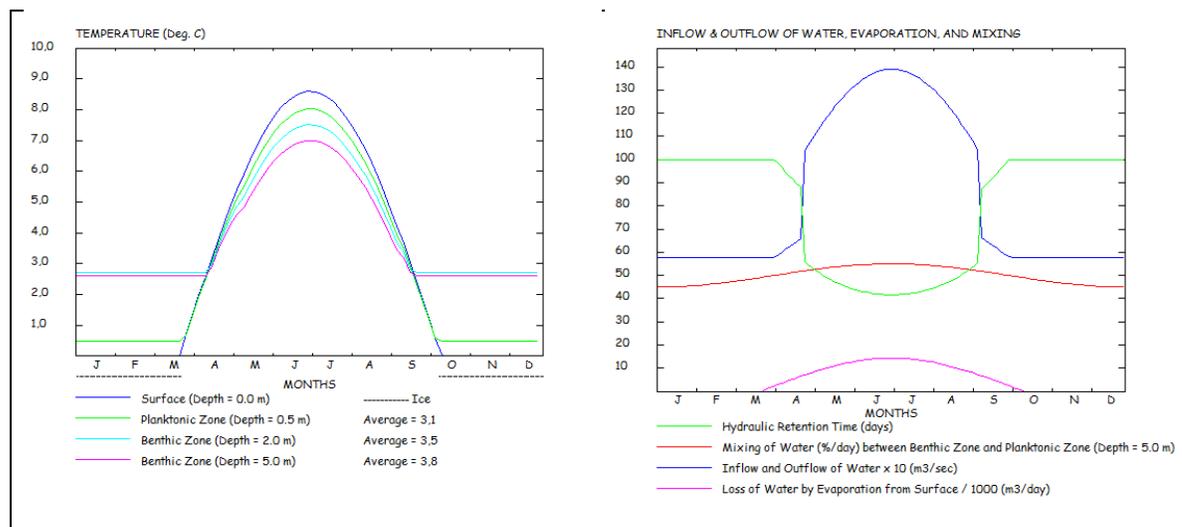

**Figura 7-1 Hidroclimatología del lago Ártico** (Randerson and Bowker, 2008). **A la izquierda la temperatura en las 3 zonas de estudio. A la derecha los flujos hidrológicos de afluente, efluente, mezcla y evaporación.**

Los nutrientes limitantes en las formas de nitratos, silicatos y dióxido de carbono están entre el 90 y 100% de disponibilidad para el fitoplancton en todo el año. El fitoplancton y perifitón están dominados por diatomeas en porcentajes de 38.6 y 45%, respectivamente. En el zooplancton dominan los herbívoros, con un 91.7% de la biomasa. En la zona bentónica, los invertebrados detritívoros alcanzan un 86.8%, y los peces piscívoros el 85.8%.

## 7.2 Lago Templado del Norte de Tierras Altas (North Higland Lake-NH).

*NH* corresponde a un ecosistema mesotrófico en una latitud templada (T° promedio de 5.3°C). Los niveles de clorofila están entre 2.2-6.2 mg/m³. La superficie está cubierta de hielo en el invierno (final de noviembre a principios de febrero). La cubierta de hielo forma una barrera para el viento, lo que minimiza la pérdida de agua por evaporación, mientras que el fondo del lago permanece sin congelar. La columna de agua no está termo-estratificada, y está permanentemente bien mezclada con niveles del 50% en verano y 90% en invierno. El flujo máximo se da en primavera y otoño (9.6 m³/s). El flujo mínimo en verano (0.6 m³/s). La evaporación se reduce debido a que el agua es más o menos fría y los gradientes de presión de vapor no son grandes (promedio= 9.262 m³/d). El tiempo de retención (*RT*) es máximo en verano con 100 días. La concentración de oxígeno es superior a 10 mg/lt en las tres capas. El *pH* ronda las 7 a 7.3 unidades, pero se mueve en rangos de 6.7 a 7.8 unidades desde la superficie al fondo.

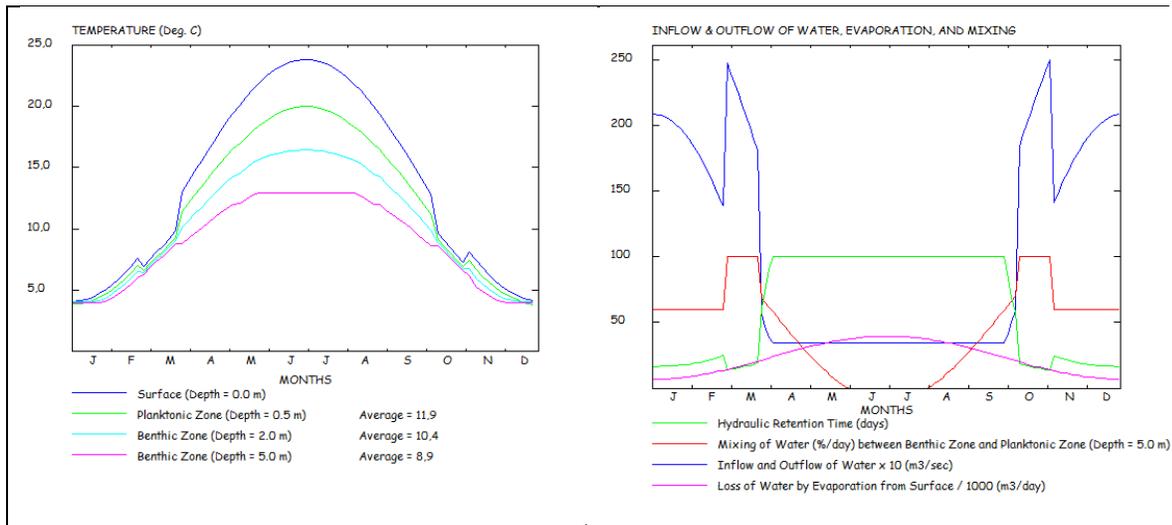

**Figura 7-2 Hidroclimatología del lago templado del norte-NH** (Randerson and Bowker, 2008)**. A la izquierda temperatura en las 3 zonas de estudio. A la derecha los flujos hidrológicos de afluente, efluente, mezcla y evaporación.**

La correlación entre variables es más estacional en *NH* que en lagos templados de tierras bajas (North Lowland-*NH*). Esto significa que el periodo de verano está relacionado con altas *RT* y altos *pH*. La estación invernal está relacionada con altos niveles de oxígeno en el flujo de entrada y salida. Sin embargo, existe una correlación más estrecha entre el oxígeno bentónico y el oxígeno del sedimento ($BO_2$, $SdO_2$).

Los nutrientes limitantes, como los nitratos y el dióxido de carbono, están en el 95% de disponibilidad para el fitoplancton. Los fosfatos y silicatos muestran variaciones y porcentajes mejores de disponibilidad. Los primeros alrededor del 80%, y los segundos cerca del 85%.

La composición de la biomasa es dominada por las diatomeas planctónicas (46.7%) y bentónicas (41%). La composición del zooplancton es dominada por el zooplancton herbívoro (91%), los carnívoros abarcan el restante 8,6%. Entre los invertebrados bentónicos, los detritívoros dominan con el 87.5%. La comunidad de peces es dominada por los peces bénticos en porcentaje de 88.9.

## 7.3 Lago Templado del Norte de Tierras Bajas (North Lowland Lake-NL).

Este lago prototipo de los lagos templados es un lago eutrófico, localizado en un clima templado del hemisferio norte. Su productividad primaria es de 6.3-19.2. Esta bajo la influencia de las estaciones de invierno, primavera, verano y otoño. En verano, las variaciones de flujo de entrada a salida caen a 3.5 de 25.2 m³/s. El tiempo de retención se incrementa en 100 días. La falta de viento y altas temperaturas (24°C), causan termo-estratificación, que se evidencia en los 24°C en la superficie, 20.6°C en la capa plantónica y 17.3°C en el bentos. Por tal motivo, el agua al ser es más densa en el fondo no se mezcla. En invierno, no se cubre de hielo la superficie, y el flujo es mínimo. En primavera y otoño la columna de agua se mezcla en un 100%, y el tiempo de retención hidráulica baja a 14 días. Esto genera un incremento en la conductividad eléctrica. En verano, las depleciones de oxígeno en las tres capas son más drásticas que en el ártico. El oxígeno está directamente correlacionado con el porcentaje de mezcla y los flujos de entrada a salida, y de forma inversa a otros parámetros cómo el *pH* y la retención hidráulica.

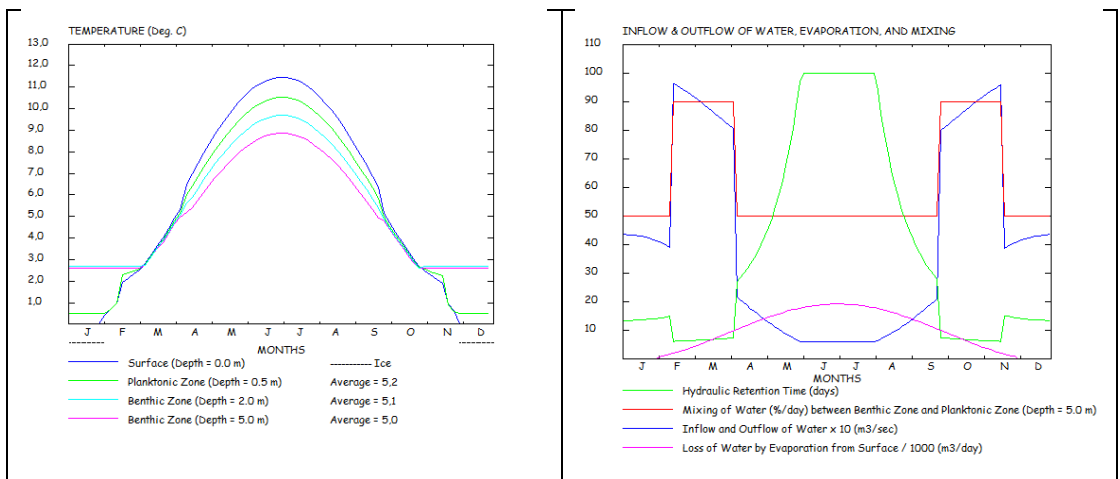

**Figura 7-3 Hidroclimatología del lago templado del norte de tierras bajas-NL** (Randerson and Bowker, 2008). **A la izquierda temperatura en las 3 zonas de estudio. A la derecha los flujos hidrológicos de afluente, efluente, mezcla y evaporación.**

Los nutrientes limitantes están por encima del 90% de disponibilidad para el fitoplancton en todas las estaciones. De acuerdo con su disponibilidad, la composición de la biomasa del fitoplancton y perifitón es dominado por diatomeas planctónicas (47%) y bentónicas (34.3%). En consecuencia, el zooplancton es 100% herbívoro. La comunidad de peces es dominada por los peces bentónicos (67%).

## 7.4 Lago Tropical (T)

*T* es un sistema híper-eutrófico (clorofila > 19.2), localizado en un clima tropical húmedo, en el norte del Ecuador. Su temperatura media es de 25°C en la capa superficial. Está sometido a una estación lluviosa y otra seca que marca los flujos máximos y mínimos, respectivamente. La alta radiación conduce a altas temperaturas en el agua y a bajas diferencias entre zonas. A pesar que se puede dar la estratificación debida al intercambio de calor entre capas, esta es menos estable que en los lagos de latitudes altas (*NH* y *NL*). Especialmente, debido a que el viento puede tener gran incidencia en la mezcla de la columna de agua.

El flujo de nutrientes y la dinámica planctónica se ven afectados por los episodios de calor y mezcla. Se destaca que la producción primaria entre lagos tropicales es de dos veces la de altas latitudes. También se ha reconocido que en *T* el factor limitante es el nitrógeno, de manera que cuando se agota los demás procesos no tienen lugar.

El equilibrio entre especies dentro del fitoplancton y perifiton es alto (33% para diatomeas, algas verdes y cianobacterias). Las poblaciones de zooplancton están dominadas por los herbívoros (90%), el bentos por invertebrados detritívoros (84%) y los peces (87%).

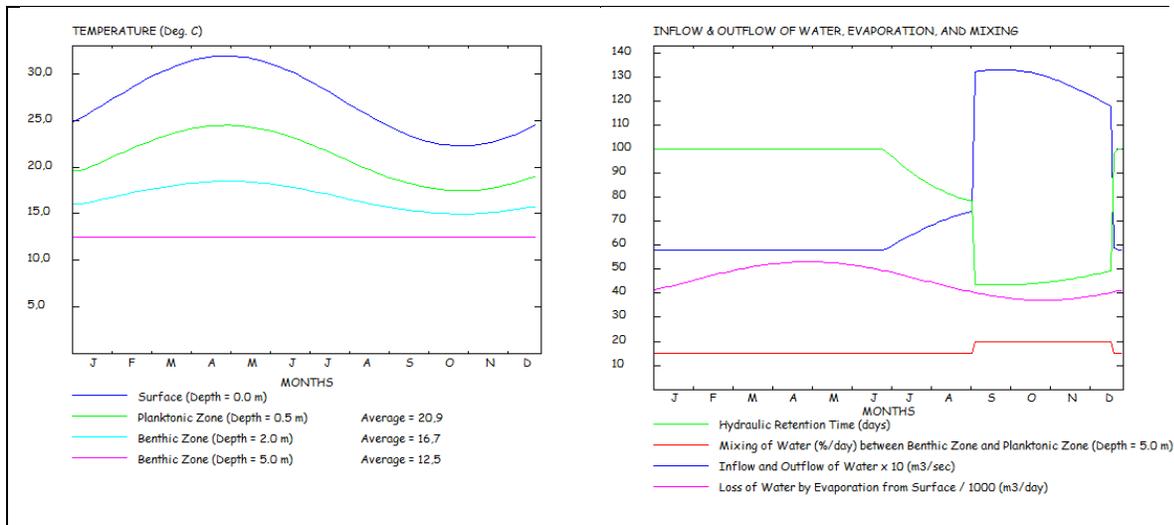

**Figura 7-4 Hidroclimatología del lago tropical. A la izquierda temperatura en las 3 zonas de estudio** (Randerson and Bowker, 2008)**.** A la derecha los flujos hidrológicos de afluente, efluente, mezcla y evaporación.

# Bibliografía